\pdfoutput=1

\documentclass[
  twocolumn,
  prb,
  showpacs,
  amsmath,
  amssymb,
  superscriptaddress
]{revtex4}

\usepackage{bm}
\usepackage{graphicx}
\usepackage{color}
\usepackage{braket}
\usepackage{ulem}
\usepackage{pifont}

\newcommand{\tr}{\text{tr}}
\newcommand{\Tr}{\text{Tr}}
\newcommand{\Sp}{\text{Sp}}

\renewcommand{\v}[1]{\textbf{\textit #1}}

\definecolor{lightgray}{gray}{.9}

\begin{document}

\title{Half-integer quantum Hall effect of disordered Dirac
fermions at a topological insulator surface}

\author{E.\ J.\ K\"onig}
\affiliation{
\mbox{Institut f\"ur Theorie der kondensierten Materie,
 Karlsruhe Institute of Technology, 76128 Karlsruhe, Germany}
}
\affiliation{
 DFG Center for Functional Nanostructures,
 Karlsruhe Institute of Technology, 76128 Karlsruhe, Germany
}

\author{P.\ M.\ Ostrovsky}
\affiliation{
Max-Planck-Institute for Solid State Research, D-70569 Stuttgart, Germany
}
\affiliation{
 L.\ D.\ Landau Institute for Theoretical Physics RAS,
 119334 Moscow, Russia
}

\author{I.\ V.\ Protopopov}
\affiliation{
 \mbox{Institut f\"ur Theorie der kondensierten Materie,
 Karlsruhe Institute of Technology, 76128 Karlsruhe, Germany}
}
\affiliation{
 Institut f\"ur Nanotechnologie, Karlsruhe Institute of Technology,
 76021 Karlsruhe, Germany
}
\affiliation{
 L.\ D.\ Landau Institute for Theoretical Physics RAS,
 119334 Moscow, Russia
}

\author{I.\ V.\ Gornyi}
\affiliation{
 Institut f\"ur Nanotechnologie, Karlsruhe Institute of Technology,
 76021 Karlsruhe, Germany
}
\affiliation{
A.\ F.\ Ioffe Physico-Technical Institute,
194021 St. Petersburg, Russia
}

\author{I.\ S.\ Burmistrov}
\affiliation{
 L.\ D.\ Landau Institute for Theoretical Physics RAS,
 119334 Moscow, Russia
}
\affiliation{
Moscow Institute of Physics and Technology, 141700 Moscow, Russia
}

\author{A.\ D.\ Mirlin}
\affiliation{
 Institut f\"ur Nanotechnologie, Karlsruhe Institute of Technology,
 76021 Karlsruhe, Germany
}
\affiliation{
\mbox{Institut f\"ur Theorie der kondensierten Materie,
 Karlsruhe Institute of Technology, 76128 Karlsruhe, Germany}
}
\affiliation{
 DFG Center for Functional Nanostructures,
 Karlsruhe Institute of Technology, 76128 Karlsruhe, Germany
}
\affiliation{
 Petersburg Nuclear Physics Institute,
 188300 St.~Petersburg, Russia.
}

\begin{abstract}

The unconventional (half-integer) quantum Hall effect for a single species
of Dirac fermions is analyzed. We discuss possible experimental measurements of
the half-integer Hall conductance $g_{xy}$ of topological
insulator surface states and explain how to reconcile Laughlin's flux insertion
argument with half-integer $g_{xy}$. Using a vortex state
representation of Landau Level wavefunctions, we calculate  current
density beyond linear response, which is in particular relevant  to the
topological image monopole effect. As a major result, the field theory
describing the localization physics of the quantum Hall effect of a single
species of Dirac fermions  is derived. In this connection, the
issue of (absent) parity anomaly is revisited. The renormalization
group (RG) flow and the resulting
phase diagram are extensively discussed. Starting values
of the RG flow are given by the semiclassical conductivity tensor which is
obtained from the Boltzmann transport theory of the anomalous Hall
effect.

\end{abstract}

\maketitle

\section{Introduction}

Topological states of matter
constitute a vibrant
field of current research. On the one hand, promising future applications---in
particular, in the fields of spintronics and quantum computation---are
expected. On the other hand, topological phases of matter provide fascinating
realizations of fundamental concepts of field theory, mathematical physics and
geometry.

Topological phases considered in the present work are fermionic topological
insulators (TIs):\cite{HasanKaneReview,QiZhang,SchnyderLudwig08,Kitaev09} materials with a band gap in
their bulk that are equipped with a ``twist'' in the structure of Bloch states.
This leads to a non-trivial topological index, and, by the bulk-boundary
correspondence\cite{Gurarie} and Callias'
theorem,\cite{Callias1978,BottSeeley1978} to protected gapless states at the
interface of two topologically distinct insulators.

The earliest example of a TI was the quantum Hall (QH) state.\cite{vonKlitzing}
The Landau-Levels (LLs) provide the bulk band gap, which is accompanied by the
topological Thouless-Kohmoto-Nightingale-den Nijs  (TKNN)
index\cite{TKNN82} and the protected chiral edge state. More recently,
time-reversal (TR) invariant two- and three-dimensional (2D and 3D) TIs were
discovered.\cite{BernevigZhang, BernevigHughesZhang, FuKaneMele, MooreBalents,
Roy, KonigMolenkampetc, Hsieh} In contrast to the TKNN integer, their
topological index takes only values in $\mathbb Z_2$. The boundary states of a
3D TI represent a single species of 2D Dirac fermions.

Alternative descriptions of TIs are topological field theories. These include,
first, the theory of electromagnetic (EM) gauge potentials, and, second,  the
diffusive non-linear sigma model (NL$\sigma$M). In contrast to the Bloch-band
description, these theories capture the general interacting problem with
quenched disorder. In the case of the integer quantum Hall effect (QHE), the
effective bulk EM theory contains a Chern-Simons (CS) term,\cite{Zhang1992,
Wen1990} whereas the field theory describing the localization physics in
the bulk is the NL$\sigma$M supplemented with a theta
term.\cite{LevineLibbyPruisken1983,Pruisken1984,PruiskeninPrangeGirvin}
At this point, it is worth reminding the reader that one of key ingredients of
the QH physics is the disorder-induced Anderson localization of
bulk states. \cite{Huckestein1995,KramerOhtsukiKettemann2005,EversMirlin2008,
AltlandSimons2010} Both theories (EM and diffusive) can be unified within the
framework of the $\mathbf{U}(1)$-gauged NL$\sigma$M. \cite{MishandlingI}

In the language of topological EM field theory, the bulk of TR invariant 3D TIs
is characterized by the $\v E \cdot \v
B$-term\cite{QiZhang,EssinMooreVanderbilt2009,EssinVanderbilt2010,
RyuMooreLudwig2012} with theta angle $\vartheta = \pi$~(mod~$2\pi$). Terminating this
bulk theory at the 3D TI boundary, one could therefore naively expect a CS
surface theory. However, this would imply
a surface Hall conductance
$\sigma_{xy} \stackrel{?}{=} \left (\frac{1}{2} \mod 1\right )
\frac{e^2}{h}$
and thus would be unphysical for two reasons:
(i) the Hall conductance should be unambiguously defined, and
(ii) its value should be zero
in the presence of TR invariance.
The absence of a
CS-term in the surface theory of TR invariant 3D TI was shown employing
BF-theory,\cite{ChoMoore2011} by straightforward integration of fermions
including the bulk states,\cite{MulliganBurnell2013} by means of the general
conjecture of cancellation of anomalies\cite{ZirnsteinRosenow2013} and by
investigating the $\mathbf{U}(1)$-gauged diffusive NL$\sigma$M, i.e. the unified
topological field theory of gauge potentials and
diffusive soft modes.\cite{KoenigMirlin2013}

In this paper we use the $\mathbf{U}(1)$-gauged NL$\sigma$M to explore the situation when
TR symmetry is broken locally\footnote{For the notion of ``local TR symmetry
breaking'', see Eq.~\eqref{eq:localTRbreaking}.} on the surface of a 3D TI, or
more generally, the QHE of a single Dirac fermion.\cite{Schakel1991} The QHE of
Dirac fermions in the context of
graphene\cite{NovoselovFirsov2005,ZhangKim2005,GusyninSharapov2005,Altland2006,
OGM2008,CastroNetoGeim2009,Goerbig2011} and 3D TI surface states
\cite{BrueneMolenkamp2011,FuKane2007,Lee2009,LiuZhang2010,NomuraNagaosa2011,
ZyuzinBurkov2011,Vafek2011,TseMacDonald2011,LiXing2011,ChuShiShen2011,
YangHan2011,SitteFritz2012,ZhangWangXie2012,BaasanjavNomura2013} has been
studied both theoretically and experimentally. The QH state is
characterized by vanishing longitudinal conductance $\sigma_{xx} = 0$ and
quantized transverse conductance taking values
\begin{equation}
\sigma_{xy} = g_D\left (\nu+\frac{1}{2}\right )\frac{e^2}{h}, \; \; \nu \in \mathbb Z. \label{eq:QHEg}
\end{equation}
Here $g_D$ denotes the number of degenerate Dirac cones, i.e. $g_D=4$ for
graphene and $g_D=2$ for thin 3D TI slabs. In particular the $\sigma_{xy} = \pm
g_De^2/2h$ states turns out to be extremely robust,\cite{NomuraFurusaki2008} they
can be observed up to room temperature\cite{NovoselovGeim2007} and can also be
induced by pure exchange coupling (quantum anomalous Hall
effect\cite{Haldane1988,OnodaNagaosa2003,LiuZhang2008,OGM2007,YuFang2010,
ChangKun2013}).

Notwithstanding the immense general interest towards the subject, the single
Dirac fermion QHE [$g_D = 1$ in Eq.~\eqref{eq:QHEg}] did not enjoy the deserved
and required attention. The following important questions were not or only
partly answered to present date:

(i) How can half-integer $g_{xy} \equiv h \sigma_{xy}/e^2$ be measured experimentally?

(ii) Does not Laughlin's flux insertion argument\cite{Laughlin1981,Halperin1982} 
forbid $\sigma_{xy} = \left (\nu+\frac{1}{2}\right )\frac{e^2}{h}$?

(iii) What is the field theory describing the localization physics of the
single-species Dirac-fermion QHE?

In this work we present a comprehensive analysis of these questions and
detailed answers to them.

While our primary consideration leading to the answers on the posed questions is
very general and is based on topology and gauge invariance, important physical
insight can be gained by a microscopic analysis of simple models. Thus, we
supplement our analysis by  two complementary semiclassical calculations of the
conductivity tensor of Dirac fermions in magnetic field. The first one is based
on the vortex-state representation of LL wavefunctions\cite{ChampelFlorens2007,
ChampelFlorens2010,HernangomezPerezFlorensChampel2014} and addresses the situation of potential disorder which is
smooth on the scale of the magnetic length. In particular, this calculation is
also applicable beyond linear response. The second one is based on the Boltzmann
transport theory of the anomalous Hall effect (AHE)\cite{Sinitsyn2008,
NagaosaOng2010} and, as usual, applies when the kinetic energy of charge
carriers exceeds the scattering rate. The semiclassical (Boltzmann) conductivity
tensor provides starting values for the RG dictated by the field theory
discussed in the context of question (iii).

The paper is structured as follows. In Sec.~\ref{sec:TME}, which concerns
question (i), we review and clarify the physics of topological
magnetoelectric effect (TME). In Sec. \ref{sec:Laughlin} we answer question (ii)
regarding Laughlin's argument. Section \ref{sec:Semiclassical} contains the
first semiclassical calculation of current density (vortex states).
In Sec. \ref{sec:NLSM} we derive the unified field theory treating both
diffusive matter fields and EM gauge potentials [question (iii)]. Subsequently,
in Sec. \ref{sec:startingvalues}, we present the second semiclassical
(Boltzmann) calculation of the conductivity tensor for gapped Dirac fermions in
magnetic field and examine the phase diagram of the problem. The renormalization
group fixed points of the field theory bring us back to the TME and question
(i), motivating a discussion of experimental conditions in
Sec.~\ref{sec:experiment}. We close the article with a summary of obtained
results and an outlook, Sec.~\ref{sec:outlook}.

\section{(Half-)integer QHE and topological magnetoelectric effect}
\label{sec:TME}

This section and section \ref{sec:experiment} are devoted to question (i) posed
in the introduction. To make the paper self-contained, we begin by
briefly reviewing and clarifying the current state of the literature.

\subsection{The QHE of a single Dirac cone in condensed matter reality: 3D TI}

The appearance of the single Dirac fermion on the 3D TI surface crucially relies
on TR symmetry. Therefore, two questions arise concerning the realization of the
single Dirac fermion QHE on the 3D TI surface:

a) Up to which magnetic field strength do surface states exist?

b) If surface states are present, does the half-integer quantization of $g_{xy}$ immediately follow?

Regarding question a), we recall that a 3D TI is characterized by the inverted
structure of the energy bands  which can be captured by the  $\vec k$-dependent
mass term
$\mathcal M\left (\vec k\right )$ in its effective 3D Dirac-like bulk Hamiltonian
\begin{equation}
\mathcal M\left (\vec k\right ) = M - B_1 k_z^2 - B_2 \v k^2.
\end{equation}
Here we follow the notation of Eq. (31) in Ref.~\onlinecite{QiZhang}, assume positive $M$,  $B_1$ and $B_2$ and denote (2D) vectors by bold italic symbols.

The 2D interface (which we assume for concreteness to occupy $z=0$ plane)
between a 3D TI and a topologically trivial  insulator (e.g. vacuum) can be
modeled by spatially dependent Dirac mass $M = M(z)$ which interpolates between
positive (topological phase) and negative (trivial phase) values changing its
sign at $z=0$.
As a consequence of the band inversion in the topological insulator the
interface supports massless Dirac fermions in the vicinity of $\v
k=0$.\cite{Callias1978,BottSeeley1978,Volkov1985}

The magnetic field $\v B=(0,0,B)$ applied to the interface can not destroy the
surface states provided the bulk gap is sufficiently large.
More precisely, for
\begin{equation}
M \vert_{z = \infty} > B_2 /l_B^2 \label{eq:localTRbreaking}
\end{equation}
the massless surface excitations give rise to a zero energy Landau level (LL)
localized at the  interface $M(z) - B_2 /l_B^2=0$.
Here $l_B = \sqrt{{\hbar}/{\left (\vert e \vert B\right )}} \approx {26
nm}/{\sqrt{B [T]}}$ is the magnetic length.
In the exemplary case of Bi$_2$Se$_3$, we can estimate\cite{QiZhang} $B_2 /l_B^2
\sim 0.6 meV \times B [T]$, while $M \sim 0.3 eV$.

In the rest of the paper (and consistently with the previous works) we will use the term  ``\textit{local} breaking of TR symmetry on the 3D TI surface'' if the magnetic field does not destroy the surface states. As we have just explained,  this does not necessarily require spatially inhomogeneous magnetic field configurations.

Let us now turn to question b). To avoid confusion,  we stress that the physics
of the half-integer QHE, which we discuss in this work, can be described in a
single particle picture and  has no direct relation  to the physics of the
fractional QHE (in the sense of St\"ormer's and Tsui's discovery)
which is a many-body phenomenon.

The half-integer QHE can be expected as soon as surface states are present.
We  emphasize, that it does not rely on a precise dispersion but rather on
the low-energy spin-texture and on the fact that there is an odd number of
Dirac fermions on the surface.
More precisely, the half-integer QHE is a manifestation of fermion-number
fractionalization in the sense of Jackiw and
Rebbi.\cite{JackiwRebbi1976,Jackiw1986} The 3D TI surface states
are topologically protected fermionic zero modes associated with a spatial
kink in a background bosonic field (the mass ``field'' in the present context).
Generally, the fermion number in the presence of a bosonic kink is shifted by
one half as compared to the situation without a kink. Specifically, if the zero
mode is filled (empty), the fermion number is $1/2$ ($-1/2$). In the presence of
the magnetic field, the zero-energy state has additional LL degeneracy
$\frac{BA}{\Phi_0}$, where $A$ is the area penetrated by the flux and $\Phi_0 =
h/e$ is the flux quantum. Consequently,  the fermion density at chemical
potential $\mu = 0^+$ is $n = N/A = \frac{B}{2\Phi_0}$. In view of the
relationship between fermion density and (the quantum part of) the transverse
conductivity,\cite{SmrckaStreda1977} this unveils the fundamental topological
reason for the half-integer QHE. The result of argument remains unchanged even
in the presence of finite but small Zeeman energy $E_Z \ll M$ (gapped 3D TI
surface states).

\subsection{Can one measure a half-integer $g_{xy}$ in a transport experiment?}
\label{sec:Notransport}

Typical transport experiments are carried out on 3D TI slabs, which have two
major surfaces (called ``top'' and ``bottom'' in what follows) with a single
Dirac fermion each. In most experimental situations, contacts are attached at or
near the side walls of the probes, and thus both major surfaces are probed
simultaneously. Therefore, quantum Hall data\cite{BrueneMolenkamp2011} of 3D TIs
displays the odd-integer series described by Eq.~\eqref{eq:QHEg} for $g_D = 2$.


\begin{figure}
\includegraphics[scale=.65]{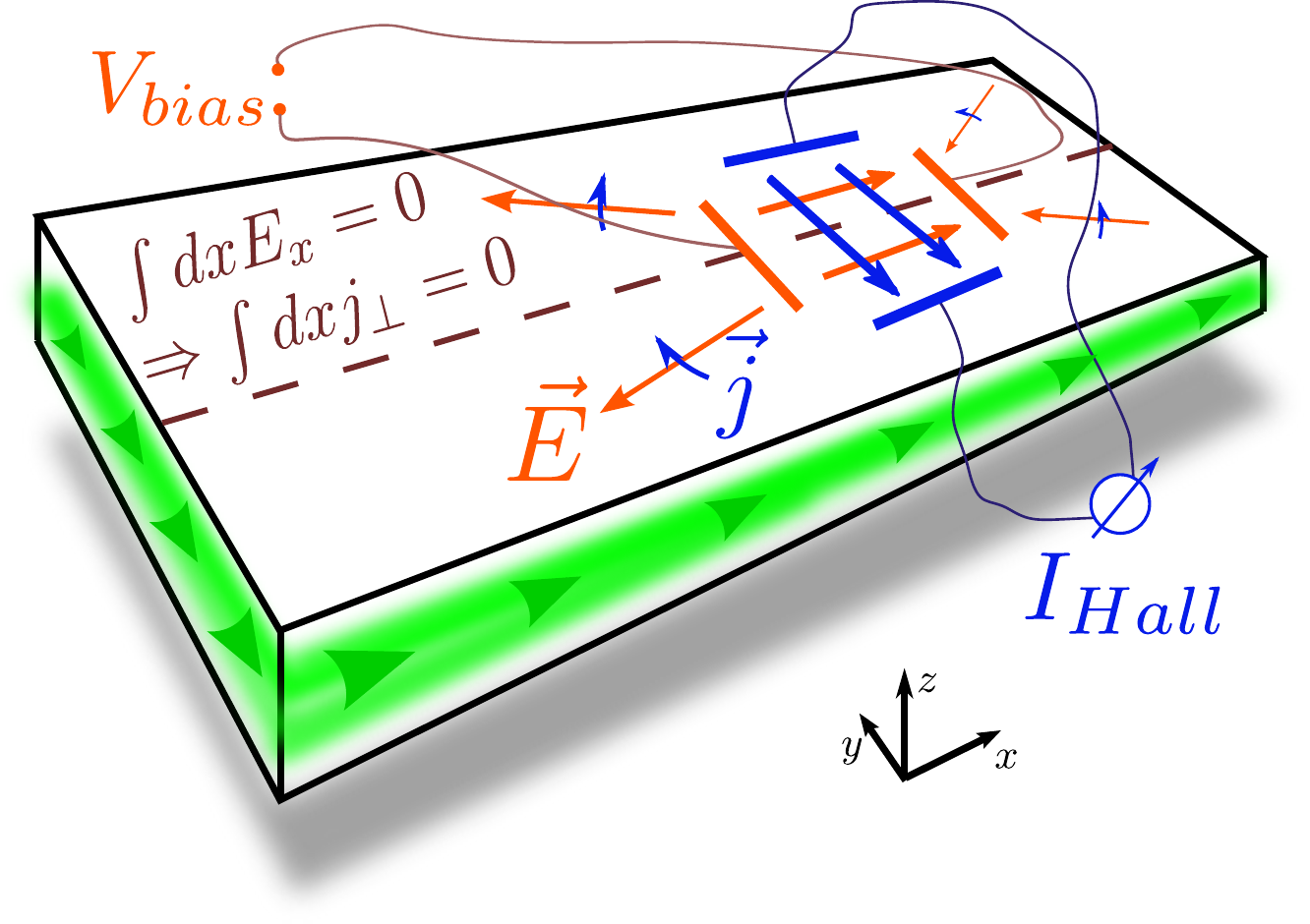}
\caption{Failure of transport measurement of half-integer Hall response. A thin 3D TI slab in a QH state $\sigma_{xy}^{top}= \sigma_{xy}^{bottom} = \frac{e^2}{2h}$ (one shared green edge channel) is probed by a local four contact measurement consisting of two opposite bias gates (orange) and, perpendicularly, two probing gates connected by an amperemeter (blue). For further explanation, see main text.}
\label{fig:NoTransport}
\end{figure}


One could
expect that it is sufficient to attach all measuring contacts
on a single surface of the 3D TI slab to measure the QHE of a
single Dirac fermion.\cite{ChuShiShen2011} If the contacts are
sufficiently far away from the sample boundaries, one might then hope to measure
a half-integer Hall response. To be specific, let us assume that a bias
voltage is applied between two electrodes attached to a TI surface, as
depicted in Fig.~\ref{fig:NoTransport}. One measures the Hall current
passing through an amperemeter connecting two perpendicular probing contacts
and hopes to extract a half-integer $\sigma_{xy}^{top}$ from
$I_{Hall}/V_{bias}$. However, this attempt will fail. Indeed, let us assume
that the surface is characterized by a half-integer quantized Hall
conductivity and zero longitudinal conductivity. In order to find the total
current between the current probes one should take into account not only the
current flowing in the part of the TI surface between the contacts but also the
current distribution in the rest of the surface. The total current can be found
by integrating the transverse current density $\int d \v l \times \v j$ along a
contour shown by the dashed line in Fig.~\ref{fig:NoTransport}. This integral
is, however, proportional to $\int d \v l \cdot \v E$ and is equal to zero,
since the surface is terminated by a metallic edge  which represents an
equipotential line. Thus, such an experiment would yield $I_{Hall} = 0$.

The above discussion assumed applying bias voltage and measuring current. One can equally analyze the reverse situation when a current is injected and the Hall voltage is probed. To this end, two metallic contacts are supposed to be attached in the central region of the TI surface. They serve as source and drain for the current. However, as in the $\sigma_{xx} = 0$ limit current always flows along equipotential lines, it is actually not possible to inject current in the middle of a QH system.  Instead, ``edge states'' circulating around the contact will be formed. Therefore, this measurement will yield a null result as well.

Thus, an attempt to measure a half-integer quantized $g_{xy}$ in a transport
experiment fails. The reason for this is as follows. To measure directly a
half-integer $g_{xy}$, one should explore local characteristics of a single
TI surface. As is clear from the above analysis, transport experiments do not
satisfy this requirement. One can, however, devise an alternative approach by
measuring an electromagnetic response of the system to a local perturbation. As
discussed below, this kind of measurement does probe local properties of the
system and, therefore, is able to yield directly a half-integer Hall
conductivity.

The case of transport experiments in non-ideal situations ($\sigma_{xx} \neq 0$) and with a more complex arrangement of contacts is left for future investigation.

\subsection{Topological electromagnetic field theory and TR invariant 3D TI}

In a series of papers,\cite{QiHughesZhang2008,QiZhangMonopole,WangQiZhang2010}
S.-C. Zhang and coworkers proposed to characterize 3D TIs by associated
electromagnetic field theories. In particular, they argued that the
corresponding bulk EM theory contains the $\v E \cdot \v B$ term (second Chern
character):
\begin{subequations}
\begin{eqnarray}
S_{\vartheta} &=& \frac{\vartheta}{2\pi} \frac{\alpha}{16 \pi} \int d^3 x dt \epsilon^{\mu\nu\rho \tau} F_{\mu\nu}F_{\rho\tau} \\
		&=& \frac{\vartheta}{2\pi} \frac{\alpha}{2 \pi} \int d^3 x dt \v E \cdot \v B \\
		&=& \frac{\vartheta}{2\pi} \frac{\alpha}{4 \pi} \int d^3 x dt \epsilon^{\mu\nu\rho \tau} \partial_\mu \left (A_{\nu}\partial_\rho A_{\tau}\right ). \label{eq:SthetaCS}
\end{eqnarray}
\label{eq:Stheta}
\end{subequations}
Here $\alpha = e^2/c\hbar$ denotes the fine structure constant of QED. If not
specified otherwise, we set the speed of light and Planck's constant to unity $c
= \hbar = 1$ in the entire paper. Greek indices label space-time coordinates.
Since this term leads to non-trivial constituent equations, the authors of
Ref.~\onlinecite{QiHughesZhang2008} coined the term ``topological
magnetoelectric effect'' (see also Sec.~\ref{sec:experiment}, below).  As can be
seen from Eq.~\eqref{eq:SthetaCS}, the $\v E \cdot \v B$ term
\begin{itemize}
\item is proportional to a quantized (topological) integral: $S_{\vartheta} =
\vartheta n$ (where $n\in \mathbb Z$) if the base manifold has no boundary.
Then, TR invariance restricts $\vartheta$ to values $0$ or $\pi$~(mod~$2\pi$).
\footnote{Even though $S_{\vartheta} \rightarrow - S_{\vartheta}$ under TR, since $S_{\vartheta} = \vartheta n$ the partition function is invariant for $\vartheta = 0$~(mod~$2\pi) \text{ or } \vartheta =\pi$~(mod~$2\pi$).}
\item is intimately related to the CS term on a possible boundary and thus to the QHE.
\end{itemize}
From the viewpoint of topological EM field theory, a TR invariant 3D TI is defined by the presence of $S_{\vartheta}$ with $\vartheta = \pi$~(mod~$2\pi$) in the bulk.

In the presence of a boundary, a naive termination of Eq.~\eqref{eq:SthetaCS}
would lead to the CS term  (for definiteness, we here consider a 3D TI in the
half space $z <0$):
\begin{equation}
S_{CS} = \frac{\vartheta}{2\pi} \frac{\alpha}{4 \pi} \int d^2 x dt \epsilon^{\nu\rho \tau} A_{\nu}\partial_\rho A_{\tau} \label{eq:CS}
\end{equation}
on the surface. The value  $\vartheta = \pi$~(mod~$2\pi$) corresponds to the
surface Hall conductivity $\sigma_{xy} = \left (\frac{1}{2} \mod 1\right )
 e^2/h$, with uncertainty in an integer multiple of $e^2/h$. Here the following
questions arise. First, the Hall conductivity is a measurable quantity and
should be defined unambiguously. Second, any non-zero Hall conductivity is in
conflict with time-reversal invariance of the system.

In a number of recent works\cite{MulliganBurnell2013, ZirnsteinRosenow2013,
KoenigMirlin2013} it was shown  that the  CS term is in fact absent on the
surface of a TR invariant 3D TI unless the TR symmetry is explicitly broken on
the surface.  We will return to this issue and the closely related  question of
parity anomaly\cite{NiemiSemenoff1983, RedlichPRL1984, RedlichPRD1984,
AlvarezGauméWitten1984} in  Sec.~\ref{sec:NLSM:parity}.

\subsection{Local TR breaking: Topological magnetoelectric effect}

While the EM theory describing a surface of a TR-invariant 3D TI does not
contain a CS term, an elegant TME description is recovered once TR invariance is
locally broken. Let us emphasize that a TME response associated
with $\v E \cdot \v B$ term is a general property of QH systems. The special
feature of 3D TI surfaces (with locally broken TR invariance) is in a
half-integer value of the associated Hall conductance.

The most prominent physical manifestations of TME include topological
Faraday and Kerr rotations\cite{HuangSikivie1985, QiHughesZhang2008, Karch09}
and the image magnetic monopole effect \cite{Sikivie84, QiZhangMonopole, Karch09, PesinMacDonald2013} (see also Sec.~\ref{sec:experiment}).
In this work we concentrate on the latter. The essence of the effect is that an
electric charge $Q$ placed above a QH system (posed in the plane $z=0$) induces
an inhomogeneous magnetic field configuration which can be described by a mirror
magnetic monopole.

To obtain the electromagnetic field developed in the system in response to the charge $Q$ we introduce  the electric and magnetic field strengths
$\v E_a$ and $\v H_a$ together with electric and magnetic inductions $\v D_a$ and $\v B_a$.  The index $a=1$ ($a=2$)
refers to  the half-space $z>0$ ($z<0$) separated by the QH system. They satisfy
the standard boundary conditions at the $z=0$ plane
\begin{equation}
\begin{array}{rcl rcl}
\left (\v D_1 - \v D_2\right )_z &=& 4\pi J_{0}, &
\epsilon_{ij} \left (\v E_2- \v E_1\right )_j &=& 0, \\
\left (\v B_1 - \v B_2\right )_z &=& 0, &
\epsilon_{ij} \left (\v H_2- \v H_1\right )_j &=&  4\pi J_{i} .
\end{array}
\label{eq:continuityoffields}
\end{equation}
Here and throughout the paper $i,j$ denote spatial indices $x$ and $y$, and
$\epsilon_{ij}$ is the antisymmetric tensor of rank two defined by
$\epsilon_{xy} = 1$. Further,
$J_0$ and $\v J$ in Eq. (\ref{eq:continuityoffields}) represent the charge
density $\rho_{3D} = J_{0} \delta \left (z\right )$ and current density $ \v J_{3D} = \v J \delta(z)$ in the QH system.

The image magnetic monopole effect can be understood from two equivalent
perspectives. One approach (which  we call the ``orthodox'' theory) utilizes the
linear response theory\cite{TseMacDonald2011} of
the QH state, while the other views the QH plane as a domain wall of $\v E \cdot
\v B$ term.  We review both these approaches below.

\subsubsection{Orthodox description of TME: surface currents.}
In the ``orthodox'' approach the inductions $\v D_a$ and $\v B_a$ are related to $\v E_a$ and $\v H_a$ via the
permittivity $\epsilon_a$ and permeability $\mu_a$ of the media surrounding the QH plane in  half-spaces $a=1,2$
\begin{equation}
 \v D_a = \epsilon_a \v E_a, \qquad\v H_a = \frac{\v B_a}{\mu_a}\,.
\end{equation}

On the other hand the linear response theory of the QH state gives
\begin{subequations}
\begin{eqnarray}
J_{0} &=& \sigma_{xy} B_z, \label{eq:QHlinearresponseK0}\\
J_{i} &=& \sigma_{xy} \epsilon_{ij} E_j. \label{eq:QHlinearresponseKi}
\end{eqnarray}
\label{eq:QHlinearresponse}
\end{subequations}
 Since $B_z$ and $\v E_{\|}$ are continuous, it does not matter whether we
associate the terms proportional to $\sigma_{xy}$ to fields stemming from region
$z>0$ or $z<0$.

The non-trivial continuity conditions can now be presented as follows
\begin{subequations}
\begin{eqnarray}
\left [\v D_1 - \left (\v D_2+4\pi \sigma_{xy} \v B\right )\right ]_z &=& 0, \label{eq:conservativeconteq1} \\
\epsilon_{ij} \left (\v H_2- 4\pi \sigma_{xy} \v E - \v H_1\right )_j &=&  0. \label{eq:conservativeconteq2}
\end{eqnarray}
\label{eq:conservativeconteq}
\end{subequations}

As we are going to discuss, these conditions imply formation of image electric
and magnetic charges whose values are controlled by the
Hall conductivity of the QH system.

\subsubsection{Theory with $\v E \cdot \v B$ term}

Instead of considering currents $J_\mu$, we can 
include a QH system into the electromagnetic theory
as a domain wall\cite{Sikivie84} of $\v E \cdot \v B$ terms with
theta angles sufficing $\vartheta_2 - \vartheta_1 = \vartheta = (2\pi)^2\sigma_{yx}/\alpha$.
In the bulk regions $a = 1,2$ we obtain the
relations \cite{QiZhangMonopole, Karch09}
\begin{eqnarray}
\v D_a &=&\epsilon_a \v E_a - \frac{\vartheta_a}{2\pi} 2\alpha \v B_a , 
\notag
\\
\v H_a &=&\frac{\v B_a}{\mu_a} + \frac{\vartheta_a}{2\pi} 2\alpha \v E_a\,,
\end{eqnarray}
leading to the same continuity conditions as Eqs.~\eqref{eq:conservativeconteq}.

As was first discovered in the eighties,\cite{Sikivie84} these continuity conditions
imply the mirror magnetic monopole effect. Assuming for simplicity  $\epsilon_1
= \epsilon_2$ and $\mu_1 = \mu_2$,
one finds the  magnetic ($g$) and electric ($q$) mirror charges
\begin{equation}
g = Q \frac{\left (\frac{\vartheta \alpha}{2\pi}\right )}{1+\left (\frac{\vartheta \alpha}{2\pi}\right )^2} , \qquad q = -Q \frac{\left (\frac{\vartheta \alpha}{2\pi}\right )^2}{1+\left (\frac{\vartheta \alpha}{2\pi}\right )^2}. \label{eq:Mirrorcharges}
\end{equation}
Physically, the inhomogeneous magnetic field is created by the non-uniform,
circular QH currents\cite{QiZhangMonopole} emerging in response to the radial
electric field in the QH-system.
This $\v B$-field induces locally varying charge density [see
Eq.~\eqref{eq:QHlinearresponseK0}] which again
leads to a radial electric field. Summing up the corresponding geometric series
one finds  both  $g$ (starting from linear order in $\vartheta$) and
$q$ (starting from quadratic order in $\vartheta$).

Contrary to transport experiments (see Sec.~\ref{sec:Notransport}), the image
charge experiment does probe directly the local value of $g_{xy}$. Therefore,
the image magnetic monopole can be used to measure a half-integer $g_{xy}$, as
was first proposed in Ref.~\onlinecite{QiZhangMonopole}.
Clearly, the monopole character of the magnetic field persists only in the 2D
``bulk'' of the QH system,
 in finite systems the magnetic field lines always close.\cite{QiZhangMonopole, SunKarch2011}
 In Sec.~\ref{sec:experiment} we will return to the image monopole effect:
we there further generalize the problem to a double layer of QH systems,
e.g. a thin 3D TI slab.

\section{Laughlin argument}
\label{sec:Laughlin}

\subsection{Phenomenology}
\label{sec:Laughlin:Phenomenology}

This section is devoted to question (ii) of the introduction: 
Is the half-integer Hall conductance of a single Dirac fermion compatible with Laughlin's flux-insertion argument,
according to which the integer QH conductance
is a direct consequence of gauge invariance?\cite{Laughlin1981,Halperin1982}

\begin{figure}
\includegraphics[scale=.6]{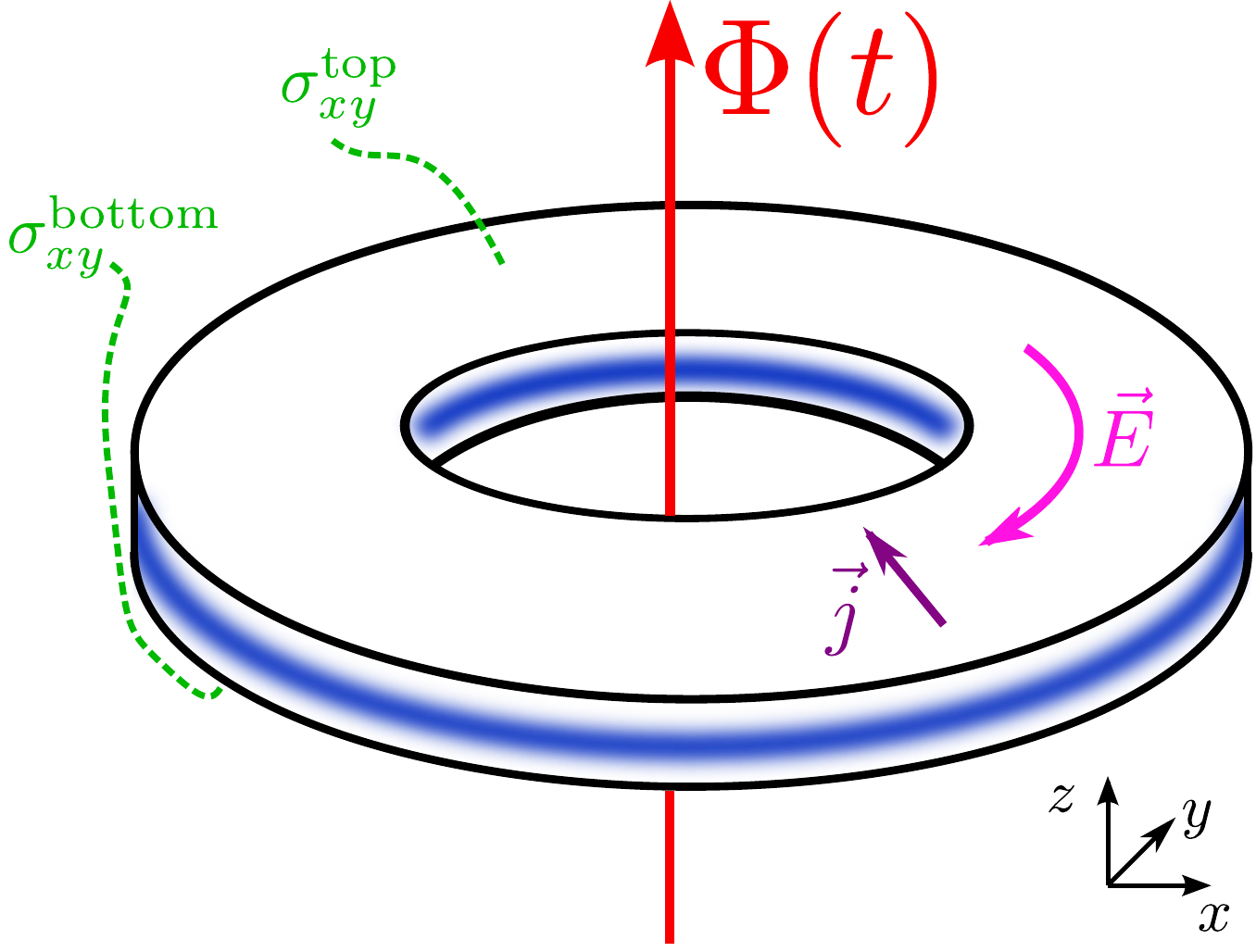}
\caption{Since the surface of a 3D TI is itself boundaryless, the modified setup for the flux insertion argument involves a torus of 3D TI surface states. (Here, $\vec{j}$ is shown for the exemplary case of $\sigma_{xy}^{top}>0$.)}
\label{fig:Laughlin}
\end{figure}

In its conventional form,\cite{Halperin1982} the argument assumes a QH film in
an annular geometry and a time dependent flux threading the ring's hole.
However, as a consequence of Nielsen-Ninomiya theorem,\cite{NielsenNinomiya1981}
a film of a single Dirac fermion cannot be realized in a condensed matter
system. 
Therefore, it is inevitable to modify the setup of the
gedanken experiment. The simplest and most direct modification is a doughnut
shaped 3D TI,\cite{Lee2009,Vafek2011} see Fig.~\ref{fig:Laughlin}. The
unavoidable change of the setup constitutes the crucial difference to the
original argument.

The setup in Fig.~\ref{fig:Laughlin} depicts the 3D TI in a QH state determined
by $\sigma_{xy}^{top}$ and $\sigma_{xy}^{bottom}$. If $\sigma_{xy}^{top} +
\sigma_{xy}^{bottom} \neq 0$,\footnote{Due to opposite orientation, the major
two 3D TI surfaces are in the same QH state if $\sigma_{xy}^{top} =-
\sigma_{xy}^{bottom}$.} chiral boundary modes appear at inner and outer
perimeters of the slab annulus (blurred blue lines). Most naturally, this occurs
if the QH state is created by an orbital magnetic field in z-direction. (The 3D
TI surface Dirac fermions are not gapped on the side walls.)

In the process of the gedanken experiment the flux passing through the hole is slowly ramped up by one
flux quantum in the period $T$, e.g. $\Phi\left (t\right ) = \frac{2\pi t}{T} \frac{\hbar}{e}$.
An azimuthal electric field and corresponding electromotive force $\mathcal E =
- \frac{d \Phi}{dt}$  are created
inducing a radial current $I = \sigma_{xy} \mathcal E$.
Over the period $T$ an overall charge $\Delta Q = \left (\sigma_{xy}^{top} +
\sigma_{xy}^{bottom}\right ) \frac{h}{e}$
is transferred between the two perimeters.

The 2D gauge potential associated to the flux piercing the hole is
 $A_i = - \frac{\Phi\left (t\right )}{2\pi} \partial_i \phi$ ($\phi$
is the azimuthal angle in 2D polar coordinates). At $t=T$ this is a pure gauge and can be
removed by a (large) gauge transformation.
\footnote{We will return to the role of large
gauge transformations below, in the context of the parity anomaly, Sec.~\ref{sec:NLSM:parity}.}
Thus, the electronic states  at $t = 0$ and at $t = T$ are  actually the states
of the same system (with $\Phi=0$)
and the charge $\Delta Q$ is the charge of its edge excitation.  In a non-interacting system,
all states have integer charge and thus $\Delta Q = \text{integer} \times e$.
As a  consequence,  $\left (\sigma_{xy}^{top} + \sigma_{xy}^{bottom}\right )$ is restricted to
integer multiples of $e^2/h$, in full accordance with the half-integer QHE on a 3D TI surface.


\subsection{Edge states, spectral flow, and microscopics}
\label{sec:Laughlin:Micro}

In Sec.~\ref{sec:Laughlin:Phenomenology} we discussed the
adiabatic flux insertion from the macroscopic point of view and came to the conclusion that
the sum  $\left (\sigma_{xy}^{top} + \sigma_{xy}^{bottom}\right )$ is quantized to integer values.
 We now turn to refinements, by means of which we can understand
half-integer quantization of $g_{xy}^{top}$ and $g_{xy}^{bottom}$.

\begin{figure}
\includegraphics[scale=.8]{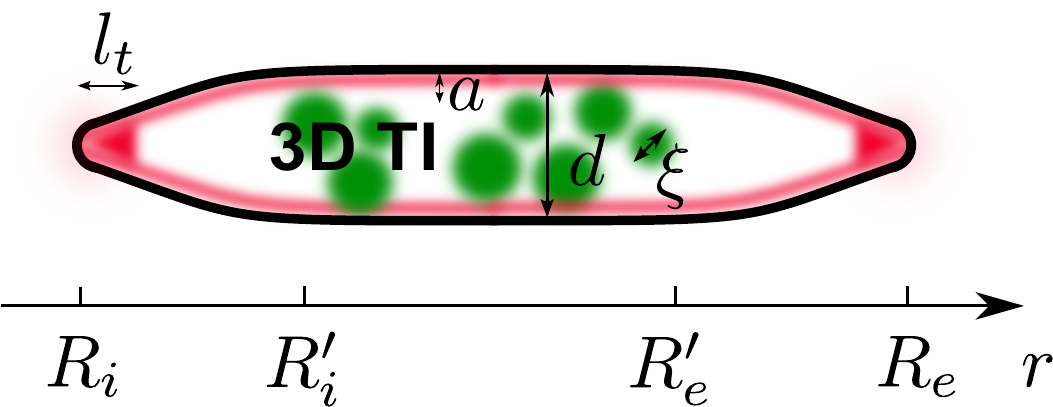}
\caption{Cross-section of the 3D TI torus depicted in Fig.~\ref{fig:Laughlin} in a plane perpendicular
to the azimuthal unit vector. Here $r = \sqrt{x^2 + y^2}$.
For the discussion of length scales and boundary conditions, see main text.}
\label{fig:BoundaryConditions}
\end{figure}

To this end, it is necessary to specify the actual nature of the edge states
 (blurred blue regions in Fig.~\ref{fig:Laughlin}),
between  which the charge $\Delta Q$ is transfered.
We remind the reader  that, due to the Klein tunneling phenomenon,  Dirac electrons cannot be confined by application of scalar potential.
A physical way to model a finite 3D TI slab is shown  schematically  in Fig.~\ref{fig:BoundaryConditions}:
In the vicinity of the perimeters of the torus ($r \approx R_{i,e}$), the 3D TI slab
gradually becomes thinner and top and bottom surfaces are strongly hybridized in the region
$\vert r - R_{i,e} \vert \ll l_t$.
This motivates
introducing the Hamiltonian
as a 4$\times$4 matrix in the space of
top/bottom and (pseudo-)spin space:
\begin{equation}
H = H^{tot}_0 + H^{tot}_{dis},
\end{equation}
with
\begin{equation}
H^{tot}_0 = \left (\begin{array}{cc}
H^{top}_0 & T(r) \\
T(r)^\dagger & H^{bottom}_0
\end{array} \right ). \label{eq:Htopbottom}
\end{equation}

In the 2D ``bulk'', we assume well defined gapless surface states with negligible penetration depth
$a\ll d$ (here $d$ is the slab thickness) and thus the intersurface hopping falls off exponentially.
In contrast, at the boundary $T(r)$ is expected
to be the dominant energy scale, which is of the order of the bulk band gap $M$:
\begin{equation}
T(r) \sim M e^{- \frac{\vert r - R_{i,e} \vert }{l_t}}
\end{equation}
Microscopically, the tunneling matrix element $T(r)$ can be determined integrating out
the side-wall states of the 3D TI. For simplicity, we assume real,
scalar $T(r) \propto \mathbf 1_\sigma$; this is not essential for conclusions
of our analysis.

Following Halperin,\cite{Halperin1982} we assume the disorder ($H^{tot}_{dis}$,
represented by green, blurry dots in Fig.~\ref{fig:BoundaryConditions}) to be
confined to the inner part of the sample $R_i' < r < R_e'$. In Fig.~\ref{fig:BoundaryConditions}
and subsequent Secs.~\ref{sec:Semiclassical}-\ref{sec:startingvalues},  for simplicity 
we assume the disorder potential
to be uncorrelated between the surfaces ($\xi \ll d$).
Qualitatively, all results of this paper are independent of this assumption,
in particular it is completely immaterial for the modified Halperin argument discussed in the present section.

The clean Hamiltonian is determined by $H^{top}_0 = - H^{bottom}_0 = H_0$ with
\begin{equation}
H_0 = v_0(\Pi_x \sigma_y - \Pi_y \sigma_x). \label{eq:H0}
\end{equation}
The symbols $\Pi_i = -i \partial_i - e \mathcal A_i \left ( \v x\right )$ denote long derivatives;
$e = - \vert e \vert$ is the electron charge.
We assume for definiteness that the magnetic field $B = \epsilon_{ij} \partial_i \mathcal A_j >0$.

The eigenstates of the clean Hamiltonian~\eqref{eq:H0} are given by LLs\cite{AharonovCasher1979,Jackiw1984} (see also App.~\ref{app:sec:Semiclassical})
\begin{equation}
\Ket{n, k}_D = \frac{1}{\sqrt{1 + \eta_n^2}} \left (\begin{array}{c}
- \eta_n \Ket{\vert n \vert - 1, k} \\
\Ket{\vert n \vert, k}
\end{array} \right ), \label{eq:Diracstate}
\end{equation}
with quantum numbers $n \in \mathbb Z$ associated to
energies 
\begin{equation}
E_n = \Omega_c \eta_n \sqrt{\vert n \vert}, \quad \Omega_c = \sqrt{2 \vert e \vert B v^2_0}.
\label{eq:DiracEOmegac}
\end{equation}
Here $\eta_n = \text{sign}(n)$ for $n \neq 0$ and $\eta_0 =0$, $\Omega_c$ is the (quantum) cyclotron frequency,
$v_0$ is the velocity of the Dirac electrons,
and $k = 1, 2, \dots , \frac{\Phi_{tot}}{\Phi_0}$ accounts for degeneracy.
The states $\Ket{\vert n \vert, k}$ describe the LLs of usual electrons with parabolic dispersion.
In this section we chose to work in  symmetric gauge and the quantum number  $k$ determines the radius $r_k$
around which the LL wave functions are peaked.\cite{Landau1930,Halperin1982}

The length scale $l_t$ of hybridization at the edges is assumed to fulfill
\begin{equation}
l_B \ll l_t \ll (R_{i,e}'-R_{i,e}).
\end{equation}
To lowest order in small parameter  $l_B /l_t$ we can neglect the mixing of Landau levels and approximate
the  Hamiltonian  $H_0^{tot}$, Eq.~\eqref{eq:Htopbottom}, by its diagonal (in LL space) blocks
\begin{equation}
H^{tot}_{0,n} = \left (\begin{array}{cc}
E_n & T(r) \\
T(r) & -E_n
\end{array} \right ).
\end{equation}
Each Hamiltonian $H^{tot}_{0,n}$ acts in the LL specific surface space spanned by $\left (\Ket{n, k}_D,0\right )^T$ and
$\left (0, \Ket{n, k}_D\right )^T$.
The Hamiltonian $H^{tot}_{0,n}$ has eigenstates
\begin{equation}
\ket{n,k,\pm} = \frac{1}{\sqrt{2\mathcal{E}_{n,\pm}(\mathcal{E}_{n,\pm}+E_n)}} \left (\begin{array}{c}
(E_n + \mathcal{E}_{n,\pm})\Ket{n, k}_D \\
T(r)\Ket{n, k}_D
\end{array} \right ) \label{eq:Hybridizationstates}
\end{equation}
with energies  $\mathcal{E}_{n,\pm} = \pm \sqrt{E_n^2 + T(r)^2}$.

Far away from the edge ($\vert r - R_{i,e} \vert\gg l_t$), $\ket{n,k,+}$ is a
state living on solely top (bottom) surface if $E_n >0$ ($E_n <0$), while $\ket{n,k,-}$ has its weight on the opposite
 bottom (top) surface. It is a crucial observation, that in contrast to the $n \neq 0$ case, the zeroth LL wave functions $\ket{0,k,\pm} $ are symmetric and antisymmetric combinations of top and bottom states without any $r$-dependent envelop.
Note that $T(r)$ drops out of Eq.~\eqref{eq:Hybridizationstates} for $n = 0$.

\begin{figure}
\includegraphics[scale=1]{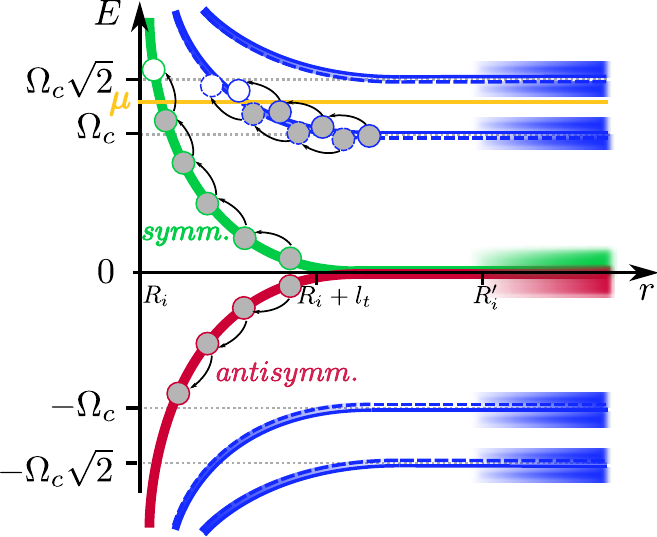}
\caption{Schematic representation of the spectrum at the inner boundary of the TI torus shown in Figs. \ref{fig:Laughlin}
and \ref{fig:BoundaryConditions}.}
\label{fig:SpectrumBoundary}
\end{figure}

Figure~\ref{fig:SpectrumBoundary} gives  a schematic representation of the LL bending around the inner perimeter
of the sample $r=R_i$.  In the 2D bulk region $ r \ll l_t + R_i$ states with $n\neq 0$ live on top  (solid lines) or bottom surface
(dashed lines). They become  hybridized (fat lines) close to the boundary.
In contrast, states of the zeroth LL always mix top and bottom surface.

The intersections of the bended LLs with the line of chemical potential define the edge states.
For the case of Fig.~\ref{fig:SpectrumBoundary} there are three of them: two originating from the
filled first LL in the two surfaces and another from the surface-symmetrized combination of the zeroth LL.
When the flux threading the hole is increased by one flux quantum, the LL-states contract and states
right above (below) the chemical potential get filled (emptied) at the internal (external) perimeter
(``spectral flow'').\cite{Halperin1982}  
In the present case (Fig.~\ref{fig:SpectrumBoundary}) the states $\ket{1,k_i,\pm}$ and $\ket{0,k_i,+}$ 
(with $r_{k_i} \approx R_i$) were filled. Similarly, the states $\ket{1,k_e,\pm}$ and $\ket{0,k_e,+}$ were emptied at the outer edge.
As a consequence of energy conservation, we conclude that during the process of flux insertion,
two electrons with energy $E = \Omega_c$ are injected into (ejected from) the disordered region of
the top surface at $r = R_e'$ ($r=R_i'$). In addition, a third electron with $E = 0$ enters (exists)
the disordered region in a symmetric superposition of top- and bottom states at the same radial positions.
By consequence, the associated current is driven through the upper and lower surface with equal weight.
\footnote{A similar argumentation for clean 3D TIs can be found in Refs.~\onlinecite{Lee2009,Vafek2011}.}
Altogether, we conclude that  $\sigma_{xy}^{top} = {3 e^2}/{2h}$ and $\sigma_{xy}^{bottom} = 3e^2/2h$.

The above analysis can be extended to a generic situation 
with the chemical potential $\mu$ located in
the mobility gaps of the two surfaces. In particular one finds half-integer values 
$\sigma_{xy}^{top} = (n+1/2)e^2/h$ for $\mu$ located between the $n$-th and the $(n+1)$-th
bulk delocalized state of the top surface and an analogous expression for the bottom surface.\footnote{The labelling of delocalized states is adiabatically connected to the labelling of LLs. In particular, the zeroth delocalized state resides per definition at zero energy. The order of all other delocalized states on the top-surface is prescribed by the associated energies.}   


\section{Semiclassical calculation of current density}
\label{sec:Semiclassical}
In the previous sections we came to the conclusion that the Dirac fermions on
the surface of a 3D TI give rise to a half-integer-quantized $g_{xy}$ provided
that the time reversal invariance is locally broken. Our  argumentation was
very general, as it relied only on gauge invariance and topology. It is
instructive, however, to have a model for which an explicit controllable
calculation of half-integer $g_{xy}$ is possible. In this section we present and
analyze  such a model consisting of
a single species of 2D Dirac fermion in the presence of orbital coupling to a
magnetic field and arbitrarily  strong but sufficiently smooth potential
landscape $V(x)$.
Apart from an externally applied electrostatic potential (caused, e.g., by a
test charge in the image magnetic  monopole experiment), the potential $V( \v
x)$ can include smooth disorder.
The externally applied potential is not required to be weak (in comparison with
the cyclotron frequency), so that our findings are valid beyond the linear
response. While the model we consider in this section is sufficiently general,
it is amenable to an analytical treatment due to semiclassical nature of the
potential. Specifically, we will calculate the Hall conductivity in this model
by using the vortex state representation of LLs.\cite{ChampelFlorens2007,
ChampelFlorens2010,HernangomezPerezFlorensChampel2014} The results of this section shed light on
the deep field-theoretical origin of the half-integer shift of Hall
conductance.

Our  fermionic Hamiltonian reads
\begin{equation}
H = H_0 + V( \v x). \label{eq:Htot}
\end{equation}
Here $H_0$ is the clean fermionic Hamiltonian introduced in Eq.~\eqref{eq:H0}.
In this section we will use symmetric gauge and the overcomplete vortex state
representation of LLs\cite{ChampelFlorens2007, ChampelFlorens2010,HernangomezPerezFlorensChampel2014} (see also
App.~\ref{app:sec:Semiclassical}). In this representation the discrete
degeneracy quantum number $k$
is replaced by the continuous  guiding center position $\v R \in \mathbb R^2$.

We are interested in the current density which couples to the macroscopic
probing  gauge potentials $A_\mu$ via  local coupling Lagrangian
\begin{equation}
\mathcal L_{\text{coupling}} =  \sum_{i = x,y} J_i  A_i  . \label{eq:couplingcurrentA}
\end{equation}
and enters subsequently the Maxwell-equations for  $A_\mu$.

We concentrate on stationary current distributions. Our  semiclassical
calculation relies on the following assumptions:

(1) The scalar potential $V\left (\v x \right )$ is smooth on the scale of the magnetic length.

(2) The macroscopic gauge potential $\v A \left (\v x\right ) $ is smooth on the scale of the magnetic length.

(3) Local thermodynamic equilibrium is maintained on the typical length scale of $V\left (\v x \right )$.

Requirement (2) is  the defining distinction between the gauge potential $\mathcal A_i$ creating the magnetic field responsible for QHE and the probing gauge potential $\v A$.
It guarantees that the electron position and the vortex (guiding center) position are indistinguishable for $\v A$.

It is convenient to combine the current densities $J_{x, y}$ into complex
combinations $J_\pm = J_x \pm i J_y$ given by
\begin{equation}
J_\pm \left (\v x\right )  = \lim_{\v x' \rightarrow \v x} \left \langle \psi^\dagger \left (t, \v x\right ) j_\pm \psi \left (t, \v x'\right ) \right \rangle \label{eq:Def_currentdensity}
\end{equation}
with
\begin{subequations}
\begin{eqnarray}
j_+ &=& -2i  e v_0\left (\begin{array}{cc}
0 & 1 \\
0 & 0
\end{array} \right ), \\
j_- &=& 2i  e  v_0\left (\begin{array}{cc}
0 & 0 \\
1 & 0
\end{array} \right ).
\end{eqnarray}
\end{subequations}
The operators $\psi^\dagger \left (t, \v x\right )$ and $\psi \left (t, \v
x\right )$ are fermionic field operators with two spinor components. Thus,
$J_\pm$ are proportional to the off-diagonal elements of the fermionic Green's
function $G_{\sigma \sigma'} \left (t, \v x; t, \v x\right )$ at equal point and
time. Since we expect ultraviolet divergences [unbounded spectrum of
Hamiltonian~\eqref{eq:H0}], we  regularize $J_\pm \left (\v x\right )$ via
point splitting. Strictly speaking, to make this procedure $\mathbf{U}(1)$ gauge
invariant, a Wilson line $e^{ie\int_{\v x'}^{\v x} (A + \mathcal A) d^2y}$
should be inserted in the end of
Eqs.~\eqref{eq:Def_currentdensity},\eqref{eq:Jsum} and \eqref{eq:Xhelp}.
However, It drops out in the limit $\v x \rightarrow \v x'$ and is thus omitted
for simplicity. The physical reason is that at small splitting $\vert \v x-  \v
x' \vert \ll l_B$, the right-hand side of Eq.~\eqref{eq:Def_currentdensity} is
invariant under the ``macroscopic'' (slow) local $\mathbf{U}(1)$ symmetry associated with
potentials $\v A$ even before taking the limit $\v x \rightarrow \v x'$.

Under the assumptions (1) - (3),  we find (see
App.~\ref{app:sec:Semiclassical})
\begin{eqnarray}
J_\pm \left (\v x\right )  &\approx & \frac{\pm i\vert e \vert }{2\pi}\lim_{\v x' \rightarrow \v x} \int d^2R \sum_{\vert n \vert = 0}^\infty \sum_{\eta_n}  n_F\left (E_{n} + V\left ( \v R\right )\right )  \notag \\
&\times & _D\braket{n, \v R \vert \v x}\braket{\v x' \vert n, \v R}_D \partial_{\pm} V\left (\v R\right ). \label{eq:Jsum}
\end{eqnarray}
Here $\braket{\v x' \vert n, \v R}_D$ are the ``vortex states'' of the Dirac
Hamiltonian defined in App.~\ref{app:sec:Semiclassical}.
This formula has the following simple physical interpretation. In order to find
the local current density $ J_\pm \left (\v x\right )$ as a response to the
electric field $\partial_\pm V\left (\v R\right )$, one should sum over all
locally filled Landau levels and perform a convolution with $\vert \braket{\v x
\vert n, \v R}_D\vert ^2$ representing the response of a single vortex state.

The integral in Eq.~\eqref{eq:Jsum} diverges at $\v x = \v x'$. However,
the point-splitting procedure, which implies the formal rule $\lim_{\v x' \rightarrow \v x} \delta(\v x - \v x') \equiv 0$,
renders Eq.~~\eqref{eq:Jsum} finite.
To make the regularization manifest, we add and subtract the zero temperature linear-response current to/from Eq.~\eqref{eq:Jsum}
\begin{gather}
X_\pm \left (\v x \right ) = \left. J_\pm\left(\v x \right)\right|_{T=0} = \frac{\pm i\vert e \vert \partial_{\pm} V\left (\v x\right ) }{2\pi}  \notag \\
\times   \lim_{\v x' \rightarrow \v x} \int d^2R \sum_{n \leq 0} \left. _D\!\braket{n, \v R \vert \v x}\braket{\v x' \vert n, \v R}_D \right.  \notag \\
= \pm i\vert e \vert \partial_{\pm} V\left (\v x\right ) \left [\lim_{\v x' \rightarrow \v x}  
l_B^2\delta\left (\v x - \v x'\right ) + \frac{1}{4\pi} \right ]
.
 \label{eq:Xhelp}
\end{gather}
The first term in the square brackets (delta function) comes from the resolution
of identity of usual (equidistant) LL [see Eq.~\eqref{eq:Resolofidentity}] and
vanishes after the point splitting. In contrast, the second term will turn out
to be responsible for half-integer $g_{xy}$. Its appearance is a direct
consequence of the definite chirality of the zeroth LL wave function (see
appendix~\ref{app:sec:Semiclassical}). Our calculation thus provides a
``pedestrian'' approach to the
Atiyah-Singer (AS) index theorem.\cite{AtiyahSinger1963,AtiyahSinger1968}

The integral determining the quantity $J_\pm - X_\pm$ is now regular, since the divergence has been shifted
entirely into $X_\pm$ and is cured by the formal point-splitting procedure:
$X_\pm=\pm i\vert e \vert \partial_{\pm} V\left (\v x\right )/4\pi.$
We can therefore take the $\v x' \rightarrow \v x$ limit, rearrange integrals and
sums and exploit once more the smoothness of $V\left (\v x\right )$
[assumption (1)] to obtain
\begin{eqnarray}
J_\pm \left (\v x\right ) &=&  \frac{\pm i\vert e \vert }{2\pi} \partial_{\pm} V\left (\v x\right )
\Big \{ \sum_{n>0}  n_F\left (E_{n} + V\left ( \v x\right )\right ) \notag
\\
&+& \sum_{n\leq0} \left [ n_F\left (E_{n} + V\left ( \v x\right )\right ) -1\right ]
+ \frac{1}{2} \Big\}. \label{eq:Jsemiclassical}
\end{eqnarray}
The local transverse conductivity 
is thus given by
\begin{eqnarray}
\sigma_{yx} \left (\v x\right )  &=& \frac{e^2 }{h}\frac{1}{2} + \frac{e^2 }{h} \Big \{  
\sum_{n>0}  n_F\left (E_{n} + V\left ( \v x\right )\right )
\notag
\\
&+& 
\sum_{n\leq0} \left [ n_F\left (E_{n} + V\left ( \v x\right )\right ) -1\right ]
\Big\}.\label{eq:sigmasemiclassical}
\end{eqnarray}

Let us reiterate that, in contrast to usual linear response calculations,
Eqs.~\eqref{eq:Jsemiclassical} and~\eqref{eq:sigmasemiclassical} are also valid
in the case of a strong static
electric field. In particular, they can be applied to study the magnetic image
monopole effect in the situation when the voltage between test charge $Q$ and
the QH system exceeds $\Omega_c/ \vert e \vert$.

\section{Field theory of localization}
\label{sec:NLSM}

This section is devoted to the field theory describing the localization physics
in the half-integer QH state. It should be emphasized that the QHE crucially
depends on the presence of disorder. Specifically, it is the disorder-induced
localization that provides mobility gaps with a finite density of states in the
bulk of a 2D system,  which in turn leads to plateaus with quantized values of
$\sigma_{xy}$ as a function of carrier density. Thus, the analysis of
half-integer QHE should contain a discussion of Anderson localization as one of
key ingredients.

On the basic level the 3D TI surface fermions are described by the Euclidean field theory
\begin{equation}
\mathcal Z = \int \mathcal D \left [\bar \psi, \psi \right ] \; e^{-S\left [\bar \psi, \psi \right ]}
\end{equation}
with the Matsubara action
\begin{equation}
S\left [\bar \psi, \psi \right ] = \int_{\tau, \v x} \bar \psi \left
(D_\tau +  H_0 +  V\left (\v x \right )  - \mu \right ) \psi + S_{\rm int}.
\label{eq:Highenergyaction}
\end{equation}
Throughout the paper we use the  notation $\int_{\tau, \v x} = \int d^2x
\int_0^\beta d \tau$;  as usual  $\beta = 1/T$ is the inverse temperature.
We compactify the space, so that the base manifold of our field theory  is
$(\mathbb R^2 \cup \lbrace \infty \rbrace) \times \mathbb S^1$. The clean, free
Hamiltonian $H_0$ was introduced in Eq.~\eqref{eq:H0}.
The long derivatives  $\Pi_i = -i \partial_i - e  \left (\mathcal A_i +
A_i\right )$  include both the  vector potential $\mathcal A_i$ responsible for
the quantizing magnetic
field $B$ and a source field $A_i$. The long Matsubara derivative is  $D_\tau =
\partial_\tau - i e \Phi$; $V(x)$ and $\mu$ represent Gaussian
$\delta$-correlated disorder potential and chemical potential respectively. The
fermionic fields $ \bar \psi \left (\v x, \tau\right ) = \left ( \bar
\psi^\uparrow, \bar \psi^\downarrow\right )$ and $ \psi  \left (\v x, \tau\right
) = \left ( \psi^\uparrow, \psi^\downarrow \right )^T$ describe the spinful
($\uparrow,\downarrow$) surface excitations. The
electron-electron interaction ($S_{\rm int}$) can also be included in our
treatment (see Refs.~\onlinecite{MishandlingI, KoenigMirlin2013}). It can be
strong ($r_s = e^2/\epsilon \hbar v_0 \sim 1$), with the only condition that it
does not induce any spontaneous symmetry breaking.

Our aim in this section is to determine the effective low-energy theory of gauge
potentials $A_\mu = \left (\Phi,A_i\right )$ in the 2D ``bulk'' of the general
interacting, disordered system without resorting to QH edge states. Let us
summarize shortly our strategy.  We  first note that there are two relevant
energy scales in this problem: the elastic scattering rate $1/\tau$ and the
(inelastic) phase breaking rate $1/\tau_\phi(T)$. At low temperatures, these
scales form the hierarchy 
\begin{equation}
 \frac{1}{\tau_\phi(T)} \ll \frac{1}{\tau}.
\end{equation}
Consequently, to get the desired theory for the gauge field,  we  integrate
out matter fields in a stepwise fashion: since electrons with quantum numbers
$n,k$ (see Sec.~\ref{sec:Laughlin:Micro}) are good excitations only above
$1/\tau$, they are integrated out first. The resulting theory then involves
gauge fields and diffusive soft modes (diffusive sigma model). To account for
the interaction of the diffusive modes (aka quantum interference effects) at
energies lower than $1/\tau$  renormalization group approach is employed. The
renormalization group flow stops  at the energy scale $1/\tau_\phi(T)$ where the
phase breaking destroys quantum interference. The remaining modes of the
matter field can then  be integrated out in the saddle-point approximation
resulting in the effective low-energy theory of the gauge potential, which was
discussed on phenomenological grounds in Sec.~\ref{sec:TME}.

\subsection{Parity anomaly}
\label{sec:NLSM:parity}

We will first review the concept of parity anomaly.

The high-energy action~\eqref{eq:Highenergyaction} is invariant under transformations of the gauge group $\mathbf G$, and in our case $\mathbf G = \mathbf U(1)$. Later we will replicate the theory $N_R$ times and $\mathbf G$ will turn out to be $\left (\mathbf U(1)\right )^{\otimes N_R}$.
In addition, in the absence of a net magnetic field or Zeeman term and after disorder average it is also invariant under the parity transformation of (2+1) dimensional space-time:
\begin{subequations}
\begin{eqnarray}
\left (x, y,\tau\right ) &\rightarrow & \left (-x, y,\tau\right ) ,\\
\psi &\rightarrow & \sigma_x \psi,\\
\bar \psi &\rightarrow & \bar \psi \sigma_x .
\end{eqnarray}
\end{subequations}
Clearly, to keep the fermionic action invariant, the vector potential should transform under $x \rightarrow -x $ as
\begin{equation}
\left ({A}_x, A_y,\Phi\right )\rightarrow  \left ( -A_x, A_y,\Phi, \right ) .\\
\end{equation}
For dynamic gauge fields we note that this transformation leaves the Maxwell term invariant. Contrary, a fixed background $B$-field does not respect this symmetry ($B$ is a pseudo scalar).

The peculiar fact about (2+1) dimensional gauge theories is that invariance
under parity transformation does not always persist to the quantized
theory.\cite{NiemiSemenoff1983, RedlichPRL1984, RedlichPRD1984,
AlvarezGauméWitten1984} Following Ref.~\onlinecite{DeserGriguoloSeminara1998} we
will however distinguish between ``parity anomaly'' and ''intrinsic parity
anomaly'' for our problem of QED$_3$ on a space-time manifold $(\v x,\tau) \in
\mathbb S^2 \times \mathbb S^1$. Of course both effects are related.

The notion of parity anomaly follows Ref.~\onlinecite{NiemiSemenoff1983} and
arises often  in the context of condensed matter physics. It boils down to
calculating $\sigma_{xy}$ for the problem of massive (2+1) dimensional Dirac
fermions in the absence of any other energy scale. The result is $\sigma_{yx} =
\frac{\text{sign}(m)}{2}\frac{e^2}{h}$. As there is no other energy scale, the
mass $m$ breaks time reversal and parity symmetries on all scales and therefore
$\sigma_{xy}(m)$ is discontinuous at $m=0$. One concludes that upon integration
of Dirac electrons and subsequent $m\rightarrow 0$ limit the effective gauge
theory contains a CS-term with prefactor $\vartheta = \pm \pi$. The notion of
parity anomaly means that the Lagrangian of fermions coupled to gauge potentials
preserves parity upon taking the massless limit, while the effective
electrodynamic Lagrangian does not. Not surprisingly, this ``anomaly''
disappears as soon as another  infrared energy scale is introduced, for example
finite temperature,\cite{DeserGriguoloSeminara1998} a finite disorder scattering
rate,\cite{OGM2007} or a finite bulk band gap $M$.\cite{MulliganBurnell2013}
Then, $\sigma_{xy}(m)$ becomes a continuous function of $m$ and $\sigma_{xy}(0)
= 0$.

The notion of intrinsic parity anomaly is more subtle. According to the early
works,\cite{RedlichPRL1984, RedlichPRD1984, Polychronakos1987} which treat the
case of strictly massless fermions, in the process of field quantization one
has two options:
\begin{enumerate}
\item[(i)] One can choose a regularization scheme in a manner preserving
parity. But then the partition function acquires a sign $(-)^k$ under large
gauge transformations.
\item[(ii)] Alternatively, one can regularize the theory in a manner preserving
gauge invariance. In this case, a CS-term with angle $\vartheta = \pi$~(mod~$2\pi$)
appears after integration of fermions. The latter breaks parity.
\end{enumerate}

A theory with anomalously broken gauge symmetry is inconsistent, therefore,
whenever $(-)^k \neq 1$ option (ii) must be chosen. A common variant of these
regularization schemes is to use regularization as in (i) and to add the CS
3-form by hand to the fermionic action\cite{RedlichPRD1984} when the latter
contributes additional factor of $(-)^k$ under large gauge transformations.

To proof assertion (i) one needs to unwind the gauge potentials $e A_n ( \v x,
\tau)= -i U^{-1}_n \nabla U_n$ associated to large gauge transformations $U_n
\in \mathbf G$. A fourth dimension (with coordinate $s$) is introduced and it
can be shown that $k$ equals the analytical index $\nu_+ - \nu_-$ of the
corresponding four dimensional Dirac operator. For non-Abelian gauge groups with
third homotopy group $\Pi_3 ( \v G) = \mathbb Z$ the AS index
theorem\cite{AtiyahSinger1963, AtiyahSinger1968} immediately implies $k = n$
($n$ is the homotopy class of $U_n$).\cite{RedlichPRL1984, RedlichPRD1984}

Contrary, for the case of QED$_3$ on $\mathbb S^2 \times \mathbb S^1$ the
topology is more complicated: as $\Pi_1 ( \v G ) = \mathbb Z$, large gauge
transformations act in the imaginary time sector. Further, topologically
distinct instanton (monopole) configurations in the spatial $\mathbb S^2 \simeq
\mathbb R^2 \cup \lbrace \infty \rbrace$ sector (i.e. field configurations with
different magnetic flux $\Phi$ through the $\mathbb R^2$ plane) have to be
treated with care. (We recall that the gauge potential, which explicitly enters
the CS term, can not be defined on the whole manifold.) Nevertheless, $k$ can
still be associated with the topological index of extended gauge fields.
Specifically, it turns out that $k = n \Phi/\Phi_0$, where $n$ is the winding in
time direction.\cite{Polychronakos1987} In the presence of time-reversal
symmetry we have $\Phi=0$ and hence $k=0$.
We thus conclude, that for the 2D theory of time-reversal-invariant surface
states of 3D TIs there is no reason for inclusion of additional CS term that
would violate the parity of the theory.

More generally, we can consider the Dirac fermions on the entire surface
wrapping the 3D TI sample.\footnote{As explained, any non-trivial $k$ is related
to the Atiyah-Singer index theorem in 4 dimensions. When the base manifold $M$
is not flat, as is the case for the 2D surface wrapping the 3D TI, the
Atiyah-Singer index theorem contains a contribution from a potentially
non-trivial Dirac genus $\widehat A$. The latter reflects the properties of the
base manifold and is constructed from Pontryagin indices of the curvature
two-form. In the present case however $\widehat A$ is trivially unity: Any
non-trivial contribution could only arise from the physical 2D surface. Since
the Pontryagin indices are per definition an even function of the curvature
2-form the leading order is one and the next order is already a four-form. As a
consequence, all of the topological findings obtained for the compactified plane
$\mathbb R^2 \cup \lbrace \infty \rbrace$ can be applied to actual, closed 3D TI
surface.} This field theory again lives on
a manifold homotopical to $\mathbb S^2 \times \mathbb S^1$. Then, since there
are no physical monopoles, the total flux through the spatial sector vanishes
even in the case of broken time reversal symmetry and hence the topological
insulator surface states do not exhibit intrinsic parity anomaly. We conclude
that additional CS terms never need to be included. Such terms will therefore
not appear in the effective electromagnetic actions to be derived in the
following sections.

Recently,\cite{MulliganBurnell2013} similar topological arguments for 3D TIs
avoiding the intrinsic parity anomaly were presented. While the topological
peculiarities of $\mathbf{U}(1)$ gauge theories were disregarded by the authors
of this work, their argument in favor of the absence
of parity anomaly in the theory of 3D TI surface states is in agreement with
our conclusion. Another line of argumentation with the same outcome is based on
the concept of cancellation of anomalies.\cite{ZirnsteinRosenow2013}

\subsection{Gauged NL$\sigma$M of integer QHE}

Before turning to Dirac fermions, we briefly review the field theoretic
description of the conventional integer QHE. The $\mathbf U(1)$-gauged
NL$\sigma$M\cite{LevineLibbyPruisken1983, MishandlingI} describing the
interaction of diffusive modes and the gauge potential has the action
\begin{eqnarray}
S&=& \frac{1}{8}\int_{\v x}  \tr \left [g_{xx}
D_i Q D_i Q +  \epsilon_{ij} \frac{\vartheta}{2\pi} Q
D_i Q D_j Q \right ]\notag \\  &&+ S_{\eta + int. + B^2}.  \label{eq:PruiskenNSLM}
\end{eqnarray}

According to the double cut-off regularization of Matsubara
frequencies,\cite{MishandlingI} the diffusive  $\left (2 N_M' \times
N_R\right ) \times \left (2 N_M' \times
N_R\right )$ matrix fields $Q\left (\v x\right )$ carry both Matsubara and
replica indices and are  typically represented as $\left (Q\right
)_{lm}^{\alpha \beta} = \left (U^{-1}\Lambda U\right )_{lm}^{\alpha \beta}$
($\alpha, \beta = 1, \dots,  N_R $ denote replicas and
$l, m \in \left \lbrace - N'_M, \dots , N'_M - 1 \right \rbrace$ Matsubara
indices). The unitary matrices  $U$ have non-trivial entries belonging to
$\mathbf U\left (2 N_M \times
N_R\right )$ in the central block $l,m \in \left \lbrace - N_M, \dots , N_M - 1
\right \rbrace$ ($N_M \ll N_M'$) and are unity outside. Recall that the
dimensionless conductances are denoted by $g_{ij} = \sigma_{ij} h/e^2$. The term
$S_{\eta + int. + B^2}$ is less important for the present discussion and we
mention it here  for completeness only. It contains frequency and interaction
contributions, as well as a term quadratic in magnetic field which renormalizes
the permeability. The kinetic term (proportional to $g_{xx}$) and the theta term
(proportional to $\vartheta$) contain long derivatives acting as
\begin{equation}
D_i Q = \partial_i Q -i e \left [\hat A_i,Q\right ] . \label{eq:longderivatives}
\end{equation}
Hatted objects are defined by $\hat a \equiv
\sum_{m,\alpha} a_m^\alpha I_m^\alpha$. In the above, we have introduced
the following $(2 N'_M \times
N_R ) \times (2 N'_M \times
N_R)$ matrices:
\begin{subequations}
\begin{eqnarray}
\Lambda^{\alpha \beta}_{lm} &=& \text{sgn}\left (\epsilon_m \right ) \delta^{\alpha \beta}  \delta_{lm} ,  \\
\left (I^{\alpha_0}_{m_0} \right )^{\alpha \beta}_{lm} &=&
\delta^{\alpha_0 \alpha} \delta^{\alpha_0 \beta}
\delta_{l-m,m_0}.
\end{eqnarray}
\end{subequations}
The limits $N_M \rightarrow \infty$, $N'_M \rightarrow \infty$ ($N'_M/N_M \rightarrow \infty$) as well as the final replica limit $N_R\rightarrow 0$ are implicitly assumed.

Differentiation of Eq.~\eqref{eq:PruiskenNSLM} with respect to the vector potential
and evaluation of the functional integral in the saddle-point approximation
leads to the identification of the NL$\sigma$M coupling constants $g_{xx}$ and
$\frac{\vartheta}{2\pi}$ with the bare longitudinal and transversal (Hall)
conductivities of the QH system (in units of $e^2/h$).\cite{MishandlingI} At the
diffusive saddle point $Q = \Lambda$ the theta term becomes the
CS-term.\cite{MishandlingI}

\subsection{Gauged NL$\sigma$M of half-integer QHE}

\subsubsection{Gauged NL$\sigma$M of Dirac fermions at $B = 0$}
\label{sec:NLSM:B=0}

We turn now to the  localization physics of a single Dirac fermion. Let us
assume that  TR symmetry is present on average (i.e., there is no net magnetic
field) but broken by a random magnetic field or random Zeeman coupling.
\footnote{The $\mathbf{U}(1)$-gauged NL$\sigma$M for Dirac fermions preserving TR
invariance in each disorder realization was recently presented in
Ref.~\onlinecite{KoenigMirlin2013}.
The  arguments concerning the absence of  parity anomaly presented in the
present work apply equally well to that situation.}
In this case, the gauged NL$\sigma$M can be derived using the non-Abelian
bosonization technique (see Ref.~\onlinecite{KoenigMirlin2013} and
appendix~\ref{sec:app:Bosonization})
\begin{equation}
S= \frac{1}{8}\int_{\v x}  \tr \left [g_{xx}
D_i Q D_i Q +  \epsilon_{ij} \frac{\theta}{2\pi} Q
\partial_i Q \partial_j Q \right ] + S_{\eta + int.},  \label{eq:B=0NSLM}\\
\end{equation}
where $\theta = \pi$~(mod~$2\pi$). It is worth emphasizing that the derivatives in
the theta term are $not$ covariant derivatives. Yet, the
action~\eqref{eq:B=0NSLM}
is gauge invariant. Indeed, local $\mathbf{U}$(1) transformations of fermionic
fields translate into the following operation on NL$\sigma$M matrices
\begin{equation}
Q \left (\v x\right )\rightarrow e^{i\hat \chi \left (\v x\right )} Q \left (\v x\right )e^{- i\hat \chi \left (\v x\right )} .
\end{equation}
The theta term in Eq.~\eqref{eq:B=0NSLM}, being quantized, is unchanged under
smooth gauge transformations.

Since the theta term does not include coupling to the
electromagnetic field, the Hall conductance of the Dirac fermions is not related
to $\theta$. Instead, $g_{xy} = 0$, which is exactly what one should expect
in the absence of a net $B$ field.

\subsubsection{Gauged NL$\sigma$M at $B \neq 0$}

The gauged NL$\sigma$M describing both electromagnetic response and localization
physics of a single Dirac fermion is
\begin{align}
S= \frac{1}{8}\int_{\v x}  \tr \Big [&g_{xx}
D_i Q D_i Q +  \epsilon_{ij} \frac{\theta}{2\pi} Q
\partial_i Q \partial_j Q \notag \\
& + \epsilon_{ij} \frac{\vartheta}{2\pi} Q
D_i Q D_j Q \Big ] + S_{\eta + int. + B^2}.  \label{eq:Bneq0NSLM}
\end{align}
The derivation of this action can be found in
appendix~\ref{sec:app:Gradientexpansion}. It is crucial to observe that only
$\vartheta$ couples to electromagnetic gauge potentials.
Thus the transverse conductivity $\sigma_{xy}$ is determined by $\vartheta$ alone, while the
localization physics is governed by the sum $\vartheta + \theta = \vartheta \pm
\pi$. 
In the renormalization group flow this will lead to an overall shift of
$g_{xy}$ by $\pm 1/2$, see Eqs. \eqref{eq:modifiedPruiskenRGequations} below.
The Matsubara NL$\sigma$M description of the QHE allows for inclusion of
electron-electron interactions.
\cite{PruiskenBaranov1995,PruiskenBurmistrov2007} The shift of the RG flow by
half a conductance quantum equally applies to the interacting case.

\subsection{RG analysis of the sigma model}

\subsubsection{RG flow and phase diagram}


\begin{figure}
\includegraphics[scale=.4]{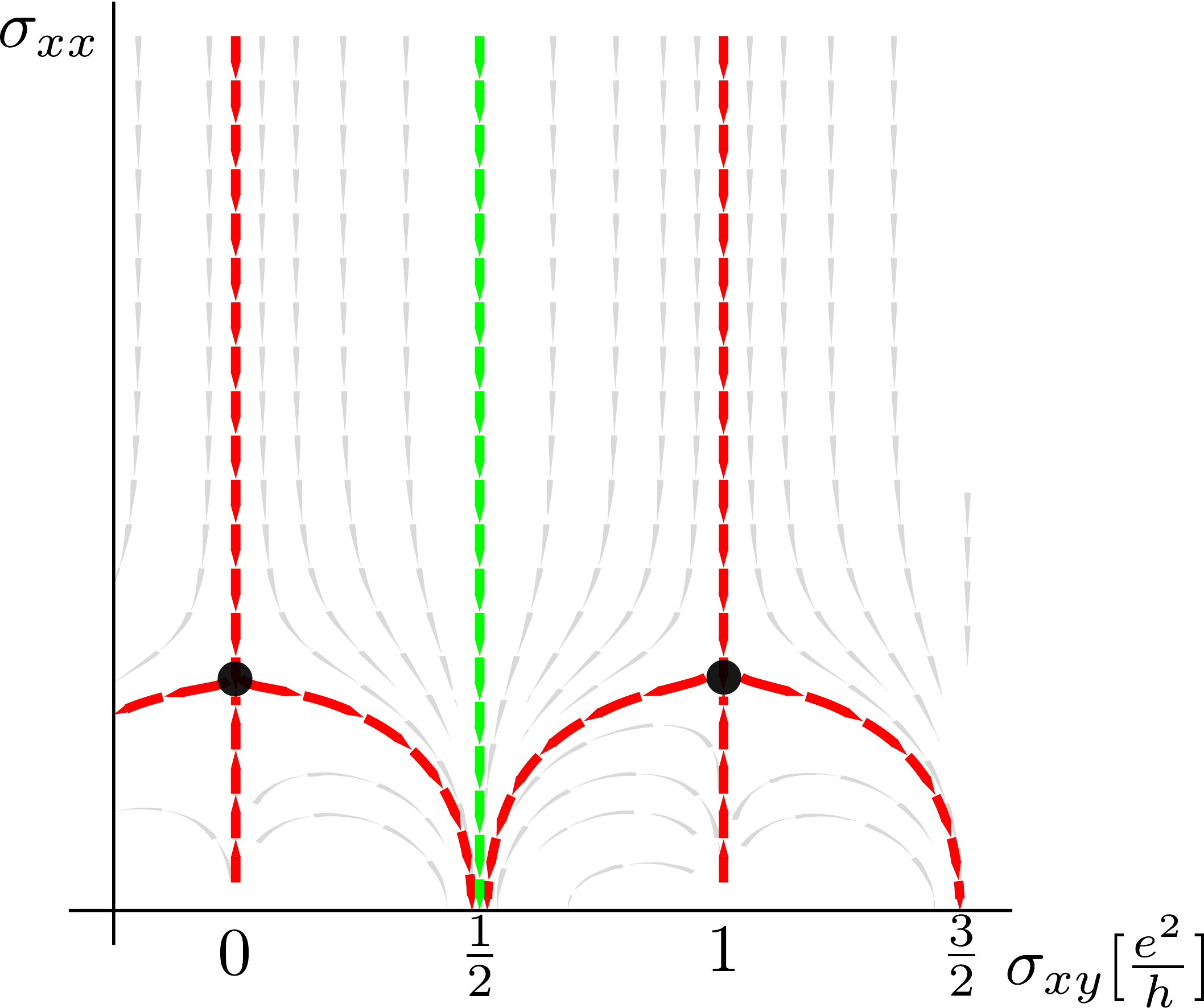}
\caption{RG flow diagram for Dirac fermions, based on the scaling proposed by Khmelnitskii. As compared to the case of fermions with parabolic dispersion, the diagram is shifted by half a conductance quantum.}
\label{fig:flowDiagramusual}
\end{figure}


Up to the important shift of the theta angle, the action~\eqref{eq:Bneq0NSLM}
corresponds to the standard Pruisken NL$\sigma$M for spinless fermions.
Therefore its renormalization\cite{LevineLibbyPruisken1983,
PruiskeninPrangeGirvin} is analogous to the conventional case. The only
modification is a connection between the theta angle and the Hall conductivity.
This implies the following RG
equations\cite{PruiskenBaranov1995,PruiskenBurmistrov2005,%
PruiskenBurmistrov2007}
\begin{subequations}
\begin{align}
\frac{dg_{xx}}{d y} = & -A-\frac{B}{g_{xx}} - C g_{xx}^2 e^{-2 \pi  g_{xx}} \cos \left[2 \pi  \left (g_{xy}\pm\frac{1}{2}\right)\right],
\\
\frac{dg_{xy}}{d y} = & -C g_{xx}^2 e^{-2 \pi  g_{xx}} \sin \left[2 \pi  \left (g_{xy}\pm\frac{1}{2}\right )\right].
\end{align}
\label{eq:modifiedPruiskenRGequations}
\end{subequations}
In these equations $y = \ln L/l$, where $L$ is the running scale and $l$ the UV
reference scale (mean free path). The equations are written with the two-loop
perturbative accuracy and contain in addition the leading non-perturbative
(instanton) contributions.
The prefactors $A,B,C$ entering these RG-equations are numerical constants.
Below we give their values both for the case of non-interacting
electrons\cite{PruiskenBurmistrov2005} and for the Coulomb
interaction:\cite{PruiskenBurmistrov2007}

\begin{subequations}
\begin{align}
A &= \left \lbrace \begin{array}{c}
0  \\
\frac{2}{\pi}
\end{array}\right .  &\begin{array}{c}
 \text{No interaction} \\
 \text{Coulomb interaction}
\end{array}   \\
B &= \left \lbrace \begin{array}{c}
{1}/{2 \pi^2} \\
\approx 0.66
\end{array}  \right .  &\begin{array}{c}
 \text{No interaction} \\
 \text{Coulomb interaction}
\end{array} \\
C &= \left \lbrace \begin{array}{c}
{4\pi} \exp(-1)  \\
4\pi \exp(1-4\gamma)
\end{array} \right . &\begin{array}{c}
 \text{No interaction} \\
 \text{Coulomb interaction}
\end{array}    \label{eq:CorrectedRGCoeffC}
\end{align}
\label{eq:CorrectedRGCoeff}
\end{subequations}
Here $\gamma\simeq 0.577$ is the Euler-Mascheroni constant.

Equations~\eqref{eq:modifiedPruiskenRGequations} lead to  the RG-flow diagram
for the half-integer QHE of Dirac
fermions,\cite{NomuraNagaosa2011,MorimotoAoki2011,MorimotoAoki2012} see
Fig.~\ref{fig:flowDiagramusual}.\cite{Khmelnitskii1983} The attractive fixed
points are now $\left (g_{xx}^*,g_{xy}^*\right ) = \left (0, \left (\nu +
{1}/{2}\right ){e^2}/{h}\right )$ while the delocalized critical state (black
dot) appears at integer valued
$g_{xy}$.\cite{BardarsonBeenakker2010,GattenlohnerTitov2014}

Starting values of the RG flow at the scale of the mean free path $l$ are
given by the Drude expression of the conductivity tensor.
We will discuss them in detail in  Sec.~\ref{sec:startingvalues} below.

\subsubsection{The $g_{xy} = 0$ transition.}

\label{sec:Fukuyama}


\begin{figure}
\includegraphics[scale=.4]{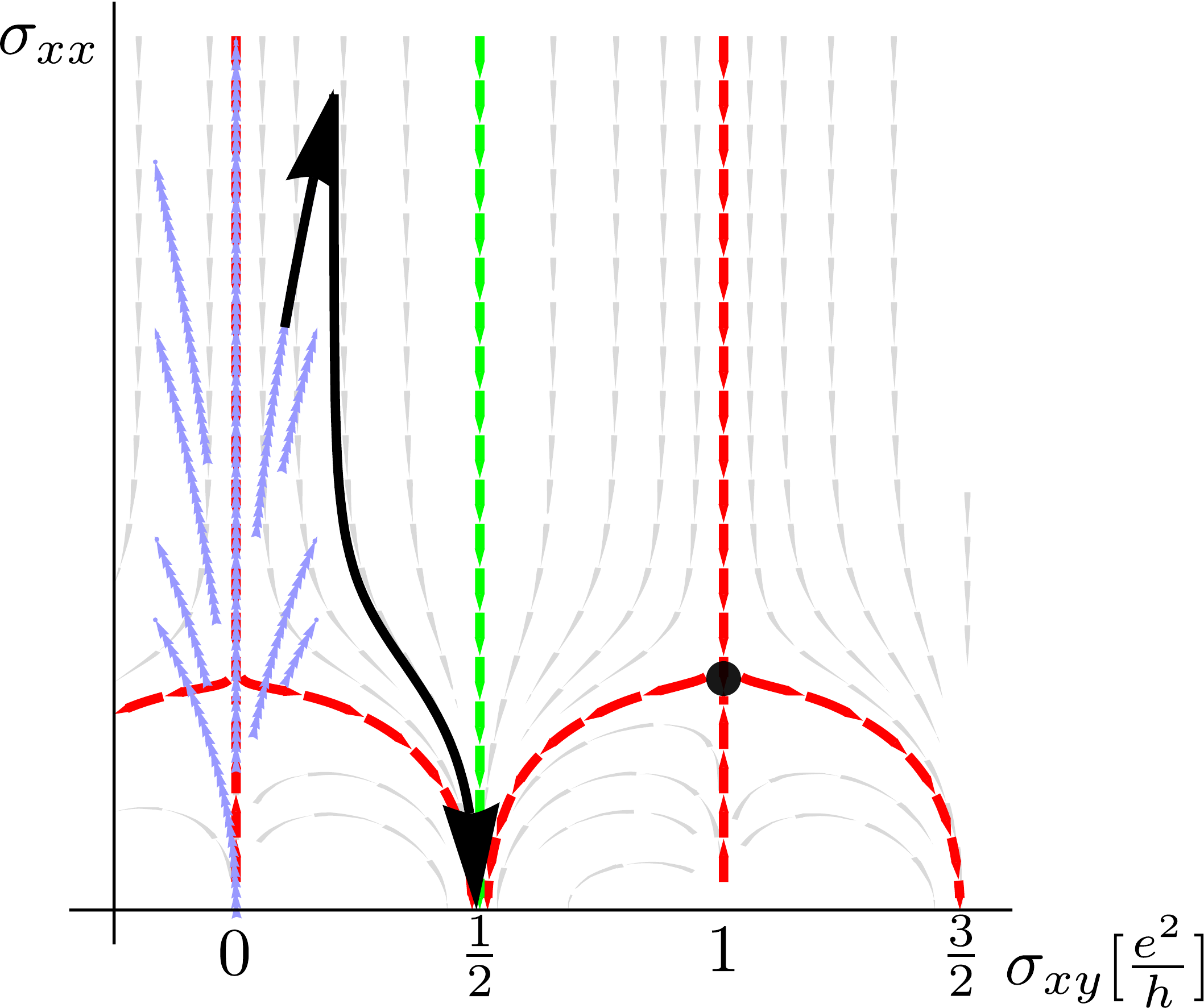}
\caption{RG flow diagram for non-interacting Dirac fermions for the case when
the magnetic field is the only source of time-reversal symmetry breaking (i.e.,
assuming no magnetic impurities and no Zeeman coupling). The QH transition
at $B=0$ ($\sigma_{xy} = 0$) is qualitatively different from all others, see
main text. The corresponding flow is depicted by black bold arrows.
It follows  the behavior of the symplectic class for $L < l_B$  and
crosses over to the unitary class at $L \sim l_B$.}
\label{fig:flowDiagramFukuyama}
\end{figure}


Generally, the universality class of the Dirac QH transition coincides with the
QH transition in parabolic
2DEG.\cite{ChalkerCoddington1988,Huckestein1995,EversMirlin2008,
PruiskenBurmistrov2007} However, if in the absence of magnetic impurities the QH
transition from $\sigma_{xy} = - {e^2}/{2h}$ to  $\sigma_{xy} = + {e^2}/{2h}$ is
driven by the variation of the magnetic field from negative to positive, then
additional soft modes, Cooperons, modify the physics at length scales smaller
then the magnetic length. This changes the nature of the transition and is
represented in Fig.~\ref{fig:flowDiagramFukuyama} by the blue upward arrows at
$\sigma_{xy} \approx 0$ for the case without electron-electron
interactions. At small length scales the systems follows the RG of symplectic
symmetry class (weak anti-localization). In the one-loop approximation the
interference corrections can be understood as a renormalization of the elastic
scattering rate, and the RG equations take the form\cite{Fukuyama1983}
\begin{subequations}
\begin{eqnarray}
\frac{d g_{xx}}{d y} &=& \frac{1}{\pi}, \\
\frac{d g_{xy}}{d y} &=& 2  \frac{g_{xy}}{g_{xx}} \times \frac{d g_{xx}}{d y} = \frac{2}{\pi}\frac{g_{xy}}{g_{xx}} .
\end{eqnarray}
\label{eq:FukuyamaRG}
\end{subequations}

\begin{figure}
\includegraphics[scale=.5]{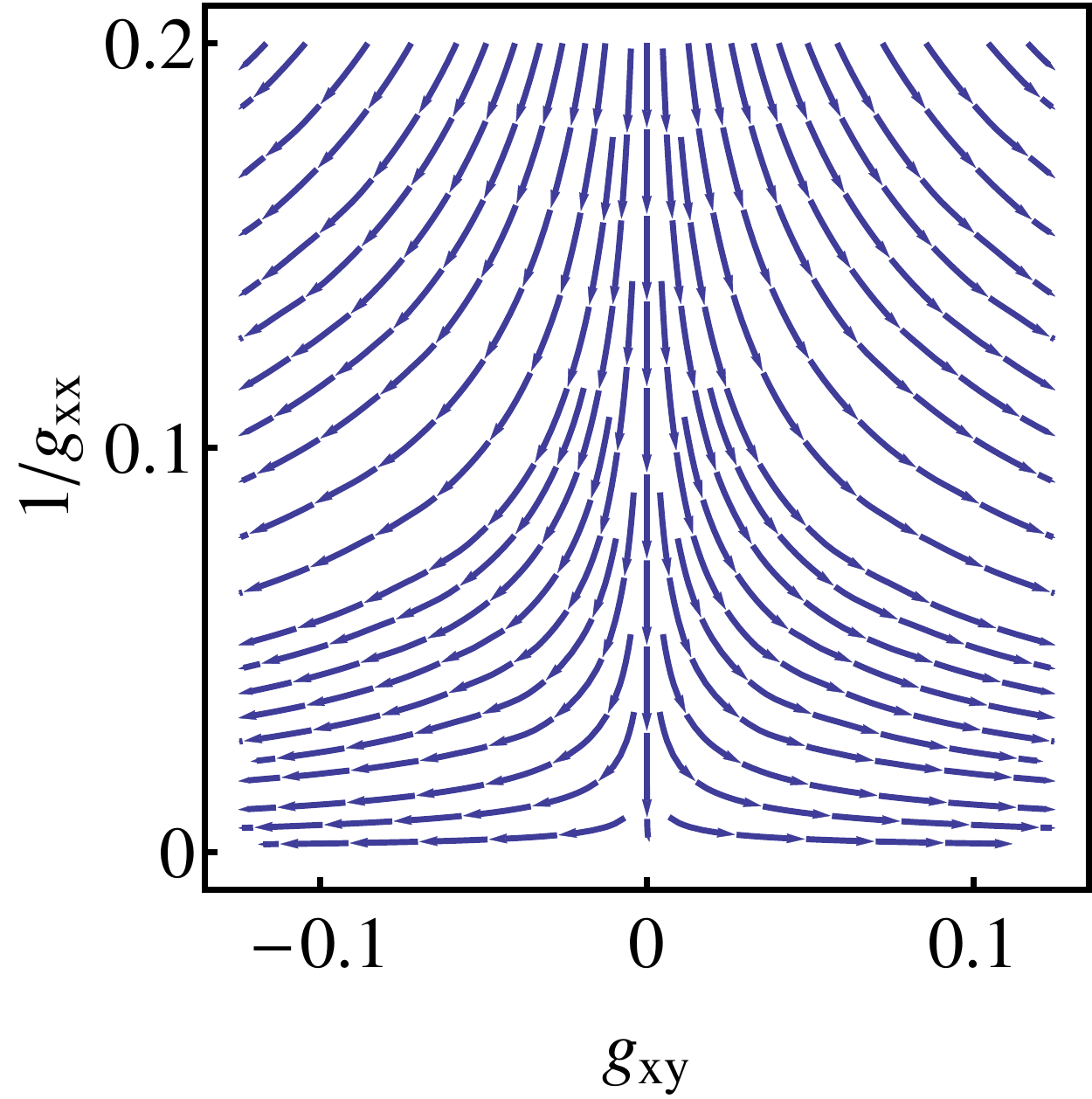}
\caption{RG flow describing the $B = 0$ transition near the ``supermetallic''
fixed point $(1/g^*_{xx},g^*_{xy}) = (0, 0)$, see Eqs.\eqref{eq:FukuyamaBKTRG}.
}
\label{fig:FukuyamaTransition}
\end{figure}

The crossover to the unitary class occurs when the running scale $L$ hits
$l_B$, and for larger length scales the flow follows
Eqs.~\eqref{eq:modifiedPruiskenRGequations}. Integrating the symplectic RG
equations up to $l_B$ we obtain
\begin{equation}
g_{xy}(l_B) \sim \frac{1}{(k_F l_B)^2} (  k_F l + \ln l_B/l)^2 \ll 1.
\end{equation}
Therefore, as long as the bare (Drude) value of the Hall
conductivity is small, 
the renormalized value
$g_{xy}(l_B)$ at the output of the symplectic stage of evolution remains small
as well, and the system flows, in the infrared limit, into one of the lowest QH
states $\sigma_{xy} = \pm {e^2}/{2h}$.

With the notation $t = 2/(\pi g_{xx})$ we rewrite the RG
equations \eqref{eq:FukuyamaRG} as follows:
\begin{subequations}
\begin{eqnarray}
\frac{d t}{d y} &=& -\frac{1}{2} t^2, \\
\frac{d g_{xy}}{d y} &=&  {g_{xy}}t .
\end{eqnarray}
\label{eq:FukuyamaBKTRG}
\end{subequations}
The RG flow dictated by Eqs.~\eqref{eq:FukuyamaBKTRG} is shown in Fig.~\ref{fig:FukuyamaTransition}.

\section{Starting values of RG: Levitation scenario and phase diagram}
\label{sec:startingvalues}

The  RG flow represented in Fig.~\ref{fig:flowDiagramusual} allows us to study
the phase diagram of the Dirac
QH effect\cite{KivelsonZhang1992,GorbarShovkovy2008}
and discuss the levitation of extended states taking place at  low magnetic
field (or, equivalently, strong impurity scattering) \cite{Laughlin1984}.

\subsection{Phase Diagram}

\begin{figure}
\includegraphics[scale=.6]{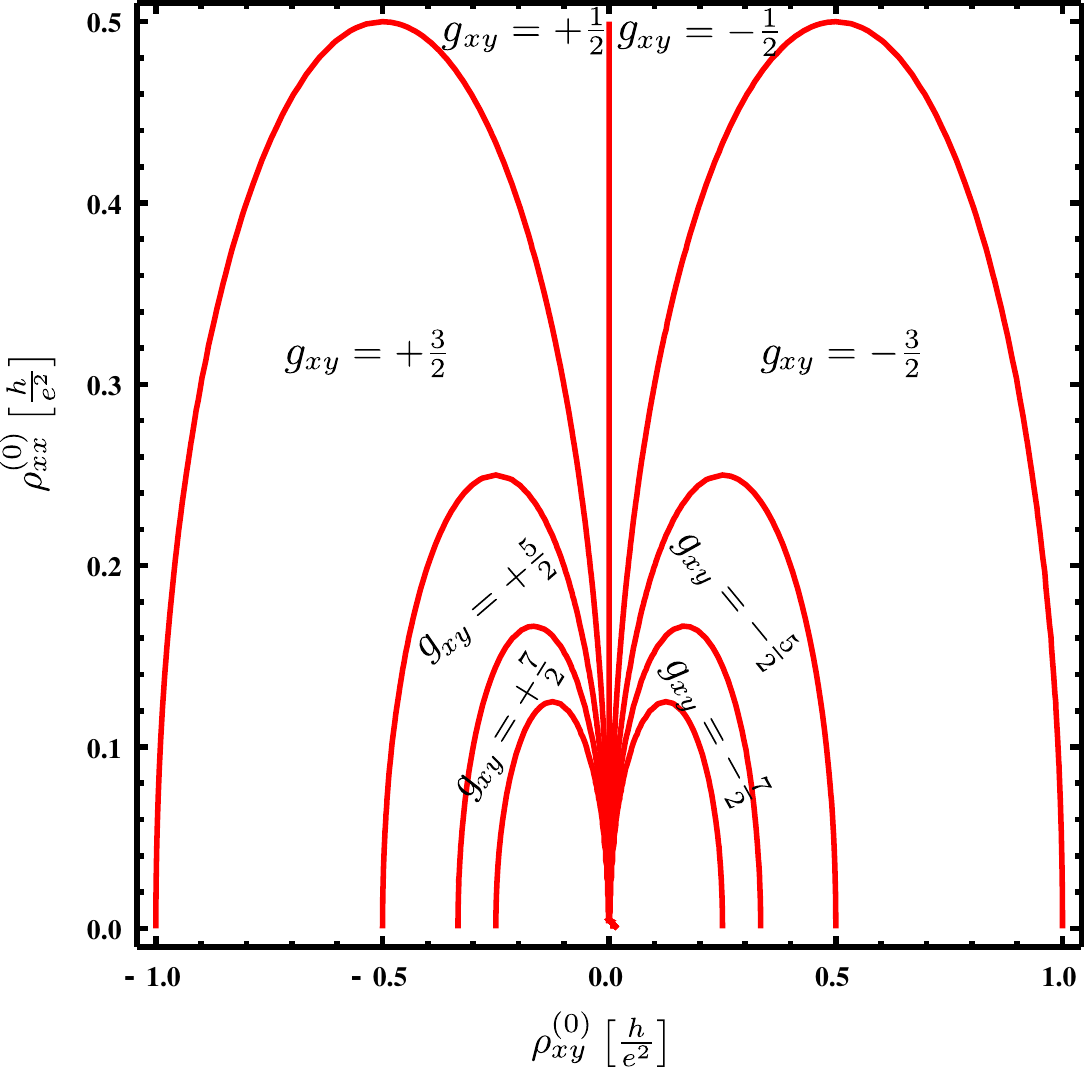}
\caption{Phase diagram for the QH of Dirac fermions in the plane of longitudinal
and Hall resistivities. }
\label{fig:PhaseDiag}
\end{figure}

The phase diagram of the Dirac quantum Hall effect can be built  by equating the
Drude value  of transverse conductance (which determines the electromagnetic
response of our system at length scales of the order of the mean free path and
constitutes the initial conditions for the RG flow discussed in the previous
section) to its values on the transition lines
\begin{equation}
\sigma_{xy}^{(0)} = n \frac{e^2}{h}, \; n \in \mathbb Z . \label{eq:Deloccondition}
\end{equation}
Figure~\ref{fig:PhaseDiag} shows the resulting phase diagram of our system  in
terms of the Drude resistivities. The major quantitative difference to the
situation of parabolic 2DEG\cite{KivelsonZhang1992} is the absence of any usual
(the one with $\sigma_{xy}=0$)
insulating phase: the diagram is covered by QH states only. In addition,
positions and radii of semicircular phase boundaries are modified.

\subsection{Drude conductance}

In order to build the phase diagram of the Dirac quantum Hall effect in terms
of the magnetic field $B$ and the chemical potential $\mu$ 
and to discuss the levitation scenario, we need to know the  Drude conductivity
tensor as a function of these parameters. In this section we present a
semiclassical derivation of the Drude conductivity tensor based on the Boltzmann
kinetic equation. We consider the general situation of Dirac fermions subject
to orbital magnetic field $B$ and Zeeman term
$H_Z=E_Z \sigma_z$ ($E_Z\ll\mu$), thus allowing for the anomalous Hall effect.
To the best of our knowledge, a comprehensive study of the Drude conductivity
tensor in these settings has not been reported in the literature so far (see
Refs. \onlinecite{SonSpivak2013,KimKimSasaki2014} for earlier work on the subject).

Our approach to the problem is justified provided that the quantum scattering
time $\tau_q\gg 1/\mu$ (which is the usual condition of applicability of
a semiclassical treatment). In the following we also assume that the classical
cyclotron frequency $\Omega_c^{\rm cl} = \left  \vert e B v_0^2/\mu c \right
\vert \ll 1/\tau_q$, which allows us to neglect the modification of the
scattering integral by the orbital magnetic field. We note that for smooth
disorder the transport scattering time $\tau_{tr}\gg \tau_q$, so that in this
case both regimes of the classically strong ($1/\tau_{tr} \gg \Omega_c^{\rm cl}
$) and classically weak ($1/\tau_{tr} \ll \Omega_c^{\rm cl} $) magnetic field
can be studied. On the other hand, for short-range impurities
$\tau_{tr}\sim \tau_q$, so that our approach does not apply to the limit of
strong fields.

The  relation between the Zeeman energy $E_Z$ (which is assumed to be small
compared to $\mu$) and the quantum scattering time $\tau_q$ controls the
importance of the coherence
in the scattering between the Zeeman-split bands. In the case of weak scattering
$E_Z\gg 1/\tau_q$  the inter-band coherence should be taken into account and
leads to the anomalous Hall effect.

To the leading order in small parameters $E_Z/\mu$ and   $\Omega_c^{\rm cl}/\mu$
the Drude conductivities are given by (we refer the reader to
appendix~\ref{sec:app:classicalTheory} for  detailed derivation)
\begin{subequations}
\begin{eqnarray}
\sigma_{xx} &=& \frac{\sigma_{xx}^{({\not B})}}{1 + (\Omega^{\rm cl}_c\tau_{tr})^2} 
\left [1 +  \frac{2\zeta \Omega^{\rm cl}_c\tau_{tr}^2/{{\tau_a} }}{1 + (\Omega^{\rm cl}_c\tau_{tr})^2} 
+ \frac{E_Z}{\mu^2 {\tau}_{{sj}}} \zeta \Omega^{\rm cl}_c\tau_{tr}\right ] , \notag \\&& \label{eq:sigmaxxAHEmaintext} \\
\sigma_{xy} &=& {-\frac{e^2}{2h}\left [\text{sign}(E_Z) \theta( E_Z^2 -\mu^2) + \frac{E_Z}{\vert \mu \vert} \theta(\mu^2- E_Z^2)  \right ]} \notag \\
 &+& \frac{\sigma_{xx}^{({\not B})}}{1 + (\Omega^{\rm cl}_c\tau_{tr})^2}  
 \Big [\zeta \Omega^{\rm cl}_c\tau_{tr}\left (1 +  \frac{2\zeta \Omega^{\rm cl}_c\tau_{tr}}{1 + (\Omega^{\rm cl}_c\tau_{tr})^2} \frac{\tau_{tr}}{\tau_a}\right ) \notag \\
 &&- \frac{\tau_{tr}}{\tau_a} -\frac{E_Z}{\mu^2 {\tau}_{sj}} \Big ] \label{eq:sigmaxyAHEmaintext} .
\end{eqnarray}
\label{eq:sigmaAHEmaintext}
\end{subequations}
Here  $\zeta = -\text{sign} (\mu B)$ and we have introduced the notations
$\nu(\mu)=\vert\mu\vert/(2\pi v_0^2) \theta(\mu^2-E_Z^2)$ and
$v(\mu)=v_0\sqrt{1-E_Z^2/\mu^2}$ for the density of states and velocity at  the
Fermi level. The classical conductance at zero magnetic field is
\begin{equation}
\sigma_{xx}^{(\not B)} (\mu) = 2\pi \frac{e^2}{h} \nu(\mu) \underbrace{\frac{v^2(\mu) \tau_{tr} (\mu)}{2} }_{D(\mu)} .
\end{equation}
At zero magnetic field, Eqs.~\eqref{eq:sigmaAHEmaintext} reproduce the results
of Ref.~\onlinecite{SinitsynSinova2007}.

\begin{figure}
\includegraphics[scale=.6]{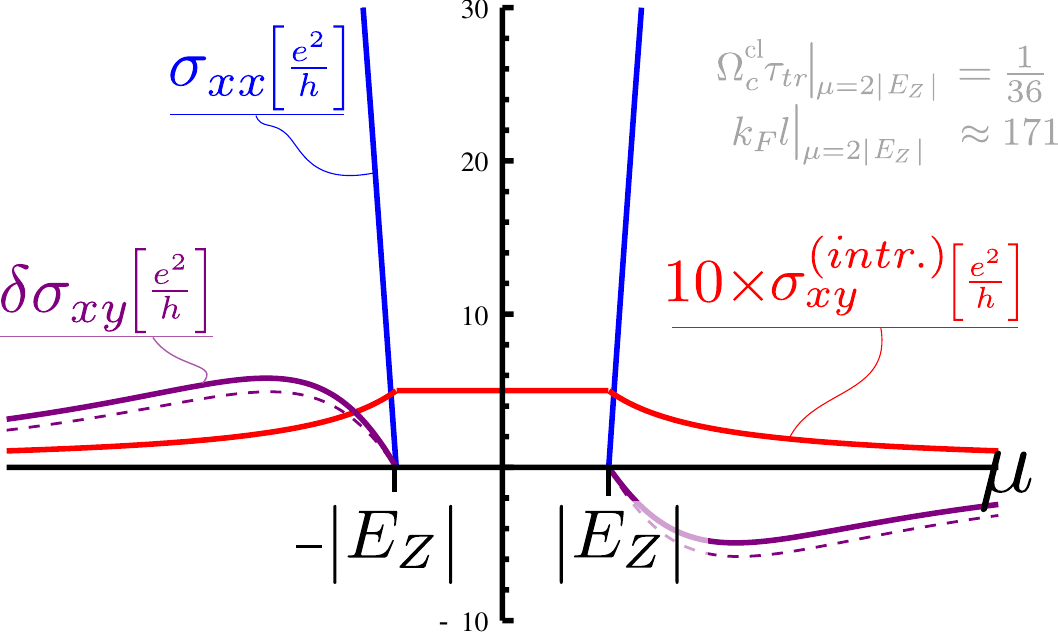}
\caption{Dependence of the classical conductivities on chemical potential.
The transverse conductivity  was split into intrinsic (red,
$\sigma_{xy}^{(intr.)}$) and Fermi-surface (violet, $\delta \sigma_{xy}$)
contributions. The dashed curves are obtained by reflection with respect to the
origin and visualize the magnitude of the AHE contributions.}
\label{fig:sigma(mu)}
\end{figure}

\begin{figure}
\includegraphics[scale=.6]{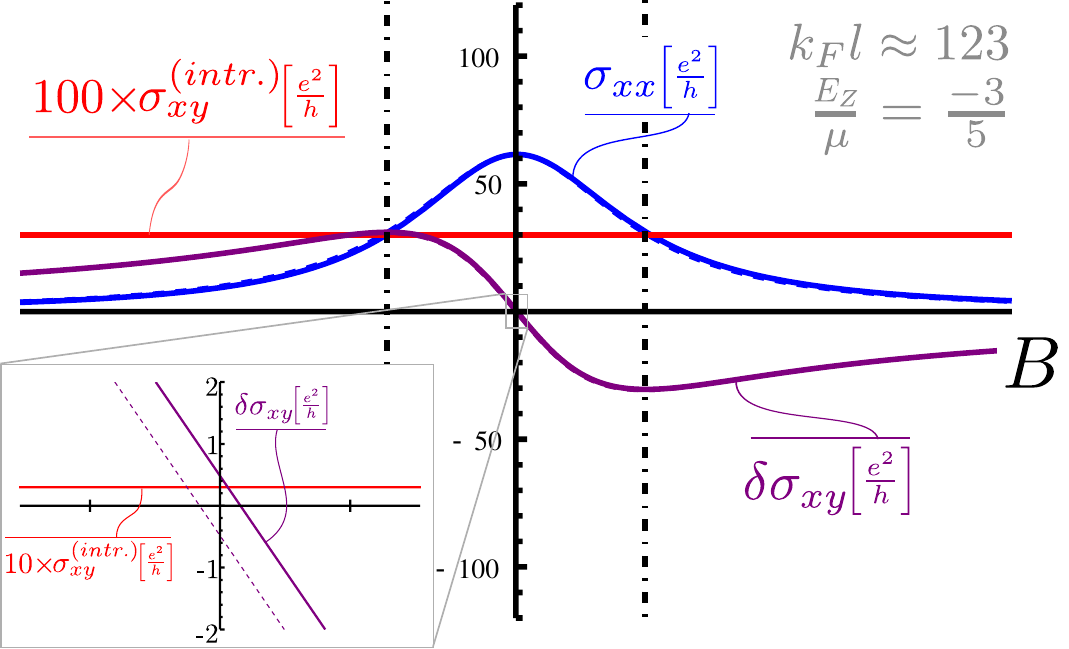}
\caption{Dependence of the classical conductivity tensor on magnetic field. The
transverse conductivity was split into intrinsic (red,
$\sigma_{xy}^{(intr.)}$) and Fermi-surface (violet, $\delta \sigma_{xy}$)
contributions. The dot-dashed vertical lines denote the position where
$\Omega_c^{\rm cl} \tau_{tr}= 1$. In the inset, the contributions to the transverse
conductance are plotted in the limit of vanishing $B$-field. The dashed
curves are obtained by reflection $B \rightarrow - B$ and visualize the
magnitude of the AHE contributions. In this plot, the Zeeman energy is assumed
to be $B$-field independent (which could result from an exchange coupling to a
nearby ferromagnetic layer).}
\label{fig:sigma(B)}
\end{figure}

The first term in the transverse conductivity in
Eq.~\eqref{eq:sigmaxyAHEmaintext} represents the so-called intrinsic Hall
conductivity\cite{XiaoChangNiu2010} related
to the modification of the  classical equations of motion for a wave packet
caused by the Berry curvature of the Dirac band.
Equations~\eqref{eq:sigmaAHEmaintext} contain also terms characterized by times
$\tau_a$ and $\tau_{sj}$. These are the scattering
times associated to the skew-scattering and the side-jump processes,
respectively.\cite{Sinitsyn2008,XiaoChangNiu2010,NagaosaOng2010} Assuming short
range impurities, one can express them, as well as the transport scattering time
$\tau_{tr}$,
in terms of disorder amplitude $V_0$
(see Ref.~\onlinecite{SinitsynSinova2007} and
appendix~\ref{app:sec:Semiclassical:EvaluationDirac}):
\footnote{For simplicity we omit the contribution to the
skew scattering time proportional to the third moment of the disorder potential,
see appendix \ref{app:sec:Semiclassical:EvaluationDirac}.
}
\begin{subequations}
\begin{eqnarray}
\frac{1}{\tau_{a}} &=&
\frac{\pi \nu(\mu)(n_i V_0^2)^2 \left[3 E_Z \left(\mu ^2-E_Z^2\right)\right]}{8 v_0^2\mu ^3},\\
\frac{1}{\tau^{sj}} &=& 2\pi n_i V_0^2 \nu(\mu),\\
\frac{1}{\tau_{tr}} &=& 2\pi n_i V_0^2 \nu(\mu) \frac{1 + 3 \left (\frac{E_Z}{\mu}\right )^2}{4}.
\end{eqnarray}
\label{eq:timesMainText}
\end{subequations}
Here $n_i$ is the concentration of impurities.
The behavior of the Drude conductivity tensor as a function of the chemical
potential and the magnetic field is illustrated in Figs.~\ref{fig:sigma(mu)} and
\ref{fig:sigma(B)}.

\subsection{Levitation of critical states}

Equating the Drude value of the Hall conductance,
Eq.~\eqref{eq:sigmaxyAHEmaintext}, with the transition lines of the RG flow
(i.e. integer $g_{xy}$), one obtains the phase boundaries of QH phases in the
$B$-$\mu $ plane.

\begin{figure}
\includegraphics[scale=.6]{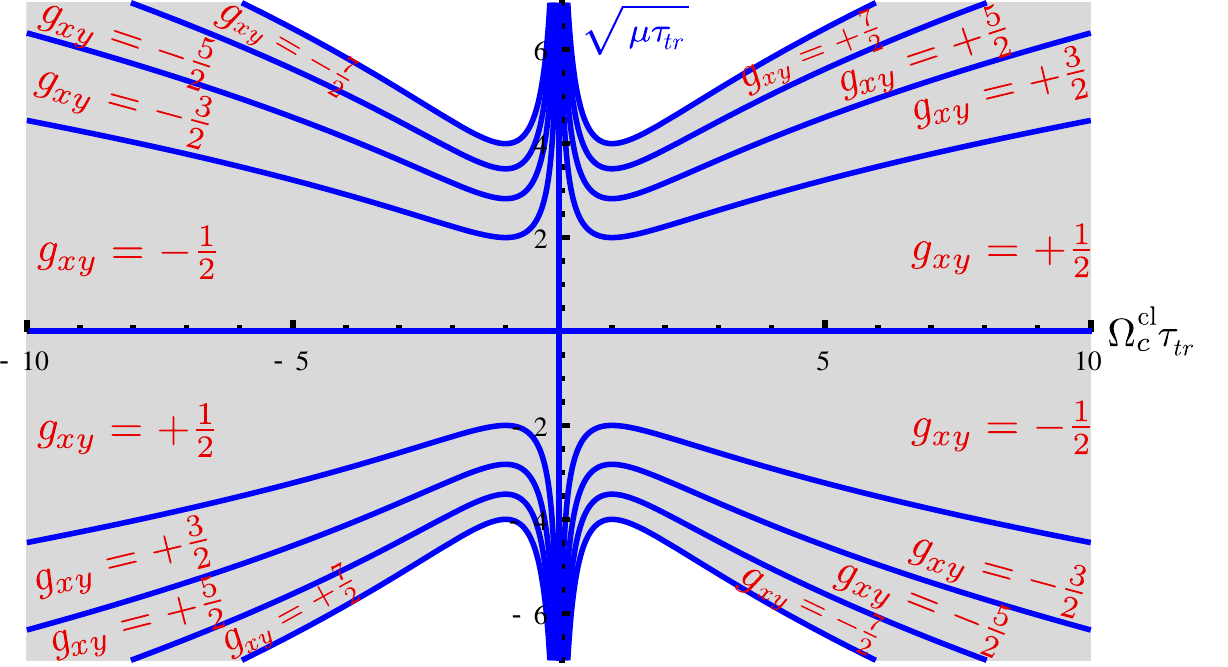}
\caption{Levitation of delocalized states of gapless Dirac fermions for Coulomb
impurities ($\mu/\tau_{tr}$ and
$\Omega_c^{\rm cl}\tau_{tr} \propto B$ are $\mu$-independent).
}
\label{fig:Levitationwomass}
\end{figure}

\begin{figure}
\includegraphics[scale=.6]{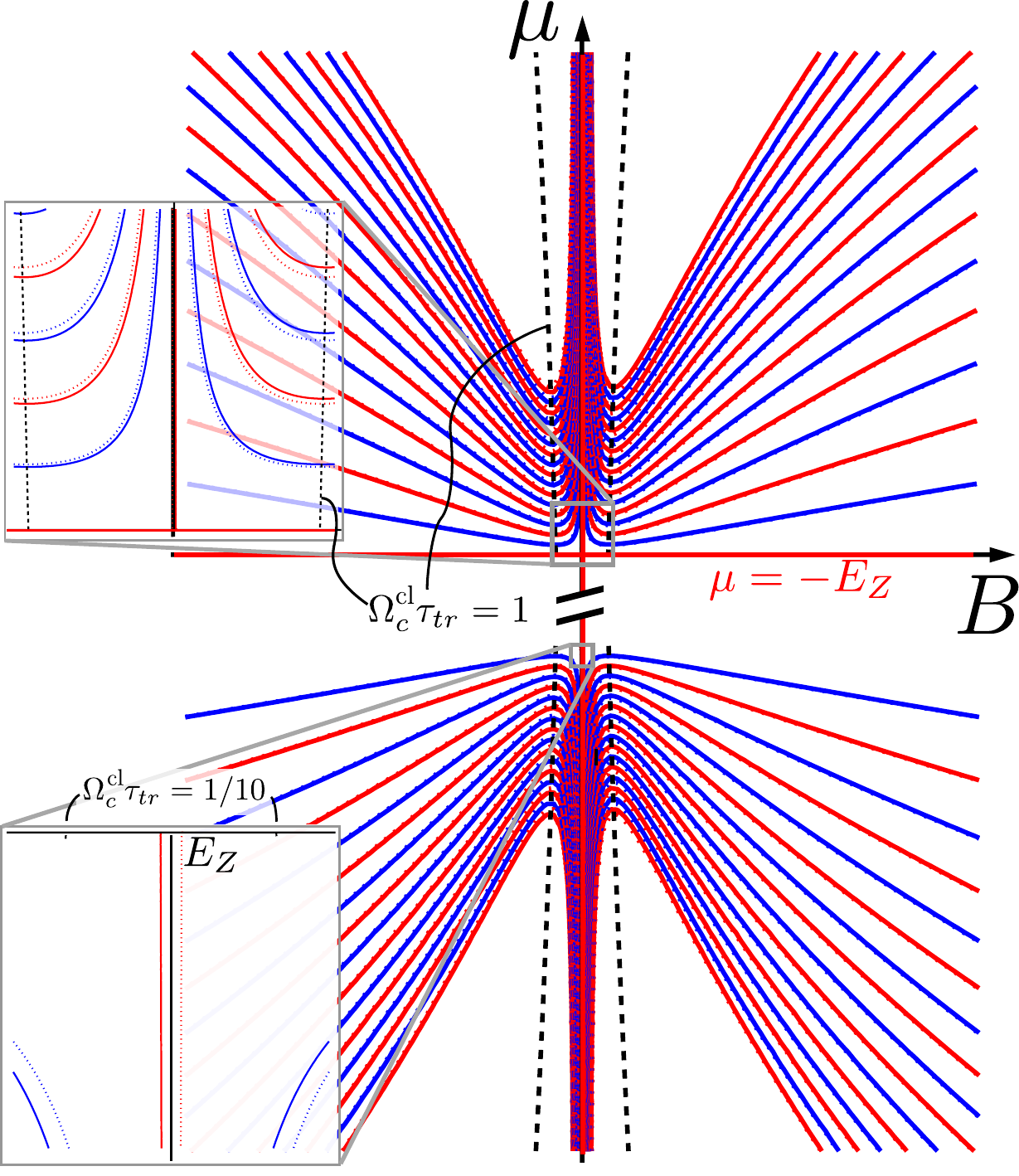}
\caption{Phase diagram of QH phases in the $B$-$\mu $ plane (``levitation
scenario'') for the case of short range impurities. Here, as in
Fig.~\ref{fig:sigma(B)}, $E_Z<0$ is assumed to be $B$-field independent. The
floating up of delocalized states with odd (even) number is depicted by
solid blue (red) curves.  The dashed lines correspond to
$\Omega_c^{\rm cl}\tau_{tr} =1$.
The insets magnify the region of weak magnetic fields.
The asymmetry under $B \rightarrow -B$ (dotted blue/red curves) is a
consequence of the AHE. Disorder strength is determined by $\vert \mu \vert
\tau_{sj} = 100$.}
\label{fig:Levitation}
\end{figure}

At small $E_Z\ll 1/\tau_q$  we can neglect all the contributions of the anomalous Hall effect in Eqs.~\eqref{eq:sigmaAHEmaintext}. Assuming further that the dominant source
of disorder are Coulomb impurities, we can deduce the dependence of the
transport scattering time on the chemical potential: $\tau_{tr}\propto \mu$.
Accordingly, the combination $\Omega_c^{\rm cl}\tau_{tr}\propto B$ is independent of
$\mu$, and the energies of critical states are given by
\begin{equation}
\mu_{deloc} = \pm \sqrt{\Omega_c^2 \vert n \vert \frac{1 +\left ( \Omega_c^{\rm cl} \tau_{tr}\right )^2}{(\Omega_c^{\rm cl}\tau_{tr})^2}+E_Z^2}, \quad n \in \mathbb Z,  \label{eq:levitation}
\end{equation}
where $\Omega_c$ is the quantum cyclotron frequency defined in Eq.~(\ref{eq:DiracEOmegac}).
For non-zero $n$, Eq.~\eqref{eq:levitation} describes the ``floating up'' or,
equivalently, ``levitation'' of delocalized critical states separating QH
phases. In the limit $\Omega_c^{\rm cl} \tau_{tr} \gg 1$ the usual LL spectrum of
gapped Dirac fermions is recovered.
For $n=0$ the solutions to be retained are $\mu_{deloc} = - \text{sign}(B)E_Z$.
This is a consequence of the AS index theorem, according to which the zeroth LL is
fully spin-polarized. The definite spin polarization predicts the sign of the
Zeeman energy and thus of the energy level. It is worth emphasizing that,
according to this result, the zeroth LL
is immune against strong scattering. As a result, in the limit of strong
scattering, $\Omega_c^{\rm cl} \tau_{tr} \ll 1$, the phases with $\sigma_{xy} = \pm
{e^2}/{2h}$ extend all the way from $\mu=0$ up to large values of $\mu$,
see Figs.~\ref{fig:Levitationwomass},~\ref{fig:Levitation}.
The robustness of the $\sigma_{xy} = \pm {e^2}/{2h}$ state against disorder
was indeed observed numerically.\cite{MorimotoAoki2009,NomuraFurusaki2008}

Our findings about the levitation of critical states in the absence of anomalous
Hall effect are summarized in Fig.~\ref{fig:Levitationwomass}. In this plot we
assumed that $E_Z=0$ and also took into account that for Coulomb impurities
$\sqrt{\mu \tau_{tr}}\propto \mu$. A generalization of this  plot to the case
of a fully developed anomalous QHE, $E_Z\gg 1/\tau_q$, is shown
in Fig.~\ref{fig:Levitation}.

\section{Experimental realization}
\label{sec:experiment}

After having derived the effective electrodynamic theory via the two-step
integration of matter fields, we return to the possibility of
experimental observation of the half-integer Hall conductivity.

\subsection{Typical experimental scales}

\begin{table}
\begin{tabular}{|l|c|c|}
\hline Quantity & Bi$_{2}$Se$_3$ & strained HgTe \\
\hline \hline Bulk band gap $M/k_B$ & 3480 K & 255 K \\
\hline Cyclotron freq. $\hbar \Omega_c/k_B$ & 210 K $\times \sqrt{B\left [T\right ]}$& 210 K $\times \sqrt{B\left [T\right ]}$\\
\hline Zeeman energy  $\vert E_Z \vert/k_B$ & 21 K  $\times B\left [T\right ]$ & 15 K  $\times B\left [T\right ]$ \\
\hline Scattering rate $\hbar/(\tau_{tr} k_B)$ & 127 K & 10 K\\ 
\hline
\end{tabular}
\caption{Typical experimental energy scales of 3D TI in Kelvin:\cite{BrueneMolenkamp2011, ShuvaevPimenov2013, LinLi2013} 
The bulk band gap $M$, the cyclotron frequency $\Omega_c = \sqrt{2 \vert e \vert B v^2_0/\hbar}$, 
the Zeeman energy $\vert E_Z \vert = g \mu_B B$ and the inelastic scattering rate $1/\tau_{tr}$. 
Bulk $g$-factors\cite{KoehlerWuechner1975,GuldnerMycielski1973} entering $E_Z$ were typically determined outside the TI regime. 
The presented values of $E_Z$ correspond to the maximal Zeeman energy, with perfect alignment of pseudospin $\sigma$ and 
electron spin $\v s$ (see main text). Here we disregard Zeeman energy due to exchange coupling.}
\label{tab:energyscales}
\end{table}

In Table~\ref{tab:energyscales}, typical energy scales of experimental setups
are presented. In the exemplary 3D TI experiments the Zeeman contribution
appears to be negligible.\cite{BrueneMolenkamp2011,ChenQi2011} This observation
is consistent with the calculated values of $\vert E_Z \vert = g \mu_B B$, see
Table \ref{tab:energyscales}.
As a side remark, we note that the spin $\sigma$ appearing in
Eq.~\eqref{eq:Htot} in general does not coincide with the physical electron spin
$\v s$.\cite{ZhangKaneMele2012,ZhangKaneMele2013} The mixing angle $\phi$
depends on how the crystal is cut and in general $E_Z \sim  g \mu_B B \cos\phi$.
In this section we neglect the possible $B$-independent Zeeman energy due
to exchange coupling and proximity to a ferromagnet.

\subsection{Image magnetic monopole effect}

\subsubsection{Magnitude of the effect}

It is useful to estimate the typical magnetic field strength associated with the
mirror monopole effect. The charge $Q_0 = U z_0$ at distance $z_0$ of the QH
system of ``filling factor'' $\nu$ is bound by the scale of ``magnetic
breakdown'' $\vert e \vert U \lesssim \Omega_c \left ( \eta_{\nu+1}\sqrt{ \vert
\nu+1 \vert}-\eta_\nu\sqrt{ \vert \nu \vert}\right )$. Using this bound and
Eq.~\eqref{eq:Mirrorcharges}, the ratio of image magnetic field and quantizing
external field can be estimated
\begin{equation}
\frac{B_{\text{image}}}{B} = \frac{g}{\vert \v r + z_0 \hat e_z \vert^2 B} \lesssim \frac{\alpha \Omega_c}{z_0 \vert e \vert B} \sim \alpha \frac{v_0}{c} \frac{l_B}{z_0}.
\end{equation}
This ratio is of the order of ${B_{\text{image}}}/{B} \sim 10^{-7}$ for the
typical magnetic field strength $B \sim 1T$ and the distance $z_0 \sim 1 \mu m$.

While in an idealized system at the QH plateau, the longitudial
conductivity $\sigma_{xx}$ is exactly zero, in a realistic situation it always
take a small but non-zero value due to a finite temperature. This allows a
rearrangement of charges in the QH system, which leads to screening of
the test charge and, as a result, destroys the driving force of ring currents
and thus the image monopole effect. In Ref.~\onlinecite{PesinMacDonald2013} the
decay rate of the image monopole effect after sudden appearance of a test charge
was found to be
\begin{equation}
\frac{1}{\tau} = \frac{2\pi \sigma_{xx}}{z_0} = \frac{\alpha c g_{xx}}{z_0} \sim 10^{12} s^{-1} \times g_{xx}.
\end{equation}
The decay of the magnetic monopole effect enforces one to perform finite
frequency measurements (or optical measurements, see below). We note,
however, that already in the early days of the QH effect the longitudinal
conductance on the QH plateau was demonstrated\cite{vonKlitzing} to be $g_{xx}
\lesssim 10^{-6}$. Thus, the decay of the monopole effect does not seem to
constitute an insuperable difficulty.

\subsubsection{Topological Magnetoelectric effect in thin 3D TI films}
\label{sec:ThinFilm}

In this section we consider the image magnetic monopole effect for a double QH
structure (i.e. a double domain wall of the theta angle multiplying the $\v E \cdot \v B$ term). This problem
is relevant for realistic 3D TI experiments: As was stated above, the electric
test charge should be placed at macroscopic distance from the QH systems. On the
other hand, typical 3D TI samples are only a few hundred \r{A}ngstr\"om thick.
Thus, the test charge simultaneously probes both TI surfaces, a double QH
structure.

\begin{figure}
\includegraphics[scale=.6]{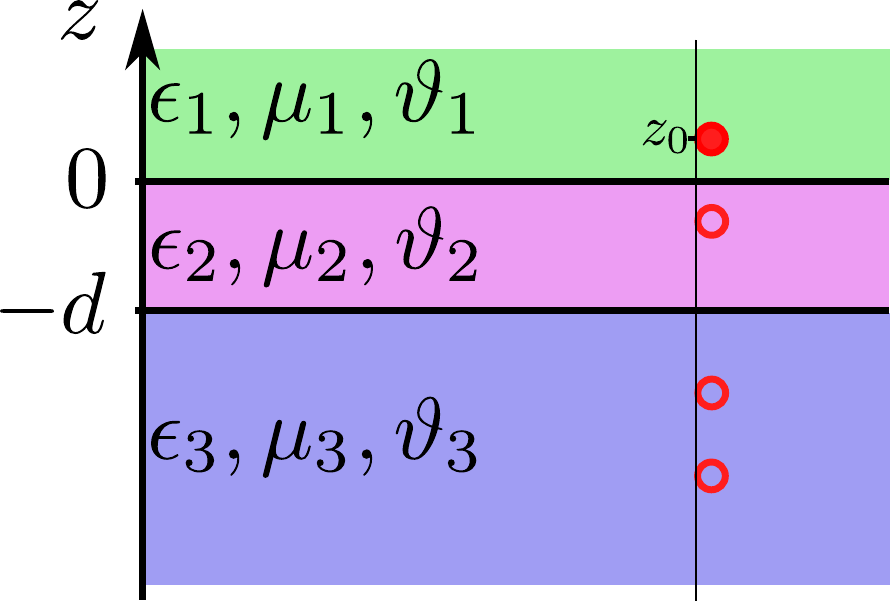}
\caption{Sketch of the setup discussed in the main text. An electric test charge
(solid dot) is placed above a double QH structure (e.g. a thin 3D TI slab) and
creates a series of magnetic and electric mirror charges (circles).}
\label{fig:Sketchimagecharges}
\end{figure}

A realistic experimental setup is shown  in Fig.~\ref{fig:Sketchimagecharges},
where the electric test charge  $Q_0$ (solid dot) is placed above a double QH
structure at position
$\left (0,0,z_0\right )$. The upper QH system, in the plane $z = 0$, has Hall
conductance $\sigma^{top}_{yx} = \left (\vartheta_2 - \vartheta_1\right
)e^2/2\pi h$, while the lower one is characterized by $\sigma^{bottom}_{yx} =
\left (\vartheta_3 - \vartheta_2\right )e^2/2\pi h$. In addition, in the three
bulk regions denoted by $a=1$ ($0<z$), $a=2$ ($-d<z<0$) and $a=3$ ($z<-d$)
localized charges might induce non-trivial electric permittivity $\epsilon_a$
and magnetic permeability $\mu_a = 1/\epsilon_a c_a^2$.


Following Ref.~\onlinecite{Karch09}, we use the unified description 
in terms of the vector $\left (\v D_a, 2\alpha \v B_a\right )^T$ which is connected to $\v E_a$ and $\v H_a$ via
\begin{subequations}
\begin{equation}
\left (\begin{array}{c}
\v D_a \\ 2\alpha \v B_a
\end{array} \right ) = \mathcal M_a \left (\begin{array}{c}
2\alpha \v E_a \\ \v H_a
\end{array} \right )
\end{equation}
with the matrix
\begin{equation}
\mathcal M_a = \frac{2\alpha}{c_a^2\epsilon_a} \left (\begin{array}{cc}
\frac{\vartheta_a^2}{4\pi^2} + (\frac{c_a\epsilon_a}{2\alpha})^2 & -\frac{\vartheta_a}{2\pi} \\
-\frac{\vartheta_a}{2\pi} & 1
\end{array} \right ).
\end{equation}
\label{eq:constitutiverelations}
\end{subequations}
The electromagnetic field above the plane $z=0$ can be expressed in terms of the
two-component potential $\underline \Phi_1= \left (\Phi_{1,E}, 2\alpha
\Phi_{1,M}\right )^T$,
\begin{equation}
\left (\v D_1, 2\alpha \v B_1\right )^T = - \nabla \underline \Phi_1.
\end{equation}
To present the potential $\underline \Phi_1$, it is convenient to perform the
Fourier transformation with respect to coordinates in the plane, $(x,y) \to
(q_x,q_y)$,
\begin{equation}
\underline \Phi_{1} \left (x,y,z,z_0\right ) = \int_0^\infty dq \;\frac{q
}{2\pi} \underline \Phi_{1}  \left (q,z,z_0\right )J_0\left (q\rho\right ).
\label{eq:DoubleQHpotential}
\end{equation}
Here $J_0\left (q\rho\right )$ is the zeroth Bessel function, $\rho =
\sqrt{x^2 + y^2}$ is the modulus of the 2D component of the position vector,
and $q = \sqrt{q_x^2 + q_y^2}$ is the norm of the 2D  component of momentum.
As shown in appendix~\ref{sec:app:finitethickness}, the Fourier transform of the
two-component potential  $\underline \Phi_1$ is given by
\begin{equation}
\underline \Phi_{1} \left (q,z,z_0\right ) = \frac{2\pi}{q} \bigg \lbrace e^{-\vert z-z_0 \vert q} +  e^{-(z+z_0)q} T_{\rm eff} \bigg \rbrace \left (\begin{array}{c}
Q_0 \\
0
\end{array} \right ). \label{eq:DoubleQHpotentialFT}
\end{equation}
Here we introduced the matrices
\begin{eqnarray}
T_{\rm eff} &=&  \left (R_{32}^+R_{21}^+e^{dq}+R_{32}^-R_{21}^-e^{-dq}\right )^{-1} \notag \\ &&\times \left (R_{32}^+R_{21}^-e^{dq}+R_{32}^-R_{21}^+e^{-dq}\right )
\end{eqnarray}
and $R_{ab}^\pm = 1 \pm \mathcal M_a \mathcal M_b^{-1}$. Each of the limits
$d\rightarrow \infty$, $d \rightarrow 0$, $\left (\epsilon_2,\mu_2,
\vartheta_2\right ) = \left (\epsilon_3, \mu_3, \vartheta_3\right )$ and $\left
(\epsilon_2,\mu_2, \vartheta_2\right ) = \left (\epsilon_1, \mu_1,
\vartheta_1\right )$ reproduces the result for a single domain wall, see
App.~\ref{sec:app:finitethickness}.

\begin{figure}
\begin{minipage}[b]{0.4\linewidth}
\includegraphics[height=3cm]{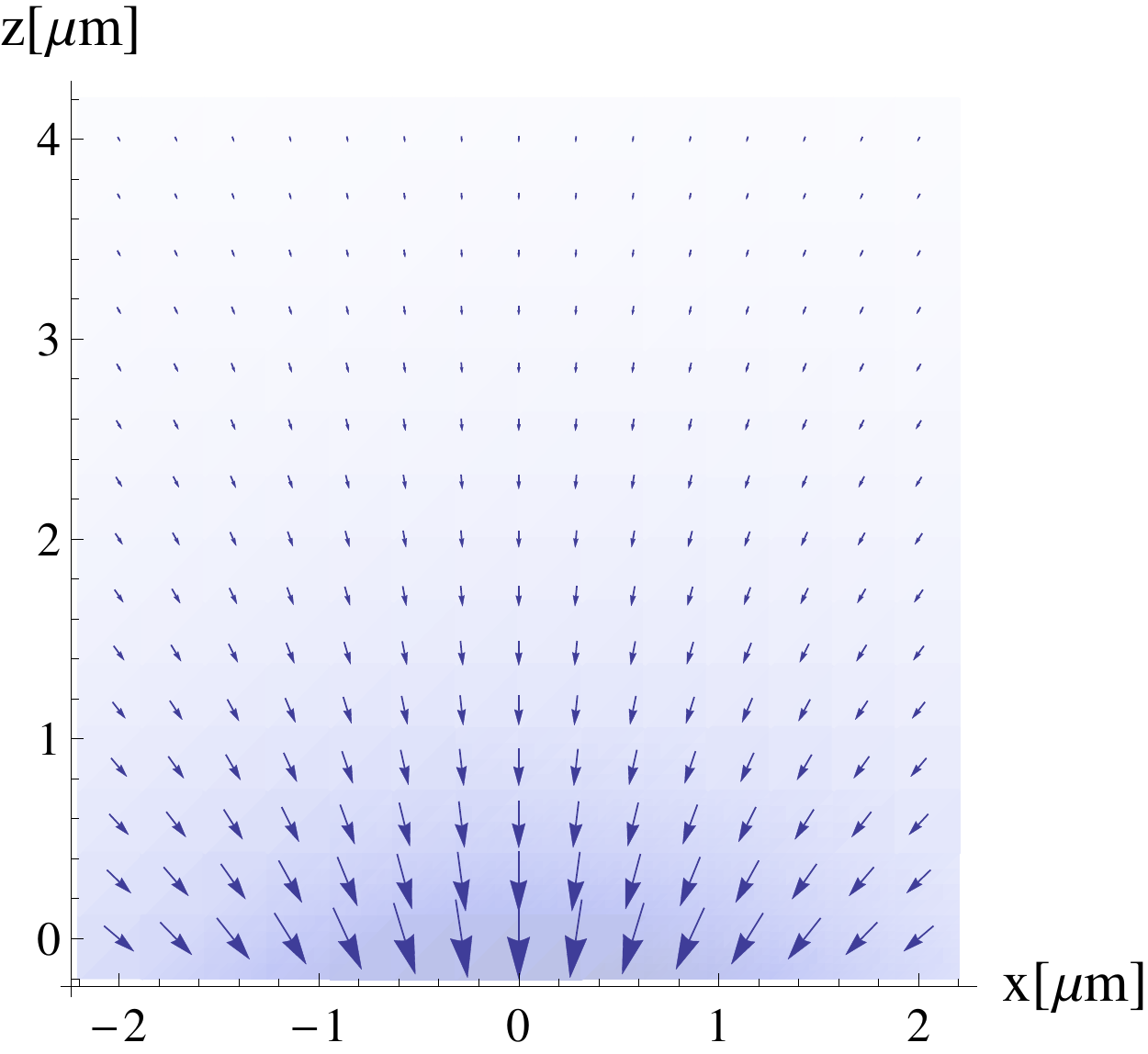}
\end{minipage}
\begin{minipage}[b]{0.6\linewidth}
\includegraphics[height=3cm]{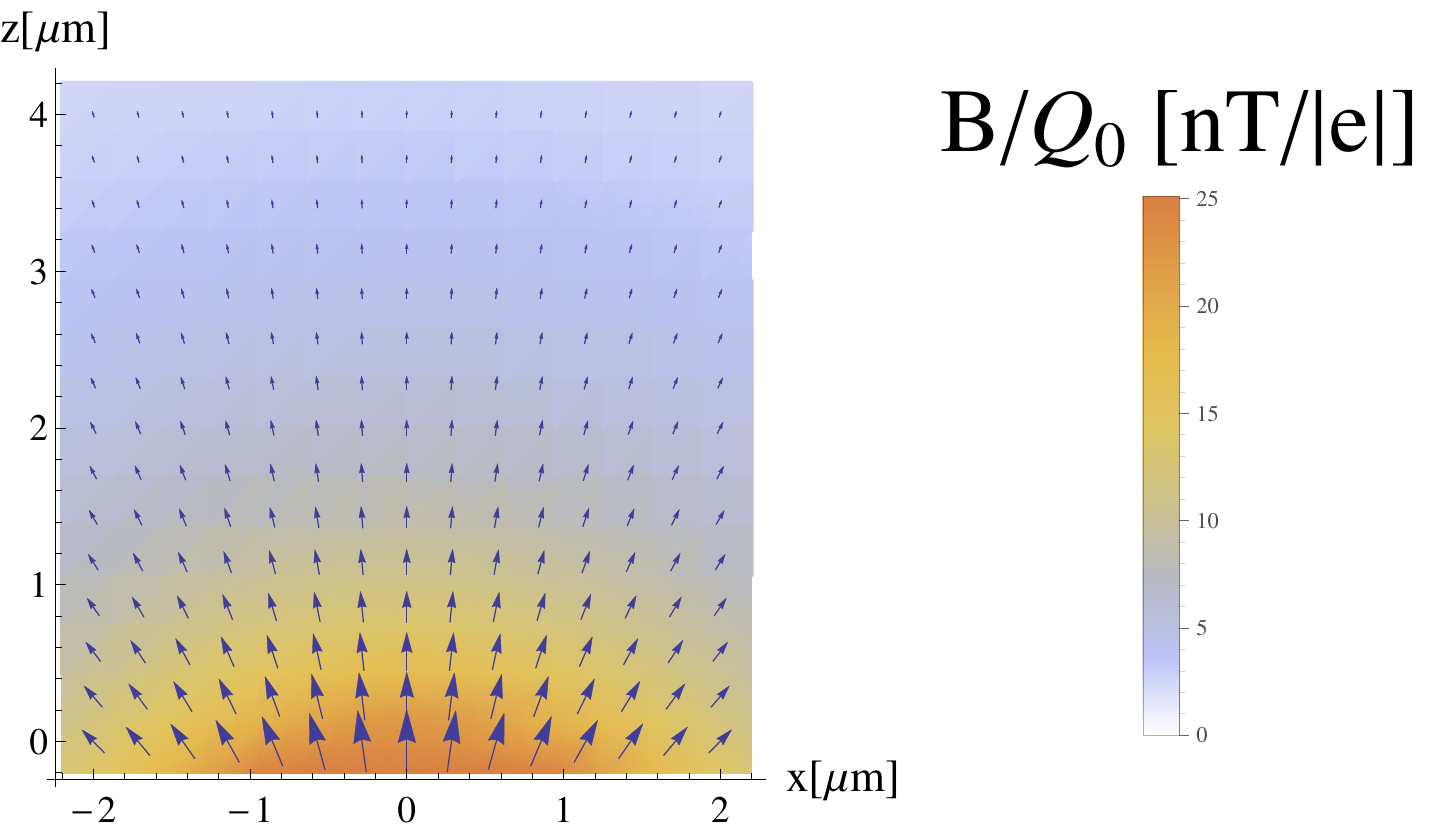}
\end{minipage}
\caption{The magnetic field configuration for 3D TI in the QH state in a setup
as depicted in Fig.~\ref{fig:Sketchimagecharges}. Left: slab of thickness $10
\mu m$. Right: slab of thickness $20 nm$. In both figures $z_0 = 2
\mu m$, $\sigma^{top}_{xy} = e^2/2h$, $\sigma^{bottom}_{xy} = - 7 e^2/2h$ and,
for simplicity, $\epsilon_1 = \epsilon_2 = \epsilon_3$ and $\mu_1 = \mu_2 =
\mu_3$.}
\label{fig:Monopoles}
\end{figure}

The two-component potential, Eq.~\eqref{eq:DoubleQHpotential}, can
also be represented as an infinite sum of mirror charges, see
App.~\ref{sec:app:finitethickness}.\footnote{Clearly, the first of these
charges corresponds to Eq.~\eqref{eq:Mirrorcharges}.} In the limit $z_0 \ll d$
the dominant contribution arises from the mirror charge, which is located in $-d
< z <0$ and solely determined by $\sigma^{top}_{xy}$. In contrast, for $z_0 \gg
d$ the double QH system behaves effectively as a single QH system with
the Hall conductivity $\sigma^{tot}_{xy}=\sigma^{top}_{xy} +
\sigma^{bottom}_{xy}$. Again the field configuration displays the mirror
monopole, but this time its strength is determined by $\sigma^{tot}_{xy}$. This
is illustrated in Fig.~\ref{fig:Monopoles} where we plot the magnetic field
corresponding to the potential, Eq.~\eqref{eq:DoubleQHpotential}, for two
otherwise identical 3D TI slabs of different thickness, $d = 10 \mu m$ and $d
= 20 nm$.
\footnote{For the actual plotting of Fig.~\ref{fig:Monopoles}, the series of
mirror charges was truncated. For the left (right) plot corresponding to $d=10
\mu m$  ($ d=20nm$) the first 21 (201) mirror charges were taken into account.}
In these plots, we have assumed that  the distance of a charge
from the top surface is $z_0 = 2 \mu m$ and the Hall conductivities
are $\sigma^{top}_{xy} = e^2/2h$ and $\sigma^{bottom}_{xy} = - 7 e^2/2h$.
Thus, for a thick slab the condition $z_0 \ll d$ is well satisfied and the
magnetic field is mainly determined by the mirror monopole corresponding to the
upper surface with $\sigma^{top}_{xy} = e^2/2h$. On the other hand, the
thickness $d$ of a thin slab is much smaller than $z_0$, so that the monopole
corresponding to the total Hall conductivity $\sigma^{tot}_{xy} = - 3e^2/h$ is
observed.

It is worth emphasizing that the magnetic field plotted in
Fig.~\ref{fig:Monopoles} only includes the field induced by the image monopole
and not the magnetic field generating the QH state.  As noted before,
typical magnetic field strengths in 3D TI QH experiments are of the order of a
few Tesla.\cite{BrueneMolenkamp2011} The induced monopole field per test charge
$Q_0$ is of the order of $10 nT/\vert e \vert$. We can estimate the voltage
associated to magnetic breakdown to be $U \sim 0.01 \dots 0.1 V$, which at a
distance of $2 \mu m$ corresponds to $Q_0 \sim 14 \dots 140 \vert e \vert$.
Therefore, the induced magnetic field can be expected to be of the order of $0.1
\dots 1 \mu T$. Measurement of such a variation of the magnetic field
is quite challenging from the experimental point of view.

\subsection{Spectroscopic measurement: Topological Faraday and Kerr rotation}

As discussed above, a measurement of the half-integer Hall conductivity should
be local and contactless. The magnetic monopole effect satisfies these
requirements but its magnitude is very small and might pose a serious
experimental obstacle. This motivates us to think about possible alternatives.

A possible experimental probe of the  QH effect is based on
the topological Faraday and Kerr rotation in spectroscopic
setups.\cite{MaciejkoZhang2010,QiZhang,TkachovHankiewicz2011,TseMacDonald2011}
In these experiments the frequency of light is typically of the order of
THz, with a wavelength $\lambda \sim 300 \mu m$.  For a sufficiently disordered
realistic system the condition
\begin{equation}
\omega \tau \equiv \frac{c}{v_0} \frac{l}{\lambda} \lesssim 1
\end{equation}
can be well satisfied. The system is then in the diffusive regime, opening a
possibility for approaching the regime of quantized Hall conductivity.

Faraday and Kerr rotation induced by surface states of 3D TI were studied in
recent spectroscopic experiments \cite{AguilarArmitage2012,ShuvaevPimenov2013}
(see also earlier works
Ref.~\onlinecite{JenkinsDrew2010,HancockMolenkamp2011}),
and magnetooscillations of conductivities were indeed observed. In these
experiments, the systems were in the diffusive regime, $\omega \tau \sim 0.1
\dots 1$. This implies that the RG
flow of conductivities, Fig.~\ref{fig:flowDiagramusual}, should be directly
observable in the frequency dependence of optical conductivity
$\sigma_{ij}\left (\omega\right )$ measured in THz spectroscopy.

There exists, however, a problem related to a small thickness $d$ of realistic
TI samples. Indeed, in order to probe separately each of the surfaces in a
spectroscopic experiment, $d$ should be larger than the wavelength $\lambda$.
On the other hand, for state-of-art structures the opposite condition is
satisfied, $d \ll \lambda$. This appears to be a serious obstacle for a
measurement of conductivities of individual surfaces by spectroscopic means.

\section{Conclusions and Outlook}
\label{sec:outlook}

In this paper, we studied the quantum Hall effect of a single-species
(2+1)-dimensional Dirac fermion. We focused on the case of 3D topological
insulator surface states, where the half-integer QHE is a manifestation of
fermion number fractionalization, see Sec.~\ref{sec:TME}. Our key results are
as follows:
\begin{enumerate}
\item We explained in
Sec.~\ref{sec:TME} why a naive attempt to measure the half-integer Hall
conductivity of a surface of a 3D TI fails.
\item
We have further reviewed in Sec.~\ref{sec:TME} the topological magnetoelectric
effect and demonstrated that it can be used for measurement of the
half-quantized Hall conductivity in topological insulator surfaces with locally
broken time-reversal invariance.
\item
Subsequently, in Sec.~\ref{sec:Laughlin} we have shown that the half-quantized
value is not in contradiction to Laughlin's flux insertion argument (which
predicts, in its conventional form, an integer value of $\sigma_{xy}$ as a
consequence of gauge invariance). Specifically, we have modified
Laughlin's argument to the case of a 3D TI and demonstrated that it leads to
half-quantized values of conductivities of each of the surfaces.
\item
Next, in Sec.~\ref{sec:Semiclassical}, we employed the vortex state basis to
calculate the conductivity tensor and explicitly uncovered the half-integer
contribution of the zeroth LL.  This calculation also allowed us to extend the
topological magnetoelectric effect beyond the linear response.
\item
In Sec.~\ref{sec:NLSM} we derived the unified field theory treating both
diffusive matter fields and EM gauge potentials. In contrast to the case of the
integer quantum Hall effect, two different theta angles appear. One of them is
associated to the Hall conductivity $\sigma_{xy}$, while the other one
(reminiscent of chiral anomaly) provides a shift of the renormalization flow
diagram.
\item
We discussed the RG flow and the phase diagram in great detail.
To this end, the semiclassical conductivity tensor of Dirac fermions in
magnetic field was derived in Sec.~\ref{sec:startingvalues}.
\item
Finally, in Sec.~\ref{sec:experiment}
we carried out an analysis of conditions for experimental observation of the
half-integer QHE. We have paid particular attention to implications of
the slab geometry characteristic for currently manufactured TI samples.
\end{enumerate}

We conclude the paper with a few remarks of a more general character:

\begin{itemize}
\item[(i)]
First, we would like to comment on the issue of a physically
observable topological $\mathbb Z_2$ invariant of 3D TIs that has been
discussed in the literature. It has been argued that the theta angle
$\vartheta$ of the bulk $\v E \cdot \v B$-term plays this role.
We have shown that in the presence of a boundary, this angle gives rise to a
topological angle $\theta = \pm \pi$ of the boundary NL$\sigma$M theory.
While this angle is indeed a $\mathbb Z_2$ invariant, it does not couple to
gauge potentials and thus does not directly represent a measurable quantity.
The situation should be contrasted to the paradigmatic case of the QH effect
where the Hall conductivity which is the topological invariant directly
corresponds to the theta angle of the theory. In the present case,
the role of the angle $\theta$ is in shifting the RG flow diagram (and thus the
fixed-point values of the Hall conductivity). Such an indirect
physical meaning of topological invariants might be applicable also to other
symmetry classes.

\item[(ii)]
We hope that our work will motivate further  experimental efforts for the
observation of half-integer QHE in surfaces of 3D TIs. It is also worth
emphasizing that the TME is not specific to to 3D TI surfaces but should also
take place in conventional QH systems. 
There, the parameter regime might be more
favorable (allowing in particular a higher number of charges on the probing
tip). 
Corresponding experimental studies would be certainly of great
interest.

\item[(iii)]
Some of our results on transport properties of 2D Dirac fermions may be
relevant also in a context more general than TI surfaces. In particular,
the peculiar critical behavior of the $\v B = 0$ QH transition
(Sec.~\ref{sec:Fukuyama}) could also be observable in doped graphene.
Further, the calculation of the semiclassical transport coefficients (that
served as starting values for the RG) in Sec.~\ref{sec:startingvalues} and
App.~\ref{sec:app:classicalTheory} is generic and applies to any 2D material
with a non-vanishing Berry curvature.

\item[(iv)]
Finally, it is worth emphasizing that even though topological insulators
avoid the fermion doubling theorem \cite{NielsenNinomiya1981} by spatially
separating two single Dirac or Majorana modes, the gauge invariance still
implies strong constraints. Specifically, the anomalous
contributions of the two surfaces mutually cancel, both in the
analysis of Laughlin's argument, Sec.\ref{sec:Laughlin}, and of the parity
anomaly, Sec.~\ref{sec:NLSM:parity}. Such an additive cancellation of anomalies
from opposite boundary states also occurs, e.g., in the context of topological
Josephson junctions.\cite{FuKane2009,CrepinTrauzettel2013} In this reasoning,
the fermion doubling plays a crucial role (even though the two species are
spatially separated). On the other hand, one could also regard the Dirac (or
Majorana) fermions on the whole boundary of TIs in dimensions larger than one as
a single mode.
In the context of 3D TIs, it is instructive to imagine a sample of, e.g.,
spherical form.
Then the notion of fermion doubling loses its meaning, as there is just a
single species of Dirac fermions on the whole closed boundary of the sample.
Nevertheless, as we discussed in Sec.~\ref{sec:NLSM:parity}, the theory is
internally consistent and the parity anomaly is avoided. Local response
properties are controlled by a theory of a single species of Dirac fermions,
and the theory is unambiguously defined due to the fact that surface is a
closed manifold.
We therefore conclude that the reason for
single-species Dirac fermions to appear only on boundaries of higher-dimensional
bulk systems (which themselves have no boundary) is very deep and ultimately
follows from gauge invariance. We expect similar arguments to
hold also for other dimensions and classes of topological insulators and
superconductors.

\end{itemize}

\section{Acknowledgements}

We acknowledge useful discussions with J.T. Chalker, T. Champel, S. Florens, Y. Gefen, T. Giamarchi, D.
Hernangomez-Perez, P. Kotetes, D.G. Polyakov, M. Titov, and P. W\"olfle. This work was supported by DFG
SPP 1666 ``Topological Insulators'', German-Israeli Foundation,
BMBF, the Council for Grant of the President of Russian Federation (Grant No. MK-4337.2013.2),
Dynasty Foundation, RAS Programs, RFBR Grant No. 14-02-00333 and
by Russian Ministry of Education and Science under the contract 14Y.26.31.0007.

\appendix

\section{Semiclassical calculation of current density}
\label{app:sec:Semiclassical}
In this appendix we present the semiclassical calculation of current density induced by external potential in QH sample with smooth disorder.

\subsection{Notation}

We consider the model of a single Dirac cone, specified by Eqs.~\eqref{eq:H0} and~\eqref{eq:Htot}. Further, we use the notation
\begin{equation}
H_0 = v_0(\Pi_x \sigma_y - \Pi_y \sigma_x) = \left (\begin{array}{cc}
0 & -iv_0\Pi_- \\
iv_0\Pi_+ & 0
\end{array} \right ).
\end{equation}
with
\begin{equation}
-i \Pi_- = -i \left (\Pi_x -  i \Pi_y\right ) , \quad
i\Pi_+ =i\left ( \Pi_x +  i \Pi_y\right ).
\end{equation}
These objects have the following commutation relation
\begin{equation}
\left [-i \Pi_-,i\Pi_+\right ] = 2i \left [\Pi_x,\Pi_y\right ] = 2 \vert e \vert \epsilon_{ij} \partial_i \mathcal A_j =  2/ l_B^2,
\end{equation}
where $l_B = \left (\vert e \vert B\right )^{-1/2}$ is the magnetic length.
Under the assumption of $B >0$ we define creation and annihilation operators
\begin{equation}
b = - \frac{l_B}{\sqrt{2}} \left (-i \Pi_-\right ), \quad
b^+ = - \frac{l_B}{\sqrt{2}} i \Pi_+ .
\end{equation}
Then, using the cyclotron frequency $\Omega_c = \frac{\sqrt{2}v_0}{l_B}$, we can rewrite Eq. \eqref{eq:H0} as
\begin{equation}
H_0 = -\Omega_c \left (\begin{array}{cc}
0 & b \\
b^+ & 0
\end{array} \right ).
\end{equation}

Independently of the gauge, this Hamiltonian has eigenstates
\begin{equation}
\Ket{n, k}_D = \frac{1}{\sqrt{1 + \eta_n^2}} \left (\begin{array}{c}
- \eta_n \Ket{\vert n \vert - 1, k} \\
\Ket{\vert n \vert, k}
\end{array} \right ), \label{eq:app:Diracstate}
\end{equation}
with eigenenergies $E_n = \Omega_c \eta_n \sqrt{\vert n \vert}$ ($\eta_n = \text{sign}(n)$ for $n \neq 0$ and $\eta_0 =0$).
The quantum number $n \in \mathbb Z$ labels the Landau level (LL) while $k = 1, 2, \dots , \frac{\Phi_{tot}}{\Phi_0}$ 
accounts for the degeneracy. The eigenstates $\Ket{\vert N \vert, k}$ constituting the spinor in Eq. \eqref{eq:app:Diracstate}  
are the conventional eigenstates of the $\vert N \vert$-th LL 
for electrons with parabolic dispersion.

\subsection{Vortex states}
For the semiclassical calculation we use the overcomplete basis of LL eigenfunctions\cite{ChampelFlorens2007,ChampelFlorens2010} (``vortex states'').  In this representation the discrete quantum number $k$ is replaced by the continuum guiding center position $\v R \in \mathbb R^2$. The wavefunction for the $n$-th Landau level 
is then given by
\begin{equation}
 \braket{\v r | n,\v R} = \frac{ e^{in \arg(\v r - \v R)}}{\sqrt{2\pi n!}l_B}\left \vert \frac{\v r - \v R }{\sqrt{2}l_B}\right \vert^n e^{-\frac{\left (\v r- \v R\right )^2 - 2i \hat{\v z}
\cdot (\v r \times \v R)}{4l_B^2}}. \label{eq:vortexstate}
\end{equation}
The vortex states are `semi-orthogonal'
\begin{equation}
\braket{n,\v R \vert n', \v R'}= \delta_{nn'} e^{-\frac{(\v R - \v R')^2 - 2 i \hat{\v z}\cdot (\v R \times \v R')}{4l_B^2}}
\end{equation}
and produce the resolution of identity
\begin{equation}
\int \frac{d^2 R}{2\pi l_B^2} \sum_{n=0}^{\infty} \ket{n, \v R}\bra{n, \v R} = \mathbf 1. \label{eq:Resolofidentity}
\end{equation}
Note that the summation  and integration in~\eqref{eq:Resolofidentity} can not be interchanged. On the other hand, for fixed $n$ and $n'$
\begin{equation}
\int d^2 R \braket{\v r \vert n, \v R}\braket{n', \v R \vert \v r'} = \braket{n, \v r \vert n' \v r'},
\end{equation}
as follows from the identity $\braket{\v r \vert n, \v R} = \braket{n, \v r \vert \v R}$.
The vortex states  $\ket{n, \v R}_D$ for the Dirac Hamiltonian are the spinors constructed out
of states (\ref{eq:vortexstate}) analogously to Eq.~(\ref{eq:app:Diracstate}).

\subsection{Current operators}
We define now the current operators $j_\pm = j_x \pm i j_y$ via $\v j = i e [H_0,\v r]$:
\begin{subequations}
\begin{eqnarray}
j_+ &=& -2i e v_0\left (\begin{array}{cc}
0 & 1 \\
0 & 0
\end{array} \right ), \\
j_- &=& 2i e  v_0\left (\begin{array}{cc}
0 & 0 \\
1 & 0
\end{array} \right ).
\end{eqnarray}
\end{subequations}
Consequently,  their matrix elements can be expressed via spinor components of $\ket{n, \v R}_D$ in the following way
\begin{eqnarray}
\left (j_+\right )_{12} &=& -2i  e v_0 \braket{n_1,\v R_1, \uparrow \vert n_2, \v R_2, \downarrow }, \\
\left (j_-\right )_{12} &=& 2i  e  v_0 \braket{n_1,\v R_1, \downarrow \vert n_2, \v R_2, \uparrow }.
\end{eqnarray}
[For any operator $O$ we use the short hand notation $\left (O\right )_{12} \equiv  \braket{1 \vert O \vert 2}$ and $\ket{1} = \ket{n_1, \v R _1}_D$ .]

\subsection{Gradient expansion}

The central assumption for the semiclassical calculation is that the potential  $V\left (\v r\right )$ is smooth on the scale of the magnetic length. In the vortex state basis we expand its matrix elements in gradients:
\begin{equation}
\left (V\right )_{12} = \left (V\right )^{(0)}_{12} + \left (V\right )^{(1)}_{12} + \mathcal{O} \left (l_B^2\partial^2 V\right ).
\end{equation}
The zeroth order is [$\v c_{12} = (\v R_2 + \v R_1)/2$]:
\begin{equation}
\left (V\right )^{(0)}_{12} = V(\v c_{12}) \braket{1 \vert 2}.
\end{equation}
The first order is [$\v d _{12} = (\v R_2 - \v R_1)/2$]:
\begin{equation}
\left (V\right )^{(1)}_{12} = \left (V\right )^{(1,0)}_{12} + \left (V\right )^{(1,+)}_{12} + \left (V\right )^{(1,-)}_{12}
\end{equation}
with
\begin{subequations}
\begin{eqnarray}
\left (V\right )^{(1,0)}_{12} & = & i \hat{\v z} \cdot \left [\nabla V\left (\v c_{12}\right ) \times \v d _{12}\right ]\braket{1 \vert 2} ,\\
\left (V\right )^{(1,+)}_{12} & = & -i \frac{E_{n_1}+E_{n_2}}{2 \vert e \vert \Omega_c^2} \left (j_+\right )_{12} \partial_- V\left (\v c_{12}\right ) ,  \\
\left (V\right )^{(1,-)}_{12} & = & i \frac{E_{n_1}+E_{n_2}}{2 \vert e \vert \Omega_c^2} \left (j_-\right )_{12} \partial_+ V\left (\v c_{12}\right ).
\end{eqnarray}
\end{subequations}

The solution of the Dyson equation for the retarded/advanced single electron Green's functions  
within the same approximation leads to:
\begin{equation}
G_{12}^{R/A}\left (\omega\right ) \approx G_{12}^{(0),R/A}\left (\omega\right ) + G_{12}^{(1),R/A}\left (\omega\right )
\label{eq:GRAExp}
\end{equation}
with
\begin{eqnarray}
G_{12}^{(0),R/A} &=& \frac{\braket{1 \vert 2}}{\omega^\pm - E_{n_2} - V\left (\v R_{2}\right ) }, \\
G_{12}^{(1),R/A} &=& \frac{\left (V\right )^{(1)}_{12} }{\left [\omega^\pm - E_{n_1} - V\left (\v R_1\right ) \right ]\left [\omega^\pm - E_{n_2} - V\left (\v R_2\right ) \right ]}. \notag \\
\end{eqnarray}
Here we use the shorthand notation  $\omega^\pm = \omega \pm i0$.

\subsection{Current density}

We base our calculation of the current density in arbitrary potential configuration on general results 
on vortex states and the semiclassical expansion reported in Ref.~\onlinecite{ChampelFlorens2007}. 

The current density [see Eq. \eqref{eq:Def_currentdensity}] is the $\v x_1 \rightarrow \v x_2$ limit of
\begin{eqnarray}
\braket{\hat j_\pm\left (\v x_1, \v x_2\right )} &=& \sum_{1,2} \int \frac{d\omega}{2\pi} i n_F\left (\omega\right )\left [G_{21}^{R}\left (\omega\right )-G_{21}^{A}\left (\omega\right )\right ] \notag \\
&& _D\braket{n_1,\v R_1\vert \v x_1} j_\pm \braket{\v x_2| n_2,\v R_2}_D . \label{eq:jx1x2}
\end{eqnarray}
Here we introduced shorthand notation
\begin{equation}
\sum_1 = \int \frac{d^2 R_1}{2\pi l_B^2} \sum_{\vert n_1 \vert  = 0}^\infty \sum_{\eta_{n_1}}.
\end{equation}

To proceed further we use the gradient expansion of the Green's functions~\eqref{eq:GRAExp}.
The important simplification comes from the fact that we are interested in the slow (on the scale of magnetic length) part of the current density (cf. discussion in Sec.~\ref{sec:Semiclassical}). We will see shortly, that most of the terms of gradient expansion~\eqref{eq:GRAExp} do not contribute in this approximation.
To demonstrate this fact we will need the Fourier representation of the vortex states
\begin{eqnarray}
\braket{\v p \vert n, \v R} &=& \frac{4l_B\pi e^{-i \v p \v R}}{\sqrt{2\pi n!}} \left [i \sqrt{2} l_B \left (p_+ + \frac{i R_+}{2l_B^2}\right )\right]^n \notag \\ &\times&  \exp\left [{\left (p_- - \frac{i R_-}{2l_B^2}\right )\left (p_+ + \frac{i R_+}{2l_B^2}\right )l_B^2}\right ] \notag \\&& \label{eq:FTofvortexstate}
\end{eqnarray}
Here, $p_\pm = p_x \pm i p_y$ and $R_\pm$ is defined analogously.\\

\subsubsection{Contributions of $V_{12}^{(0)}$ and $V_{12}^{(1,0)}$}
The simplest terms of the gradient expansion are those involving $V_{12}^{(0)}$ and $V_{12}^{(1,0)}$.
Writing down the corresponding current densities $J_\pm\left (\v q\right )$ in momentum representation  we find
[see Eq. \eqref{eq:Def_currentdensity} and the $\v x_1 \rightarrow \v x_2$ limit of Eq. \eqref{eq:jx1x2}]
\begin{eqnarray}
J_\pm\left (\v q\right )\vert_{V_{12}^{(0)},V_{12}^{(1,0)}} &\propto & \int \left (d \v p\right ) \braket{2 \vert 1} \braket{1\vert \v p - \v q} j_\pm \braket{\v p \vert 2} \notag \\
&\approx &e^{-i\v q \v R_{1}}\int \left (d \v p\right ) \braket{2 \vert 1} \braket{1\vert \v p} j_\pm \braket{\v p \vert 2} \notag \\
& = & e^{-i\v q \v R_{1}}\braket{2 \vert 1} \left (j_\pm \right )_{12}= 0 \label{eq:vanishingdueto2}
\end{eqnarray}
In this expression $\v q$ is the slow momentum associated with the macroscopic vector potential $ A_\mp \left (- \v q\right )$  and $ \braket{1\vert \v p - \v q} \approx e^{-i\v q \v R_{1}}\braket{1\vert \v p}$ to zeroth order in $q l_B \ll 1$ [see Eq. \eqref{eq:FTofvortexstate}].
Similar analysis shows that only $V_{12}^{(1,\mp)}$ and not $V_{12}^{(1,\pm)}$ contribute to  $J_\pm$.

\subsubsection{Leading contribution}
The leading contribution to the current densities can now be presented as
\begin{eqnarray}
\braket{\hat j_\pm\left (\v x_1, \v x_2\right )} &=& \sum_{1,2}  \frac{\pm i n_F\left (E_{n_1} + V\left ( \v R_1\right )\right )}{2 \vert e \vert \Omega_c^2}  \frac{E_{n_1}+E_{n_2}}{E_{n_1} -E_{n_2}}\notag \\
&\times &\Big [\partial_{\pm} V\left (\v c_{12}\right ) \left (j_\mp\right )_{21} \braket{1\vert \v x_1}j_\pm \braket{\v x_2 \vert 2} \notag \\ &+& \partial_{\pm} V\left (\v c_{12}\right ) \left (j_\mp\right )_{12} \braket{2\vert \v x_1}j_\pm \braket{\v x_2 \vert 1} \Big ]. \notag \\
&&
\label{eq:jd}
\end{eqnarray}
This expression can be further simplified by employing the relation: 
\begin{equation}
\sum_{2} \frac{E_{n_1}+E_{n_2}}{E_{n_1} -E_{n_2}} \braket{\v x_2 \vert 2}  \left (j_+\right )_{21} = -\left (2 \vert n_1 \vert +1\right ) j_+ \braket{\v x_2 \vert 1}.
\end{equation}
Here we used $\left (j_+\right )_{21} \propto \delta_{\vert n_2 \vert -1, \vert n_1 \vert}$ and the following identities
\begin{equation}
\sum_{2} \left \lbrace \begin{array}{c}
1 \\
\eta_{n_2}
\end{array} \right \rbrace \braket{\v x \vert 2}\braket{ 2 \vert \v x'} = \left \lbrace \begin{array}{c}
\delta(\v x- \v x') \mathbf 1_{\sigma} \\
0
\end{array} \right \rbrace.
\end{equation}
As a result, we obtain
\begin{eqnarray}
\braket{\hat j_\pm\left (\v x_1, \v x_2\right )} &=&\pm i\vert e \vert l_B^2 \sum_{1}  n_F\left (E_{n_1} + V\left ( \v R_1\right )\right ) \partial_{\pm} V\left (\v R_1\right ) \notag \\
&\times &\Big [ \braket{1\vert \v x_1}\braket{\v x_2 \vert 1}- 2\vert n_1 \vert\braket{1\vert \v x_1}\sigma_z\braket{\v x_2 \vert 1}\Big ]. \notag \\ &&
\end{eqnarray}
The term proportional to $\vert n_1 \vert$ does not contribute to the slow part of the current.
Indeed, consideration similar to that of  Eq. \eqref{eq:vanishingdueto2}  leads to
\begin{equation}
\int \left (d \v p \right )\vert n_1 \vert\braket{1\vert \v p - \v q}\sigma_z\braket{\v p \vert 1} \stackrel{ql_B \ll 1}\approx \vert n_1 \vert\braket{1 \vert \sigma_z\vert 1} = 0.
\end{equation}

\subsubsection{Regularizing the divergence}
The expression~\eqref{eq:jd} is singular in the limit ${\v x}_1\rightarrow {\v x}_2$.
The singularity can be regularized by  adding and subtracting the following term (corresponding to the linear-response current at zero temperature and chemical potential $\mu = 0^+$):
\begin{eqnarray}
X_\pm \left (\v x \right ) &=& \frac{\pm i\vert e \vert \partial_{\pm} V\left (\v x\right ) }{2\pi}  \notag \\ 
&\times &  \lim_{\v x' \rightarrow \v x} \int d^2R \sum_{n \leq 0} \left . _D\braket{n, \v R \vert \v x}\braket{\v x' \vert n, \v R}_D \right . \notag \\
&=& \frac{\pm i\vert e \vert \partial_{\pm} V\left (\v x\right ) }{2\pi}  \notag \\ &\times & \lim_{\v x' \rightarrow \v x} \int d^2R \sum_{n \leq 0} \Big \lbrace \frac{\braket{\v x' \vert \vert n \vert, \v R} \braket{\vert n \vert, \v R \vert \v x}}{1 + \vert \eta_n \vert}\notag \\ && + \frac{\eta_n^2\braket{\v x' \vert \vert n \vert -1, \v R} \braket{\vert n \vert-1, \v R \vert \v x}}{1 + \vert \eta_n \vert}  \Big \rbrace \notag \\
&=& \pm i\vert e \vert \partial_{\pm} V\left (\v x\right ) \notag \\ &\times &\lim_{\v x' \rightarrow \v x} \left [l_B^2\delta\left (\v x - \v x'\right ) + \frac{\braket{0 \v x' \vert 0, \v x } }{4\pi} \right ]
\end{eqnarray}
To get the $\delta$-function contribution we have used 
the resolution of identity, Eq. \eqref{eq:Resolofidentity}, for up and down components separately.
However, the double weight of down component of the zeroth LL generates the second contribution in 
the angular bracket: this is where half-integer $g_{xy}$ comes from.

\section{Derivation of the NL$\sigma$M describing the half-integer QHE}
\subsection{No net $B$-field: Non-Abelian bosonization.}
\label{sec:app:Bosonization}
The first step  of the derivation of   NL$\sigma$M for the QH problem is to apply non-Abelian bosonization to the system of disordered Dirac fermions which is TR invariant on average but contains a random Zeeman term.

The model under consideration is Eq. \eqref{eq:Highenergyaction}, with the following white-noise  scalar disorder potential
\begin{equation}
\left \langle V\left (\v x\right ) V\left (\v x'\right ) \right \rangle = \frac{1}{\pi \nu \tau_{sc}} \delta(\v x - \v x'),
\end{equation}
and a random Zeeman term $H_Z = m \sigma_z$
\begin{equation}
\left \langle m\left (\v x\right ) m\left (\v x'\right ) \right \rangle = \frac{1}{\pi \nu \tau_{Z}} \delta(\v x - \v x').
\end{equation}
After  disorder  averaging, the Matsubara action of our system receives an additional contribution
\begin{eqnarray}
S^{dis} &=& -\frac{1}{2\pi \nu} \int_{\v x} \Big [\frac{1}{\tau_{sc.}}\left (\bar \psi(\v x) \psi(\v x)\right ) \left (\bar \psi(\v x) \psi(\v x)\right ) \notag \\
&& + \frac{1}{\tau_{Z}}\left (\bar \psi(\v x) \sigma_z \psi(\v x)\right ) \left (\bar \psi(\v x)\sigma_z \psi(\v x)\right ) \Big ]. \label{eq:SeffDisorder}
\end{eqnarray}

\subsubsection{SCBA.}

On the  mean-field level the fermionic Green's functions are given  
by the self consistent Born approximation (SCBA). The SCBA  equation for the  self energy reads
\begin{equation}
\Sigma_{n} = -\frac{1}{\pi \nu \tau} \mathcal G_n\left (\v x, \v x\right ). \label{eq:SCBA}
\end{equation}
Here the scattering rate  $1/\tau = 1/\tau_{sc} + 1/\tau_{Z}$.  The solution of Eq. \eqref{eq:SCBA} is (in the limit $k_F l \gg 1$)
\begin{equation}
\Sigma_n = \frac{i}{2\tau} \text{sign}(n).
\end{equation}

\subsubsection{Non-Abelian bosonization.}
In order to go beyond the mean-field treatment, we derive the NL$\sigma$M from the fermionic action.
We will employ the double cut-off truncation scheme in Matsubara space\cite{MishandlingI} and use non-Abelian bosonization,\cite{Witten1984} with the dictionary for the $\v U(2N_M'N_R)\times \v U(2N_M'N_R)$ invariant model being \cite{KnizhnikZamolodchikov1984,Tsvelik2007}
\begin{subequations}
\begin{eqnarray}
\psi^\uparrow \otimes \bar \psi^\downarrow &\leftrightarrow & \frac{1}{4\pi v_0}  U^\dagger \partial_+ U , \\
\psi^\downarrow \otimes \bar \psi^\uparrow &\leftrightarrow & \frac{1}{4\pi v_0}  U \partial_- U^\dagger , \\
\psi^\uparrow \otimes \bar \psi^\uparrow & \leftrightarrow & - \lambda  U^\dagger ,\\
\psi^\downarrow \otimes \bar \psi^\downarrow & \leftrightarrow & \lambda  U.\\
\end{eqnarray}
\end{subequations}
Here $U \in \v U(2N_M'N_R)$ is a unitary matrix field. Typically it is decomposed in a phase (Abelian bosonization) and a special unitary part 
$$U = e^{i \sqrt{\frac{4\pi}{2N_M'N_R}} \Phi} \tilde U.$$ 
The dimensionful constant $\lambda$ is of the order of the UV-cutoff. In the presence of disorder and a finite chemical potential it turns out to be of the order of the density of states, see below, App.~\ref{sec:app:Bosonization:SCBA}.

The kinetic part of the action can now be rewritten as \cite{Faddeev1976,Gerasimov1993,LosevShatashvili1995,Smilga1996}
\begin{eqnarray}
S &=&\int_{\v x} \frac{1}{2} (D_i \Phi)^2 = \int_{\v x} \frac{1}{8\pi} \Tr D_i {\tilde U}^\dagger D_i {\tilde U} \notag \\
&+& \int_{\v x, w} \frac{-i}{12\pi} \epsilon_{ijk} \Tr \left ({\tilde U}^\dagger D_i {\tilde U}\right )\left ({\tilde U}^\dagger D_j {\tilde U}\right )\left ({\tilde U}^\dagger D_k {\tilde U}\right ) \notag \\
&+& \int_{\v x, w} \frac{i}{8 \pi} \epsilon_{ijk} \Tr F_{ij} \left ({\tilde U}^\dagger D_k {\tilde U} +  D_k {\tilde U}{\tilde U}^\dagger\right )\\
&\doteq &\int_{\v x} \frac{1}{2} (D_i \Phi)^2 + \int_{\v x} \frac{1}{8\pi} \Tr \partial_i {\tilde U}^\dagger \partial_i {\tilde U} \notag \\
&+& \int_{\v x, w} \frac{-i}{12\pi} \epsilon_{ijk} \Tr \left ({\tilde U}^\dagger \partial_i {\tilde U}\right )\left ({\tilde U}^\dagger \partial_j {\tilde U}\right )\left ({\tilde U}^\dagger \partial_k {\tilde U}\right ) \notag \\
&+& \int_{\v x} \frac{-ie}{4\pi} \Tr\left [A_- {\tilde U}^\dagger \partial_+ {\tilde U} + A_+ {\tilde U}\partial_- {\tilde U}^\dagger \right ] \notag \\
&+& \int_{\v x} \frac{-e^2}{4\pi} \Tr\left [A_- {\tilde U}^\dagger A_+ {\tilde U} - A_+ A_-\right ] . \label{eq:Bosonizedkineticterm}
\end{eqnarray}
Here the symbols $D_i$ denote long derivatives,  $D_i =\partial_i - ie A_i^{(0)}$ when acting on a scalar field and $D_i =\partial_i - ie \left [A_i, \cdot \right ]$ when acting on a matrix field. The gauge potentials are arbitrary $\v U(2N_M'N_R)$ gauge potentials [split in traceless (traceful) components $A_i$ ($A_i^{(0)}$)], $F_{ij}$ is the corresponding field strength tensor. For the problem of disordered Dirac fermions coupled to $\mathbf{U}(1)$ gauge potentials, we will set $A_i = \hat A_i$ in the end [see Eq. \eqref{eq:longderivatives}]. The symbol $\doteq$ here denotes equality for all cases when gauge fields are non-topological (recall, that we are interested in situations without net magnetic flux through the spatial plane). 

The expressions containing integrals over the variable $w$ involve the extension of the base manifold [$(\v x, w) \in (\mathbb R^2 \cup \lbrace \infty \rbrace, [0,1])$]. In these terms $\tilde U$ implicitly denotes a different function $\tilde U(\v x, w)$ which coincides
with the physical field on the physical space $\tilde U(\v x, 0) = \tilde U(\v x)$ while  taking a uniform fixed value at $w =1 $, e.g. $\tilde U(\v x, 1) = \v 1$.

\subsubsection{Bosonized SCBA.}
\label{sec:app:Bosonization:SCBA}

The SCBA equation  \eqref{eq:SCBA} can be rederived in the bosonic language
\begin{eqnarray}
\Sigma &=& \frac{1}{\pi \nu \tau} \langle \psi \otimes \bar \psi \rangle_{SCBA} \notag\\
&\leftrightarrow&  \frac{1}{\pi \nu \tau}\left  \langle \left (\begin{array}{cc}
- \lambda U^\dagger & \frac{1}{4\pi v_0} U^\dagger \partial_+ U \\
\frac{1}{4\pi v_0} U \partial_- U^\dagger & \lambda U
\end{array} \right )_{\sigma} \right \rangle_{SCBA} . \label{eq:SCBABoson}
\end{eqnarray}
The symbol $\langle \dots \rangle_{SCBA}$ denotes self-consistent SCBA average.
Equations \eqref{eq:SCBABoson} are consistent with the previous solution provided $U = i \Lambda$ and $\lambda = \nu \pi/2$ ($\Lambda_{nn'} = \delta_{nn'} \text{sign}(n)$).

\subsubsection{Bosonized effective action.}

We now return to Eq. \eqref{eq:SeffDisorder}. We bosonize both channels of possible soft modes

\begin{eqnarray}
S^{dis} &\leftrightarrow & \frac{-1}{2\pi \nu\tau_{sc}}  \int_{\v x} \Big [\left \lbrace \tr  \left (\begin{array}{cc}
- \lambda U^\dagger & \frac{U^\dagger \partial_+ U}{4\pi v_0}  \\
\frac{U \partial_- U^\dagger}{4\pi v_0}  & \lambda U
\end{array} \right )_{\sigma}\right \rbrace^2 \notag \\
&& - \tr\left \lbrace \left (\begin{array}{cc}
- \lambda U^\dagger & \frac{U^\dagger \partial_+ U}{4\pi v_0}  \\
\frac{U \partial_- U^\dagger}{4\pi v_0}  & \lambda U
\end{array} \right )_{\sigma}^2\right \rbrace\Big ] \notag \\
&+&  \frac{-1}{2\pi \nu \tau_{Z}}    \int_{\v x}   \Big [\left \lbrace\tr \left (\begin{array}{cc}
- \lambda U^\dagger & \frac{U^\dagger \partial_+ U }{4\pi v_0} \\
\frac{-U \partial_- U^\dagger}{4\pi v_0}  &- \lambda U
\end{array} \right )_{\sigma}\right \rbrace^2 \notag \\
&&- \tr\left \lbrace \left (\begin{array}{cc}
- \lambda U^\dagger & \frac{U^\dagger \partial_+ U}{4\pi v_0}  \\
\frac{-U \partial_- U^\dagger}{4\pi v_0}  &- \lambda U
\end{array} \right )_{\sigma}^2\right \rbrace\Big ] \notag \\
&\doteq & \frac{\lambda^2}{2\pi \nu} \int_{\v x} \Big [\frac{\tr \left ( [U^\dagger + U]^2\right )}{\tau} \notag \\
&& - \frac{ \left (\tr \left [U^\dagger - U\right ]\right )^2}{\tau_{sc}}-\frac{\left ( \tr \left [U^\dagger+ U\right ]\right )^2}{\tau_{Z}}\Big ] .\label{eq:disordermassterms}
\end{eqnarray}
Here, the sign $\doteq$ indicates that in this formula we omitted the gradient terms which renormalize the kinetic part of the action as well as a constant.

\subsubsection{Saddle-point equations.}

By infinitesimal left rotation of spatially constant $U$ we determine the saddle point equations for the disorder induced potential
\begin{eqnarray}
0 &=& \frac{i \lambda ^2}{\pi \nu} \Big [\frac{ \left (U^2 - [U^\dagger]^2\right )}{\tau}  + \frac{\left (U^\dagger  - U\right )\tr \left [U^\dagger +	 U\right ] }{\tau_{sc}} \notag \\
&&+\frac{\left (U^\dagger  - U\right ) \tr \left [U^\dagger+ U\right ] }{\tau_{Z}}\Big ] .
\end{eqnarray}
We see that  the SCBA solution  $U = i \Lambda$ solves the saddle point equation.

\subsubsection{Goldstone manifold and field theory.} We will now rotate the bosonic fields by slow, small, unitary rotations: $U \rightarrow   U^\dagger_{\text{soft}} U U_{\text{soft}}$.  For the saddle point solution $U = i \Lambda$ these fields equally annihilate the disorder induced mass terms of Eq. \eqref{eq:disordermassterms}. Thus the effective field theory will be constructed on a saddle point manifold, namely the coset space formed by the $Q = U^\dagger_{\text{soft}} \Lambda U_{\text{soft}}$ which is $\mathbf U(2N_MN_R)/\mathbf U(N_MN_R)\times \mathbf U(N_MN_R)$. To derive the effective field theory, Eq. \eqref{eq:B=0NSLM} of the main text, the following steps are in order: (i) The prefactor of the gradient term is renormalized by integration of the $U$ fields in SCBA approximation.\cite{KoenigMirlin2013} (ii) Upon restriction to the coset space, the Wess-Zumino-Novikov-Witten term in third line from the bottom of Eq. \eqref{eq:Bosonizedkineticterm} becomes the theta term with short derivatives and angle $\theta = \pi$~(mod~$2\pi$).\cite{BocquetSerbanZirnbauer2000,Altland2006} (iii) The last two lines of the same Eq. \eqref{eq:Bosonizedkineticterm} provide the gauge potentials entering the long derivatives of the gradient term. It is an important observation that terms containing $Q^\dagger \partial_\pm Q$ and $\epsilon_{ij} \tr A_i Q^\dagger A_j Q$ drop out in view of the hermiticity and unitarity of $Q$. Therefore the theta term has short derivatives. (iv) Frequency and interaction terms were not discussed in this appendix, but can be equally included following Ref.~\onlinecite{KoenigMirlin2013}. (v) The subscript $_{\text{soft}}$ is omitted in all other parts of this paper.

\subsection{Finite net magnetic field: Gradient expansion.}
\label{sec:app:Gradientexpansion}

We now turn to the derivation of the NL$\sigma$M describing disordered Dirac fermions in strong magnetic field ($\Omega_c \tau \gg 1$). The fermionic action on saddle point level is
\begin{eqnarray}
S\left [\bar \psi ,\psi \right ] &=& \int_{\v x} \bar \psi \Big [-i \hat \epsilon - i e \hat \Phi + H_0\left( (\v p - e \left [{\mathcal A} + \hat{\v A}\right ]\right ) \notag \\
&& - \mu - i(\Sigma^R) '' Q \Big ] \psi
\end{eqnarray}
The SCBA is justified in the center of LLs with large index $\vert n \vert \gg 1$.  
For the present case of Dirac fermions, the imaginary part of retarded self energy $(\Sigma^R )'' = 1/2\tau$ 
is energy dependent and non-trivial (trivial) in spin space for the zeroth (all other) LLs, see, e.g., 
Ref.~\onlinecite{OGM2008} for more details.

Just as in the previous section, we set $Q = U^\dagger \Lambda U$ with slow  unitary $(2N_MN_R)\times(2N_MN_R)$ matrix field  $U$. 
Note that at this stage, all sums and traces over Matsubara indices go from negative to positive infinity. Thus $\Lambda$ is an infinite matrix with only diagonal entries $\Lambda_{nn'} = \delta_{nn'}\text{sign}(n)$. In order to obtain finite-dimensional $\Lambda$ (and thus $Q$), a second cut-off will be introduced at the end of this section.\cite{MishandlingI}

In order to perform an accurate gradient expansion of the action, it is convenient to re express the partition function as
\begin{eqnarray}
\mathcal Z &=& \int \mathcal D Q \int \mathcal D \left [\bar \psi, \psi\right ] e^{-S\left [\bar \psi ,\psi \right ]} \notag \\
&=& \int \mathcal D Q  \mathcal J\left [U, \hat \Phi, \hat A_i\right ] \int \mathcal D \left [\bar \psi ', \psi '\right ] e^{-S\left [\bar \psi ',\psi '\right ]}.
\end{eqnarray}
Here the rotated fields $\psi '  = U \psi$ and $\bar \psi' = \bar \psi U^\dagger$ were introduced at the expense of the Jacobian $\mathcal J\left [U, \hat \Phi, \hat A_i\right ]$.
The action for the rotated fermions reads
\begin{equation}
S\left [\bar \psi ',\psi '\right ] = \int_{\v x} \bar \psi' \left [- \mathcal G^{-1} - i e {\mathbf{\Phi}} + \v J \cdot \mathbb{A} \right ] \psi' .\label{eq:Actionpsiprime}
\end{equation}
where  we use the notation $J_i = \delta H_0/\delta A_i$ and the rotated gauge potentials are
\begin{eqnarray}
{\mathbf{\Phi}} &=& \frac{1}{e} U\left [\hat \epsilon, U^\dagger \right ] + U \hat \Phi U^\dagger \notag \\
\mathbb A_i &=& -\frac{1}{e} U\left [-i \partial_i, U^\dagger \right ] + U \hat A_i U^\dagger .
\end{eqnarray}
We denote the SCBA Green's function by
\begin{eqnarray}
\mathcal G_{mm'}^{\alpha \alpha'} { \left (\v x, \v x'\right )} &=& \Big (\big [i \epsilon_m - H_0\left (\v p - e {\mathcal A} \right )  + \mu \notag \\
&+&  i(\Sigma^R) '' \text{sign}(\epsilon_m)\big ]^{-1}\Big )_{\v x, \v x'} \delta_{mm'}^{\alpha \alpha'}.\notag\\
\end{eqnarray}
As we are working in the limit $\epsilon_n \ll(\Sigma^R) '' $, we will partly drop the frequency dependence below. 

We will further use the notation 
$$- \mathbf G ^{-1}= - \mathcal G^{-1} - i e {\mathbf{\Phi}} + \v J \cdot \mathbb{A}. $$ 
Since we are interested in the topological theta term involving spatial derivatives only, we omit $\mathbf \Phi$ in what follows.
Integrating fermions out, we get
\begin{equation}
\mathcal Z  = \int \mathcal D Q \mathcal  J\left [U, \hat \Phi, \hat A_i\right ] e^{-S_{eff}},
\end{equation}
with
\begin{equation}
S_{eff} = - \Tr \ln \left [- \mathbf G^{-1}\right ]. \label{eq:trln}
\end{equation}
Here and below $\Tr$ includes also the spatial integration. 
The expansion of $S_{eff}$ in $\mathbb A$ (omitting $\mathbf \Phi$ and the constant term) yields
\begin{eqnarray}
S_{eff} & \approx & \Tr \left [\mathcal G_0 J_i \mathbb A_i\right ] + \frac{1}{2} \Tr \left [\mathcal G_0 J_i \mathbb A_i \mathcal G_0 J_j \mathbb A_j\right ].
\end{eqnarray}

\subsubsection{The RR- and AA-correlators in the term $\mathcal O \left (\mathbb A^{2}\right )$. }

First, we will disregard the diffusive fields and set $\mathbb A = \hat A$. 
Recall that we are working with infinite Matsubara sums. 
The expansion contains the standard conductivity term
\begin{equation}
\frac{1}{2} \Tr \left [\mathcal G J_i \hat A_i \mathcal G J_j \hat A_j\right ] {=} e^2 \sum_{m>0,\alpha}\int_{\v x} m (A_i)_{-m}^\alpha g_{ij} (m) (A_j)_{m}^{\alpha},
\end{equation}
with
\begin{equation}
g_{ij} (m) = \frac{1}{e^2 A m} \sum_{k} \Sp \left[  J_i  \mathcal G_{k+m} J_j \mathcal G_k\right ].
\end{equation}
Here $A$ denotes the sample area. The symbol $\Sp$ involves trace in spin and real space only. 
The $RR + AA$ contribution to $g_{xx}$ for Dirac fermions is non-zero but negligible as compared to the $RA$ contribution. 
In contrast, for the transverse DC conductivity we find the standard $g_{xy}^{II}$ contribution:
\begin{eqnarray}
&& \frac{1}{2} \Tr \left [\mathcal G J_{[i} \hat A_i \mathcal G J_{j]} \hat A_j\right ] \notag \\ &&\stackrel{RR+AA}{=}  e^2 g_{xy}^{II} \epsilon_{ij}\sum_{m>0,\alpha}\int_{\v x} m (A_i)_{-m}^\alpha  (A_j)_{m}^{\alpha}.
\end{eqnarray}
(Square brackets in the indices denote antisymmetrization.) Now we return to the full $\mathbb A$ which we write as $\mathbb A = \Delta \mathbb A + \hat A$. Clearly, $\Delta \mathbb A = \mathbb A - \hat A$ is a finite $(2N_MN_R)\times (2N_MN_R)$ matrix. We will show that also for the full $\mathbb A$ we have
\begin{eqnarray}
&& \frac{1}{2} \Tr \left [\mathcal G J_{[i} \mathbb A_i \mathcal G J_{j]} \mathbb A_j\right ] \notag \\ &&\stackrel{RR+AA}{=}  e^2 g_{xy}^{II} \epsilon_{ij}\sum_{m>0,\alpha}\int_{\v x} m (A_i)_{-m}^\alpha  (A_j)_{m}^{\alpha}.
\end{eqnarray}
Indeed, all terms linear or quadratic in $\Delta \mathbb A_I$ involve traces over the finite $(2N_MN_R)\times (2N_MN_R)$ space. All of these finite traces vanish by symmetry, for example
\begin{eqnarray}
&&\frac{1}{2} \Tr \left [\mathcal G J_{[i} \Delta\mathbb A_i \mathcal G J_{j]} \Delta \mathbb A_j\right ] \stackrel{RR}{=} \notag \\
&&\frac{1}{2A} \Sp \left [\mathcal G^R J_{[i}  \mathcal G^R J_{j]}\right ] \Tr \left [\Delta\mathbb A_i \frac{1 + \Lambda}{2} \Delta \mathbb A_j\frac{1 + \Lambda}{2} \right ] \notag \\
&&= 0.
\end{eqnarray}

\subsubsection{The RA-correlator in the term $\mathcal O \left (\mathbb A^{2}\right )$.}

For the RA-correlator we obtain the standard result:
\begin{eqnarray}
\frac{1}{2} \Tr \left [\mathcal G J_i \mathbb A_i \mathcal G J_j \mathbb A_j\right ] &\stackrel{RA}{=}& 	\sum_i \Tr\left [\mathbb A_i \mathbb A_i - \mathbb A_i \Lambda\mathbb A_i \Lambda\right ] \notag \\
&& \times \frac{1}{8A} \Sp \left [  \mathcal G^A J_x \mathcal G^R J_x  +\mathcal G^R J_x  \mathcal G^A J_x \right ] \notag \\
& +& 2 \Tr\left [\Lambda \left (\mathbb A_x \mathbb A_y - \mathbb A_y\mathbb A_x\right )\right ] \notag \\
&& \times \frac{1}{8A} \Sp  \left [  \mathcal G^R J_x  \mathcal G^A J_y - \mathcal G^A J_x \mathcal G^R J_y  \right ]. \notag \\
\end{eqnarray}
In what follows we use the notation
\begin{eqnarray}
g_{xx} &=&\frac{1}{e^2 A} \Sp \left [ \mathcal G^R J_x  \mathcal G^A J_x \right ], \notag \\
g_{xy}^I &=& \frac{-1}{2 e^2 A} \Sp \left [  \mathcal G^R J_x  \mathcal G^A J_y - \mathcal G^A J_x \mathcal G^R J_y \right ].
\end{eqnarray}

\subsubsection{Term of $\mathcal O \left (\mathbb A^{1}\right )$. }

We follow the steps presented in Ref.~\onlinecite{Pruisken1984} and use
\begin{equation}
\frac{\partial}{\partial \mu} \mathcal G\left (\v x, \v x'\right ) = - \int d^2 x'' \mathcal G\left (\v x, \v x''\right )\mathcal G\left (\v x'', \v x'\right )
\end{equation}
to rewrite the  $\mathcal O \left (\mathbb A^{1}\right )$ part of the action as
\begin{eqnarray}
\Tr \mathcal G \v J \cdot \mathbb{A} &=&  - \int_{- \infty}^\mu d \tilde \mu \int_{\v x, \v x'}  \tr \left [\mathcal G\left (\v x, \v x' \right ) \mathcal G\left (\v x', \v x \right ) J_i \mathbb A_i \left ( \v x\right )\right ] \notag \\
&=& - \frac{1}{2} \frac{\partial}{\partial B} \int_{-\infty}^\mu d \tilde \mu \tr ^\sigma  \mathcal G^{R-A}\left (0,0\right ) \Tr \left [\epsilon_{ij} \partial_i \mathbb A_j \Lambda \right ]\notag \\
&& - \frac{1}{2} \frac{\partial}{\partial B} \int_{-\infty}^\mu d \tilde \mu \tr ^\sigma \mathcal G^{R+A}\left (0,0\right ) \Tr \left [\epsilon_{ij} \partial_i \mathbb A_j \right ] .\notag \\
\end{eqnarray}
The term $\Tr \left [\epsilon_{ij} \partial_i \mathbb A_j \right ] = 0$ vanishes, since it contains commutators of small matrices and the only non-commuting term is $\Tr \hat A = 0$ by assumption of purely dynamic gauge fields. In contrast, the $\Tr \left [\epsilon_{ij} \partial_i \mathbb A_j \Lambda \right ]$ term plays an important role: its prefactor $\partial n/\partial B$ is related to $g_{xy}^{II}$ by the Smrcka-Streda formula.\cite{SmrckaStreda1977}

\subsubsection{Collecting all terms.}
We are now in position to present the full gradient expansion of $S_{eff}$
\begin{eqnarray}
S_{eff} &=& \frac{g_{xx}e^2 }{4}\Tr\left [\mathbb A_i \mathbb A_i - \mathbb A_i \Lambda\mathbb A_i \Lambda\right ]   \notag \\
& -&  \frac{g_{xy}^Ie^2 }{2} \epsilon_{ij} \Tr\left [\Lambda \left (\mathbb A_i \mathbb A_j\right )\right ]  \notag \\
&- & \frac{g_{xy}^{II} e^2}{2}  \epsilon_{ij} \Big [ \frac{-i}{e}\Tr\left [\partial_i \mathbb A_j \Lambda \right ] \notag \\
&&+\sum_{m,\alpha}\int_{\v x} m (A_i)_{m}^\alpha  (A_j)_{-m}^{\alpha}\Big ].
\end{eqnarray}

In order to rewrite $S_{eff}$ in a more compact way we will introduce a second cut-off $N_M'$ in Matsubara space.\cite{MishandlingI} In particular, now also $\hat A$ and $\Lambda$ are \textit{finite} matrices of size $(2N_M'N_R)\times (2N_M'N_R)$. We assume $N_M'/N_M \rightarrow \infty$. Then we can use the notation (from now on $\Tr$ denotes finite traces)
\begin{equation}
D_i Q \equiv \partial_i Q - i e \left [\hat A_i , Q \right ] = -i e U^\dagger \left [ \mathbb A_i, \Lambda \right ] U
\end{equation}
to express
\begin{eqnarray}
\Tr D_i Q D_i Q &=& 2 e^2 {\Tr} \left [\mathbb A_i^2 - \left (\mathbb A_i \Lambda \right )^2\right ], \notag \\
\epsilon_{ij}\Tr Q D_i Q D_j Q &=& 4 e^2 \epsilon_{ij} \Tr\left [\Lambda \mathbb A_i \mathbb A_j\right ] \notag \\
&=&4e\epsilon_{ij}\Big [-i {\Tr} \left ( \partial_i \mathbb A_j \Lambda\right ) \notag \\
&&+ e\int_{\v x} \sum_{n, \alpha} n (A_i)_{n}^\alpha(A_j)_{-n}^\alpha\Big ].
\end{eqnarray}

Including  now the contribution of the Jacobian of the transformation from initial to rotated fermions we find the sigma model action
\begin{eqnarray}
S &=& \frac{1}{8} \left (g_{xx}  \Tr \left [D_i Q\right ]^2 -g_{xy} \epsilon_{ij}\Tr Q D_i Q D_j Q\right ) \notag \\
&& - \ln \mathcal  J\left (U, \hat A\right )
\end{eqnarray}

Finally, we will discuss the Jacobian $\mathcal J\left (U, \hat A\right )$ in more detail. 
Generally speaking, its precise value depends on the regularization of the functional integral
measure of the initial fermionic field theory. The same ambiguity is generally present in the 
microscopic calculation of $g_{xy}$ due to the unbounded spectrum of Dirac fermions.
The final result should however be independent of the regularization. 
We have learned in Sec. \ref{sec:NLSM:parity} that one can regularize the fermionic theory in such a manner
that the  parity symmetry  is preserved. In our present problem this does not contradict gauge invariance or any other fundamental principle.
Choosing such a regularization, we see that  $g_{xy}$ vanishes for $B \rightarrow 0$. On the other hand, in the $B \rightarrow 0$ limit, the action should reproduce the result~\eqref{eq:B=0NSLM}.  We therefore conclude that
the Jacobian $- \ln  \mathcal  J\left (U, \hat A\right )$ equals the theta term with short derivatives. This concludes the derivation of Eq. \eqref{eq:Bneq0NSLM} of the main text.

\section{Classical conductivity tensor}
\label{sec:app:classicalTheory}

In this appendix we present the semiclassical Boltzmann calculation of the conductivity tensor in the presence of both orbital magnetic field and a Zeeman term $H_Z = m v_0^2 \sigma_z$.

\subsection{Semiclassical theory of anomalous Hall effect.}
The basic concepts of anomalous Hall effect (AHE) are reviewed in Refs.~\onlinecite{NagaosaOng2010,Sinitsyn2008}. 
The contributions to the AHE are threefold [we denote the two bands by $\xi=\pm$ with dispersion relation 
$\epsilon_\xi (\v p) = \xi \sqrt{v_0^2 \v p^2 + (m v_0^2)^2}$]: 

(i) Intrinsic AHE, which is the contribution of integral over Berry connection $\Omega_\xi  = - \frac{m v_0^4}{2 \epsilon_{\xi}^3}$; 

(ii) Skew scattering $\omega_{ll'} \neq \omega_{l'l}$ contribution, which splits into a) conventional ($\omega_{ll'} \propto V^3$) and b) intrinsic ($\omega_{ll'} \propto V^4$). [Here $\omega_{l'l}$ is the squared scattering amplitude from state $l = \left (\v p, \xi\right )$ to $l' = \left (\v p', \xi'\right )$.]

(iii) Side-jump contributions, which again are twofold, including a) side jump accumulation and b) modification of collision integral in view of work performed due to the side jump at a single scattering event.

These contributions are reflected in the equations of motion\cite{ChangNiu1996,SundaramNiu1999,SinitsynNiuMacDonald2006}
\begin{eqnarray}
\dot r_i&=& \stackrel{\equiv v_i^{(\xi)}}{\overbrace{\frac{ \partial \epsilon_\xi \left (\v p\right )}{\partial p_i}}} {- \epsilon_{ij} \dot p_j \Omega_\xi} {+ \sum_{l'} \omega_{l'l}  \left ( \delta\v r_{l'l}\right )_i },\\
\dot p_i &=& F_i = q \left ( E_i + \epsilon_{ij} \dot r_j B/c\right ),
\end{eqnarray}
as well as in the collision integral of the Boltzmann equation
\begin{equation}
\partial_t f + \dot{\v r} \cdot \partial_{\v x} f + \dot{\v p} \cdot \partial_{ \v p} f = \text{St}\left [f\right ] ,\label{eq:BoltzmannEq}
\end{equation}
where
\begin{equation}
{\text{St}\left [f\right ]} = - \sum_{l'} \left [{\omega_{l'l}} f_l - {\omega_{ll'}} f_{l'} \right ]. \label{eq:BoltzmannCollin}
\end{equation}

The precise modification of the collision integral, which will involve the work $W_{1 \rightarrow 2} = \v F \delta \v r_{l_2l_1}$, will be presented below, Sec.~\ref{sec:WorkCollisionKernel}.

It is worth to notice that the collision integral for elastic scattering does not contain Pauli blocking terms (which would change the results in view of skew scattering). The reason~\cite{Sinitsyn2008} is that, in contrast to the case of inelastic scattering, the incoming and outgoing states $l$ and $l'$ should be considered as a single scattering state, and thus Pauli blocking factors [e.g. $f_l (1-f_{l'})$] are superfluous. This can also be understood in the derivation of the Boltzmann equation from Schwinger-Keldysh quantum field theory. Since elastic scattering is evoked by a static disorder potential (it only couples to $\gamma_{\rm cl}$ in Keldysh space), the collision integral in the quantum kinetic equation $\Sigma^K - (\Sigma^R \circ F - F \circ \Sigma^A)$ contains only a single Keldysh Green's function/self energy and thus only a single distribution function.

The side-jump shift of the trajectory is expressed as
\begin{eqnarray}
\delta \v r_{l_2l_1} &=& \braket{u_{\xi_2,\v p_2} \vert i \partial_{\v p_2} u_{\xi_2,\v p_2}} - \braket{u_{\xi_1,\v p_1} \vert i \partial_{\v p_1} u_{\xi_1,\v p_1}} \notag \\ && - (\partial_{\v p_1}+\partial_{\v p_2}) \text{arg}(V_{l_2, l_1}) \label{eq:generalsidejumpformula}
\end{eqnarray}
even in the presence of smooth electromagnetic fields [we denote by $e^{i \v p \v r}\ket{u_{\xi, \v p}}$ the eigenstates to the $B=0$ limit of Hamiltonian \eqref{eq:H0}]. We will use the notation
\begin{equation}
\sum_{l'} \omega_{l'l}  \left ( \delta\v r_{l'l}\right ) = (1 - \frac{qB \Omega_\xi}{c})\frac{\Omega_\xi \underline \epsilon \v p}{\tau^{sj}} . \label{eq:sidejumptimes}
\end{equation}
with the mean side-jump time $\tau^{sj} (\epsilon_\xi)$.
We can diagonalize the equations of motion as follows:\cite{XiaoShiNiu2005,DuvalStichel2006,SonSpivak2013,KimKimSasaki2014}
\begin{equation}
\left (\begin{array}{c}
\dot r_i \\
\dot p_i
\end{array} \right ) = \frac{ \left (\begin{array}{c}
v_i^{(\xi)} + \sum_{l'} \omega_{l'l}  \left ( \delta\v r_{l'l}\right )_i  - \epsilon_{ij} \Omega_\xi q E_j \\
\epsilon_{ij}(\v v^{(\xi)} + \sum_{l'} \omega_{l'l}  \left ( \delta\v r_{l'l}\right ))_j\frac{qB}{c} + q  E_i
\end{array} \right )}{1 - \frac{\Omega_\xi qB}{c}}. \label{eq:Eqofmotion}
\end{equation}
At
\begin{equation}
1 = \frac{\Omega_{\xi} qB}{c} = - \zeta \Omega_c^{\rm cl} \frac{m v_0^2}{2 \epsilon_\xi^2}
\end{equation}
the clean classical equations of motion correspond to pure Hall response.\cite{DuvalStichel2006} However,  in parameter space this point lies outside the region of validity of the Boltzmann equation. [We introduced $\zeta \Omega_c^{\rm cl} = \frac{ v_0^2 qB}{ \epsilon_\xi c}$, $\zeta = \text{sign}(qB\epsilon_\xi)$.]

It is worth noticing that equations of motion \eqref{eq:Eqofmotion}, which contain the disorder-induced side jump terms, 
do not correspond to a Hamiltonian flow. Without the side-jump terms, these equations are perfectly Hamiltonian, 
but with a modified Poisson bracket (we are not using canonical coordinates).\cite{DuvalStichel2006}
Therefore, the invariant phase space volume element acquires an additional term\cite{XiaoShiNiu2005}
\begin{equation}
dV = \left (1-\frac{q\Omega_\xi B}{c}\right ) d^2 p d^2 x.
\end{equation}

We hence deduce that in the Boltzmann equation and in the equations determining current we need to use
\begin{equation}
\sum_{l'} \doteq \sum_{\xi '}\int \left (dp'\right ) \left (1- \frac{qB\Omega_{\xi'}}{c}\right ).
\end{equation}
In the following, we employ polar coordinates determining each momentum $\v p$ by modulus of kinetic energy and angle $\left (\epsilon, \phi\right )$ with $\epsilon = \sqrt{v_0^2 p^2 + (m v_0^2)^2}$ .
In this notation we write 
\begin{equation}
\hat p = (\cos \phi, \sin \phi), \hat e_{\phi} = (- \sin \phi, \cos \phi) = - \underline \epsilon \hat p
\end{equation}
and
\begin{equation}
\nabla_{\v p } f_l  = \hat p \frac{\partial \epsilon}{\partial p} \partial_\epsilon f_l+ \frac{\hat e_{\phi}}{p} \partial_\phi f_l  = \v v^{(\xi)} \partial_{\epsilon_\xi}f_l+ \frac{\hat e_{\phi}}{p} \partial_\phi f_l.
\end{equation}
Here we introduced the matrix representation of the 2D Levi-Civita symbol $(\underline \epsilon)_{ij} = \epsilon_{ij}$.

\subsection{The Collision kernel and Side step}

\label{sec:WorkCollisionKernel}
As explained above, upon a scattering event $l_1 \rightarrow l_2$ the final state corresponds to a trajectory
that is shifted as compared to the initial state by
\begin{equation}
\delta \v r_{l_2l_1} \simeq \v r_{l_2}\left (t =0\right ) - \v r_{l_1}\left (t =0\right ).
\end{equation}
If the scattering event takes place in an external electric (but not magnetic) field, the kinetic energy is not conserved, 
as the potential energy changes at the scattering event\cite{NagaosaOng2010,Sinitsyn2008}
\begin{equation}
\Delta \epsilon^{\delta \v r_{l_2l_1}} \simeq U\left (\v r_{l_2}\right ) - U\left (\v r_{l_1}\right ) 
= \nabla U \delta \v r_{l_2l_1} = - q \v E \delta \v r_{l_2l_1}
\end{equation}

More generally (in the presence of both $\v E$ and $B$ fields), we can say that there is a work to be performed at a scattering event $l_1 \rightarrow l_2$ with side jump. Energy conservation $\epsilon_{initial} = \epsilon_{final}$ implies
\begin{equation}
\xi_1 \epsilon ( \v p_1) = \xi_2 \epsilon(\v p_2) - W_{1\rightarrow 2}
\end{equation}
where $W_{1\rightarrow 2} = \v F \delta \v r_{l_2l_1}$. 
A priori it is not clear, whether one should use $\v F = \dot{\v p}_1$ or $\v F = \dot{\v p}_2$. 
We will fix this question below, App.~\ref{sec:app:ConsLaws}.

The contribution of out-processes ($l \rightarrow l'$) to the collision integral is not altered.
Contrary, for in-processes ($l' \rightarrow l$), energy conservation implies
\begin{equation}
f_{l'} = f\left (\xi, \epsilon- \xi W_{l' \rightarrow l}, \phi' \right ) \approx f_{l'} - \partial_{\epsilon_\xi} f_{l'} W_{l' \rightarrow l}.
\end{equation}
Here we expanded the distribution function under the assumption of small work $W = {W}_{l' \rightarrow l} = {{\v F \delta \v r_{ll'} }}$ as compared to the chemical potential. We will find below, that this assumption is justified, see Eq. \eqref{eq:WorkbyBfield}.

\subsection{Full Boltzmann equation}

Let us return to the Boltzmann equation presented above \eqref{eq:BoltzmannEq}. 
It is worth splitting the contribution of $\dot{\v p} \nabla_{\v p} f$
into two terms, as follows
\begin{equation}
\dot{\v p} \partial_{\v p} f = \dot{\v p}_{clean} \partial_{\v p} f + \frac{qB/c}{1- qB\Omega/c} (\partial_{\v p} f) \underline \epsilon \left [\sum_{l'} \omega_{l'l} \delta r_{l'l}\right ] .
\end{equation}

Bringing the last term to the right hand side of the Boltzmann equation leads to
\begin{equation}{
\partial_t f + \dot{\v r} \partial_{\v r} f + \dot{\v p}_{clean} \partial_{\v p} f = \left . St[f]\right \vert_{full}} \label{eq:BoltzmannFull}
\end{equation}
with
\begin{equation}
\left . St[f]\right \vert_{full}  = \left . St[f]\right \vert_{(s)}+\left . St[f]\right \vert_{(a)}+\left . St[f]\right \vert_{W_E}+\left . St[f]\right \vert_{W_B},
\end{equation}
where
\begin{eqnarray}
\left . St[f]\right \vert_{(s)} &=& - \sum_{l'}  \omega_{l'l}^{(s)} \left (f_l - f_{l'}\right ) ,\\
\left . St[f]\right \vert_{(a)} &=& - \sum_{l'}  \omega_{l'l}^{(a)} \left (f_l + f_{l'} \right ) ,\\
\left . St[f]\right \vert_{W_E} &=& - \sum_{l'}  \omega_{ll'} \partial_{\epsilon_\xi} f_{l'}\delta \v r_{ll'} \frac{q \v E}{1- \frac{qB\Omega}{c}} , \\
\left . St[f]\right \vert_{W_B} &=& - \sum_{l'}  \omega_{ll'} \partial_{\epsilon_\xi} f_{l'}\delta \v r_{ll'} \frac{\underline \epsilon \tilde v \frac{qB}{c}}{1- \frac{qB\Omega}{c}} - \dot{\v p}_{sj} \partial_{\v p} f \notag \\
                                &=&    - \sum_{l'}  \omega_{ll'} \partial_{\epsilon_\xi} f_{l'}\delta \v r_{ll'} \frac{\underline \epsilon \tilde{\v v} \frac{qB}{c}}{1- \frac{qB\Omega}{c}} \notag \\
                                &&  - \sum_{l'}  \omega_{l'l} \partial_{\epsilon_\xi} f_{l} \delta \v r_{l'l} \frac{\underline \epsilon^T \v v^{(\xi)} \frac{qB}{c}}{1- \frac{qB\Omega}{c}}.
\end{eqnarray}
Here we have introduced the notation $\omega_{l'l}^{(s)} = (\omega_{l'l} + \omega_{ll'})/2$ and $\omega_{l'l}^{(a)} = (\omega_{l'l} - \omega_{ll'})/2$ and neglected accumulation of side-jump and skew-scattering effects (higher orders in $mv_0^2/\mu$). In the last line we used $\sum_{l'} \omega_{l'l} \delta r_{l'l} \propto \underline \epsilon \v v$ and dropped terms $\mathcal O (\omega_{ll'}^2)$ .
The velocity $\tilde{\v v}$ is a placeholder for $\v v^{(\xi)}$ or ${\v v^{(\xi)}}'$, because at this point, it is unclear whether $W = \dot{\v p} \delta \v r_{ll'}$ or $W = \dot{\v p}' \delta \v r_{ll'}$ should be chosen.

\subsection{Conservation laws at $\v E = 0$.}

\label{sec:app:ConsLaws}

Clearly, for $\v E = 0$ 
\begin{equation}
\sum_l \left . St[f]\right \vert_{full} = 0 \text{ and } \sum_l \epsilon(\v p) \left . St[f]\right \vert_{full} = 0
\end{equation}
provided
\begin{equation}
\delta \v r_{ll'} \underline \epsilon\left (\tilde{\v v} - {\v v^{(\xi)}}'\right ) = 0. \label{eq:conditionForConservation}
\end{equation}

Under the assumption that the side jump $\v r_{ll'}$ contains only terms proportional 
to $\v p - \v p'$ and $ (\v p + \v p') \v p \underline \epsilon \v p'$ [see Eqs. \eqref{eq:generalsidejumpformula} and \eqref{eq:SidejumpDirac} below] we find that the two solutions $\tilde{\v v} = -{\v v^{(\xi)}} $ and $\tilde{\v v} =  {\v v^{(\xi)}}'$ are legitimate. Both possible solutions lead to the same collision integral $\left . St[f]\right \vert_{full}$ [see again Eq. \eqref{eq:conditionForConservation} for the solution $\tilde{\v v} = - {\v v^{(\xi)}}$]. We will use $\tilde{\v v} = -{\v v^{(\xi)}} $, then
\begin{equation}
\left . St[f]\right \vert_{W_B} = - \sum_{l'}  \omega_{ll'}^{(s)} \delta \v r_{ll'} \frac{\underline \epsilon {\v v}^{(\xi)} \frac{qB}{c}}{1- \frac{qB\Omega_\xi}{c}} \left (\partial_{\epsilon_\xi} f_{l}- \partial_{\epsilon_\xi} f_{l'}\right ).
\end{equation}

\subsection{Back to the Boltzmann Equation }

The Boltzmann equation \eqref{eq:BoltzmannFull} at zero $\v E$ field is solved by any isotropic function $f_l = f_0$. The physical solution is the Fermi-Dirac distribution function. To access the static, homogeneous non-equilibrium distribution function we restrict ourselves to linear response and use the expansion in harmonics:

\begin{equation}
f_l = \sum_n f_n e^{in\phi}\; \leftrightarrow\; f_n = \int \frac{d\phi}{2\pi} f_l e^{-in\phi}.
\end{equation}

The left-hand side (LHS) of Eq. \eqref{eq:BoltzmannFull} becomes (using $E_\pm = E_x \pm i E_y$.)

\begin{eqnarray}
\dot{\v p}_{clean} \partial_{\v p } f &=& \sum_n \Big \lbrace e^{i (n+1)\phi} 
\left [(\xi v \partial_{\epsilon_\xi}f_n - nf_n/p) \frac{q E_-}{2}\right ] \notag \\
&& \phantom{\sum_n \Big \lbrace}+ e^{i (n-1)\phi} \left [(\xi v \partial_{\epsilon_\xi}f_n + nf_n/p) \frac{q E_+}{2}\right ] \notag \\
&& \phantom{\sum_n \Big \lbrace}+ e^{i n\phi} (- \zeta \Omega_c^{\rm cl} i n f_n) \Big \rbrace \frac{1}{1 - \frac{qB\Omega_\xi}{c}}.
\end{eqnarray}

The right-hand side (RHS) becomes

\begin{eqnarray}
\left . St[f]\right \vert_{(s)} =  - \left (1 - \frac{q B \Omega_\xi}{c}\right ) \sum_n \frac{f_n}{\tau^{(s)}_n}e^{in\phi}
\end{eqnarray}
with
\begin{equation}
\frac{1}{\tau^{(s)}_n} = \int \left (d p'\right ) \omega_{l'l}^{(s)} \left (1 - e^{i n (\phi' - \phi)}\right ).
\end{equation}
The first symmetric scattering rate $\frac{1}{\tau^{(s)}_1}$ is the transport rate $\frac{1}{\tau^{(s)}_1} = \frac{1}{\tau_{tr}}$.

The skew scattering contribution to the collision term is
\begin{equation}
\left . St[f]\right \vert_{(a)} = - \left (1 - \frac{q B \Omega_\xi}{c}\right ) \sum_n i\frac{f_n}{\tau_n^{(a)}}e^{in\phi}.
\end{equation}
It contains the skew-scattering rates
\begin{equation}
\frac{1}{\tau^{(a)}_n} = -i \int \left (d p'\right ) \omega_{l'l}^{(a)} \left [e^{i n (\phi' - \phi)} + 1\right ].
\end{equation}
The last term ``$+1$'' in the square bracket drops out in view of the definition of $\omega_{l'l}^{(a)}$.

The contribution of work by electrical field is (no accumulation of skew scattering and side jump)
\begin{eqnarray}
\left . St[f]\right \vert_{W_E} &=& - \sum_n \partial_{\epsilon_\xi} f_n \frac{\Omega_\xi p }{\tau_{sj}^{(n)}} \frac{i q}{2} \times \notag \\ && \times \left (e ^{i(n+1)\phi} E_- -e ^{i(n-1)\phi} E_+\right )
\end{eqnarray}
Higher harmonics of the mean side-jump time $\tau_{sj}^{(n)}$ are defined analogously to \eqref{eq:sidejumptimes}
\begin{equation}
\sum_{l'} \omega_{l'l}  \left ( \delta\v r_{l'l}\right )e^{in(\phi' - \phi)} = (1 - \frac{qB \Omega_\xi}{c})\frac{\Omega_\xi \underline \epsilon \v p}{\tau_{sj}^{(n)}} .
\end{equation}
and thus $\tau_{sj}^{(0)} = \tau^{sj}$ from Eq. \eqref{eq:sidejumptimes}.

The last contribution is the side-jump work by $B$-field. It reads
\begin{eqnarray}
\left . St[f]\right \vert_{W_B} &=& \sum_n \partial_{\epsilon_\xi} f_n e^{in \phi} \left (1 - \frac{q B \Omega_\xi}{c}\right ) \frac{\langle W_{l' \rightarrow l}^{(B)} \rangle}{\tau^{(W)}_{n}} . \notag \\
\end{eqnarray}
We introduced the averaged power
\begin{equation}
\frac{\langle W_{l' \rightarrow l}^{(B)} \rangle}{\tau^{(W)}_{n}} = \int (d\v p') \omega_{ll'} \delta \v r_{ll'} \frac{\underline \epsilon (-{\v v}^{(\xi)}) \frac{qB}{c}}{1- \frac{qB\Omega_\xi}{c}} \left (1 - e^{i n (\phi'-\phi)}\right ).
\end{equation}
The subscript $_{l' \rightarrow l}$ will be mostly omitted in the following.

In the linear response approximation, the Boltzmann equation involves only the $0$th and ($\pm 1$)st harmonics:

\begin{equation}
\frac{\tau_{tr}\xi v m_1 }{\left (1 - \frac{q B \Omega_\xi}{c}\right )^2} \left ( - \partial_{\epsilon_\xi} f_0 \right ) \frac{q E_-}{2} = m_2^{-1} f_1 - \partial_{\epsilon_\xi} f_1 \frac{\langle W^{(B)} \rangle \tau_{tr}}{\tau^{(W)}}. \label{eq:Boltzmannf1}
\end{equation}
In this equation, we introduced the complex functions $m_{1,2} = m_{1,2} (\epsilon_\xi) $ with
\begin{subequations}
\begin{eqnarray}
m_1 &=& \left [1+i \left (1 - \frac{q B \Omega_\xi}{c}\right ) \frac{\Omega_\xi k}{\tau_{sj} v \xi}\right ], \\
m_2 &=& \left\{1+i \left [\frac{\tau_{tr}}{\tau_a} - \frac{\zeta \Omega_c^{\rm cl} \tau_{tr}}{\left (1 - \frac{q B \Omega_\xi}{c}\right ) ^2}\right ]\right\}^{-1} .
\end{eqnarray}
\end{subequations}
For simplicity, we dropped the subscript $_{1}$ in all of the scattering rates, thus $\tau^{(W)}_{1} =\tau^{(W)}$, $\tau_1^{(a)} = \tau_a$ and we introduced the transport time $\tau_s^{(1)} = \tau_{tr}$.

\subsection{Solution of the Boltzmann equation}

\subsubsection{A representation of the delta function.}

\begin{figure}
\includegraphics[scale=.45]{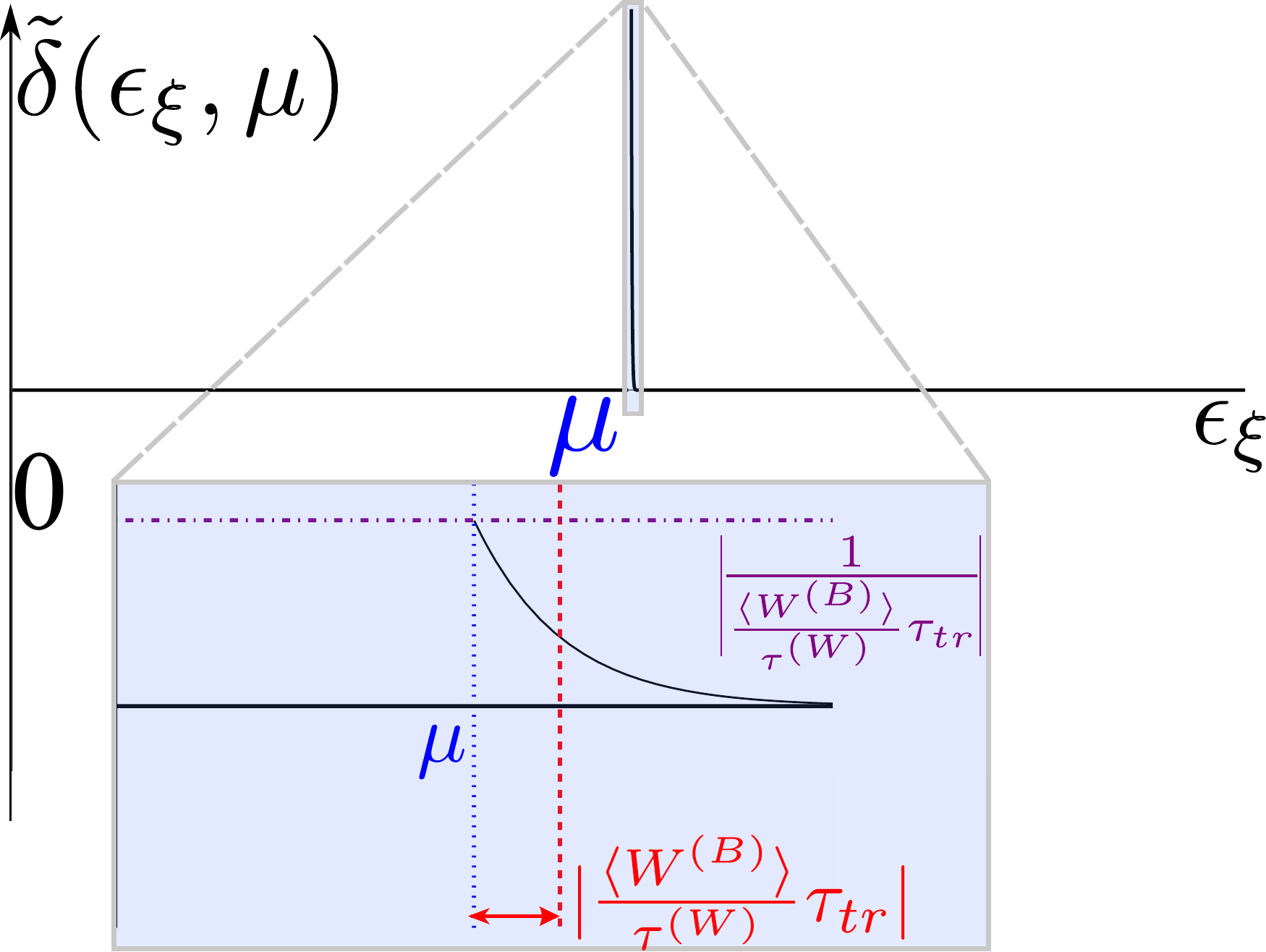}
\caption{The broadened delta function entering the general solution of the kinetic equation. In the inset, the same function is shown in the vicinity of the chemical potential, where it takes the value $\left \vert \tau_{tr}\langle W^{(B)} \rangle /\tau^{(W)}\right \vert^{-1}$.}
\label{fig:vardelta}
\end{figure}

For the solution of the kinetic equation, the following broadened delta function will be needed:
\begin{eqnarray}
\tilde{\delta} ( \epsilon_\xi, \epsilon'') &=& \text{sign}\left(\mathfrak{Re}\left[\frac{\langle W^{(B)} \rangle }{\tau^{(W)}}\right]\right) \notag \\
&\times & \frac{e^{- \int_{\epsilon_{\xi}}^{\epsilon''}\frac{d \epsilon'}{[\langle W^{(B)} \rangle m_2 \tau_{tr}/\tau^{(W)} ]\vert_{\epsilon'}}}}{[\langle W^{(B)} \rangle m_2 \tau_{tr}/\tau^{(W)} ] \vert_{\epsilon''}} \notag \\
&\times & \theta \left (\text{sign}(\mathfrak{Re}[\frac{\langle W^{(B)} \rangle }{\tau^{(W)}}])(\epsilon''- \epsilon_\xi)\right ). \label{eq:vardelta}
\end{eqnarray}
As a function of $\epsilon_\xi$ it is peaked at $\epsilon''$ and asymmetrically exponentially decaying into the direction prescribed by $\text{sign}(\mathfrak{Re}[\langle W^{(B)} \rangle /\tau^{(W)}])$. It is assumed, that this quantity is energy independent within a given band $\xi$, see Eq. \eqref{eq:WorkbyBfield} below for the case of Dirac electrons. Equation \eqref{eq:vardelta} is applied to the AHE; in the regime of applicability, ${\langle W^{(B)} \rangle }/{\tau^{(W)}}$ is smooth on the scale on which $\tilde \delta$ decays, see Fig. \ref{fig:vardelta}.


The broadened delta function leads to the following approximate convolutions for functions $f(\epsilon)$ which are smooth on the scale of $\tau_{tr}\langle W^{(B)}\rangle/\tau^{(W)}$:
\begin{eqnarray}
\int d \epsilon_\xi f(\epsilon_\xi) \tilde{\delta} ( \epsilon_\xi, \epsilon'') &\approx& f(\epsilon'') - \left (f \langle W^{(B)} \rangle m_2 \tau_{tr}/\tau^{(W)} \right )'_{\epsilon''} ,\notag \\&& \label{eq:tildedeltaapprox1} \\
\int d \epsilon'' f(\epsilon'') \tilde{\delta} ( \epsilon_\xi, \epsilon'') &\approx& f(\epsilon_\xi) + \left (f \langle W^{(B)} \rangle m_2 \tau_{tr}/\tau^{(W)} \right )'_{\epsilon_\xi} .\notag \\
\end{eqnarray}

\subsubsection{General linear response solution.}

The general linear response solution for equation \eqref{eq:Boltzmannf1} is
\begin{eqnarray}
f_1 ( \epsilon_\xi ) &=& \int_{-\infty} ^{\infty} d\epsilon'' \Big \lbrace \Big [\frac{\tau_{tr}}{\left (1 - {q B \Omega_\xi}/{c}\right )^2} \xi v m_1 m_2 \left ( - \partial_{\epsilon''} f_0 \right )\Big ]_{\epsilon''} \notag \\ && \times \tilde \delta (\epsilon_\xi, \epsilon'') \Big \rbrace \frac{q E_-}{2}.
\end{eqnarray}
Formally, there is also an exponentially growing solution which has been dropped for obvious physical reasons.
In the limit $T\gg \vert \langle W^{(B)} \rangle \vert $ the approximate solution is
\begin{eqnarray}
f_1 ( \epsilon_\xi ) &=& \left [\frac{\tau_{tr}}{\left (1 - \frac{q B \Omega_\xi}{c}\right )^2} \xi v m_1 m_2 \left ( - \partial_{\epsilon_\xi} f_0 \right )\right ]_{\epsilon_\xi} \frac{q E_-}{2} \notag\\
&+&  \left [\langle W^{(B)} \rangle m_2 \tau_{tr}/\tau^{(W)} \right ]_{\epsilon_\xi}  \notag \\
& \times & \partial_{\epsilon_\xi}\left [\frac{\tau_{tr}}{\left (1 - \frac{q B \Omega_\xi}{c}\right )^2} \xi v m_1 m_2 \left ( - \partial_{\epsilon_\xi} f_0 \right )\right ]_{\epsilon_\xi} \frac{q E_-}{2}. \notag \\ && \label{eq:f1TllW}
\end{eqnarray}
This solution can also be obtained by iteratively solving Eq. \eqref{eq:Boltzmannf1}.
In the limit when temperature $T$ is smaller than all other scales, we can use the zero temperature solution
\begin{equation}
f_1 ( \epsilon_\xi ) = \tilde \delta (\epsilon_\xi, \mu) \left [\frac{\tau_{tr}}{\left (1 - \frac{q B \Omega_\xi}{c}\right )^2} \xi v m_1 m_2 \right ]_{\mu} \frac{q E_-}{2}.
\end{equation}
When convoluted with a function $f ( \epsilon_\xi )$ which is smooth on the scale of the magnetic work (for example the current), $f_1 ( \epsilon_\xi )$ will be approximated according to Eq. \eqref{eq:tildedeltaapprox1}. By comparison with Eq. \eqref{eq:f1TllW} we see, that the results for the current in the limits $ \vert \langle W^{(B)} \rangle \vert \ll T \ll\mu  $ and $T\ll \vert \langle W^{(B)} \rangle \vert \ll\mu $ coincide.

\subsection{Conductivity at $T=0$ and $\mu \gg T\gg \vert \langle W^{(B)} \rangle \vert $}

\subsubsection{Intrinsic contribution.}
As explained, the total current density also has a contribution of the filled bands (intrinsic AHE):

\begin{equation}
\v j^{intr.} = \sum_l \frac{- \Omega_{\xi} q^2 \underline \epsilon \v E}{1 - \frac{\Omega_{\xi} q B}{c}} f_{0,l} = \sigma_{xy}^{intr.} \underline{\epsilon} \v E, \label{eq:intrinsicHallgeneral}
\end{equation}

where in the case of Dirac fermions\cite{LudwigGrinstein1994,SinitsynSinova2007}
\begin{eqnarray}
\sigma_{xy}^{intr.} &=& -\frac{q^2}{4\pi}\Big [\text{sign}(m) \theta( m^2 v_0^4 -\mu^2) \notag \\ &&+ \frac{mv_0^2}{\vert \mu \vert} \theta(\mu^2- m^2 v_0^4)  \Big ]. \label{eq:intrinsicHallconductivity}
\end{eqnarray}

\subsubsection{Non-equilibrium contribution.}
The longitudinal and transverse conductivity are
\begin{equation}
\sigma_{xx} = \mathfrak{Re}\sigma(\mu) \text{ and } \sigma_{xy} = \mathfrak{Im}\sigma(\mu),
\end{equation}
where the complex function $\sigma(\mu)$ is
\begin{eqnarray}
{\sigma(\mu)}&=& {\Big [ \sigma_{xx}^{(\not B)} \frac{m_1^2m_2}{\left (1 - \frac{q B \Omega_\xi}{c}\right )^2} }\notag \\
&&{-  \partial_\mu \left (\sigma_{xx}^{(\not B)} \frac{\langle W^{(B)}\rangle}{\tau^{(W)} \xi v}  m_1 m_2 \right ) \frac{v \xi \tau_{\tr} m_1 m_2}{\left (1 - \frac{q B \Omega_\xi}{c}\right )^2}\Big ]}. \notag \\ \label{eq:sigmaofmu}
\end{eqnarray}
In this expression we used the expression for the conductivity in zero magnetic field:
\begin{equation}
\sigma_{xx}^{(\not B)} (\mu) = q ^2 \nu(\mu) \underbrace{\frac{v^2(\mu) \tau_{tr} (\mu)}{2}}_{D(\mu)} , 
\end{equation}
where $\nu(\mu)$ is the density of states.

\subsection{Evaluation for Dirac fermions}
\label{app:sec:Semiclassical:EvaluationDirac}

While the solution given in Eqs.~\eqref{eq:intrinsicHallgeneral} and~\eqref{eq:sigmaofmu} is a priori general (not restricted to the situation of Dirac fermions) we now return to the case of 3D TI surface states.
The various Fermi surface contributions are
\begin{eqnarray}
m_1 
&=& 1-i \left (1 + \frac{1}{2} \frac{mv_0^2}{\mu} \frac{\zeta \Omega_c^{\rm cl}}{\mu} \right )  \frac{1}{2}\frac{mv_0^2}{\mu} \frac{1}{\mu\tau_{sj}} \notag \\
&\approx & 1 -i  \frac{1}{2}\frac{mv_0^2}{\mu} \frac{1}{\mu\tau_{sj}}.
\end{eqnarray}
The approximation $\approx$ keeps only the leading order $\mathcal O(\frac{mv_0^2}{\mu},\frac{\Omega_c^{\rm cl}}{\mu})$. Note that the imaginary part (the side jump contribution) is small in $1/k_Fl$. Next,
\begin{eqnarray}
m_2  
&=& \frac{1 - i \left [\frac{\tau_{tr}}{\tau_a} - \frac{\zeta \Omega_c^{\rm cl} \tau_{tr}}{\left (1 - \frac{q B \Omega_\xi}{c}\right ) ^2}\right ]}{1 +\left [\frac{\tau_{tr}}{\tau_a} - \frac{\zeta \Omega_c^{\rm cl} \tau_{tr}}{\left (1 - \frac{q B \Omega_\xi}{c}\right ) ^2}\right ]^2} \notag \\
&\approx&  \frac{1 + i \zeta \Omega^{\rm cl}_c\tau_{tr}}{1 + (\Omega^{\rm cl}_c\tau_{tr})^2} \left [1 + 2\frac{\zeta \Omega^{\rm cl}_c\tau_{tr}}{1 + (\Omega^{\rm cl}_c\tau_{tr})^2} \frac{\tau_{tr}}{\tau_{a}}\right ] \notag \\ &&- i \frac{\frac{\tau_{tr}}{\tau_{a}}}{1 + (\Omega^{\rm cl}_c\tau_{tr})^2} .
\end{eqnarray}

\subsubsection{Scattering rates in leading approximation: Short-range impurities.}

The symmetric scattering matrix element is
\begin{eqnarray}
\omega_{ll'}^{(s)} &=& 2\pi n_i V_0^2 \delta(\epsilon_\xi (\v p) - \epsilon_\xi (\v p'))\notag \\
&\times&\left [\cos^2 \left (\frac{\phi- \phi'}{2}\right ) + \left (\frac{mv_0^2}{\epsilon_\xi}\right )^2\sin^2 \left (\frac{\phi- \phi'}{2}\right ) \right ],\notag \\
\end{eqnarray}
where $n_i$ and $V_0$ are concentration respectively strength of short ranged impurities.
The transport rate evaluated at the chemical potential immediately follows,
\begin{eqnarray}
\frac{1}{\tau_{tr}} &=& \int (dp') \omega_{ll'}^{(s)} \left [1 - \cos( \phi' - \phi)\right ] \notag \\
&=& 2\pi n_i V_0^2 \nu \frac{1 + 3 \left (\frac{mv_0^2}{\mu}\right )^2}{4}.
\end{eqnarray}
According to Ref.~\onlinecite{SinitsynSinova2007} the side jump is
\begin{equation}
\delta\v r_{l'l} = \frac{\Omega_\xi \underline \epsilon (\v p - \v p') }{\vert \braket{u_{\xi, \v p}\vert u_{\xi, \v p'}}\vert^2} \label{eq:SidejumpDirac}
\end{equation}
and thus the side jump rate follows to be
\begin{eqnarray}
\frac{1}{\tau^{sj}} &=& 2\pi n_i V_0^2 \nu.
\end{eqnarray}
This is the same as the quantum rate in a normal material (the quantum rate is different for the Dirac problem).
We also refer to Ref.~\onlinecite{SinitsynSinova2007} for the skew scattering rate, which is
\begin{eqnarray}
\frac{1}{\tau_{a}} &=& \frac{\pi \nu(\mu)}{2} \Big[ \frac{ n_i V_1^3 m \left(\mu^2-m^2v_0^4\right)}{2 \mu ^2} \notag \\
&&+ \frac{(n_i V_0^2)^2 \left(3  m \left(\mu ^2-m^2v_0^4\right)\right)}{4 \mu ^3}\Big].
\end{eqnarray}
Both terms in the square bracket are manifestly beyond Born approximation (the first term involves the third moment of the disorder potential $V_1^3$.)

The power provided by the B-field is
\begin{eqnarray}
\frac{\langle W^{(B)} \rangle}{\tau^{(W)}} &=& 2\pi n_i V_0^2  \nu \frac{ \frac{qB\Omega_\xi}{c} }{1- \frac{qB\Omega_\xi}{c}} \xi v p \frac{3}{2} \notag \\
&=& \frac{3 v_0^2 p^2}{2\mu \tau_{sj}} \left (-\frac{\frac{1}{2} \frac{mv_0^2}{\mu} \frac{\zeta \Omega_c^{\rm cl}}{\mu}}{1 + \frac{1}{2} \frac{mv_0^2}{\mu} \frac{\zeta \Omega_c^{\rm cl}}{\mu}} \right ). \label{eq:WorkbyBfield}
\end{eqnarray}
In the case of short-range impurities we can omit the contribution of $\langle W^{(B)} \rangle$ to the conductivity since
\begin{equation}
\frac{\langle W^{(B)} \rangle}{\tau^{(W)}} \frac{\tau_{tr}}{\mu} \sim \frac{ v_0^2 p^2}{\mu^2} \frac{\tau_{tr}}{\tau_{sj}} \frac{mv_0^2}{\mu} \frac{\Omega_c^{\rm cl}}{\mu} \approx 0
\end{equation}
is beyond leading order in $\mathcal O(\frac{mv_0^2}{\mu},\frac{\Omega_c^{\rm cl}}{\mu})$.

\section{Magnetic mirror charge for a double QH structure}
\label{sec:app:finitethickness}

\begin{figure}
\includegraphics[scale=.45]{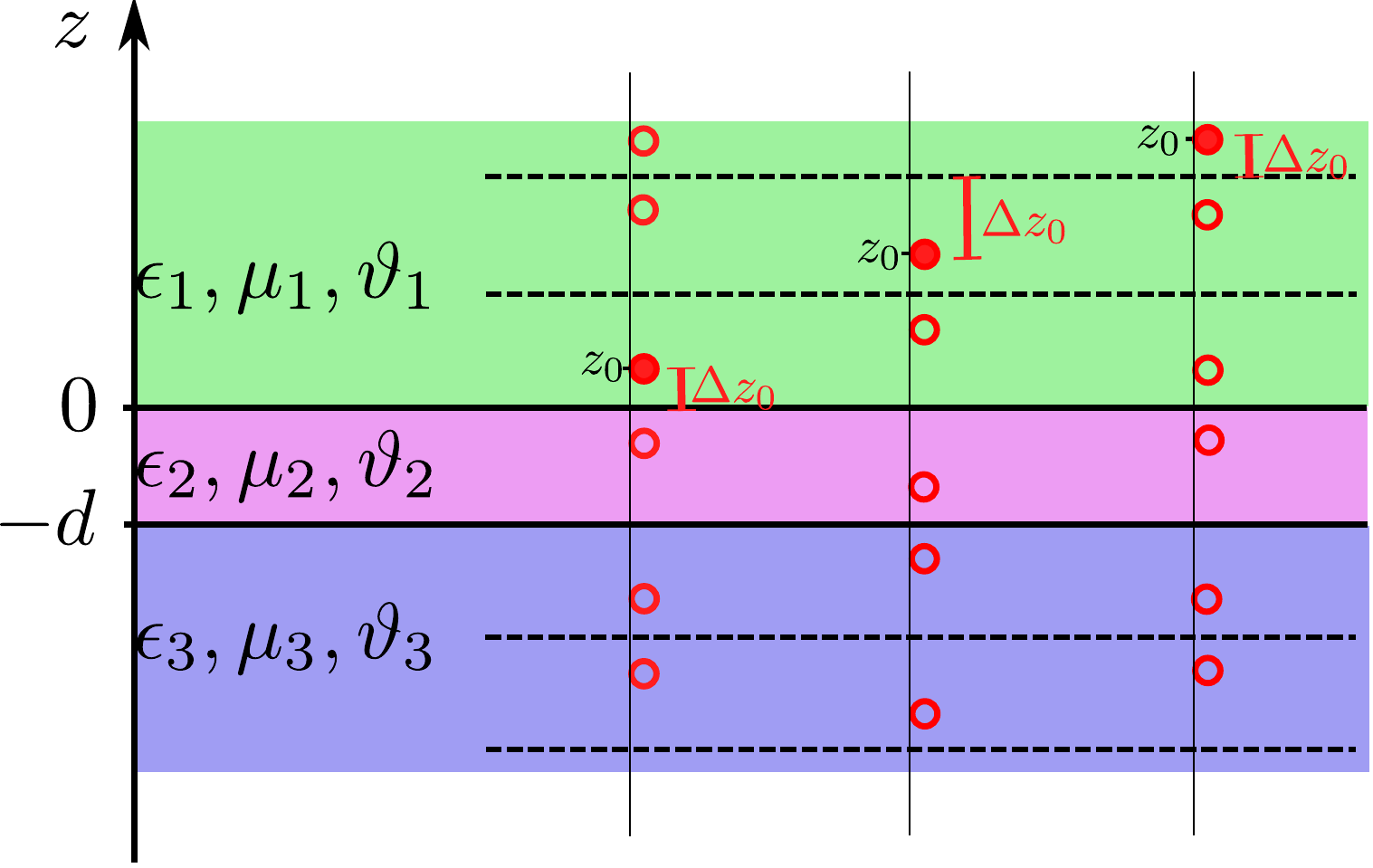}
\caption{Sketch of three different scenarios for the setup discussed in appendix \ref{sec:app:finitethickness}. The left and right scenario correspond to case I, the middle one to case II.}
\label{fig:app:Sketchimagecharges}
\end{figure}

In this appendix we consider the image magnetic monopole effect for a double QH structure (a double domain-wall of $\v E \cdot \v B$ states).  We consider the setup as in Fig. \ref{fig:app:Sketchimagecharges} and define the following three regions in real space
\begin{itemize}
\item[\ding{192}] $= \left \lbrace \v r \in \mathbb R^3 \vert 0 < z \right \rbrace$,
\item[\ding{193}] $= \left \lbrace \v r \in \mathbb R^3 \vert -d \leq z \leq 0\right \rbrace$,
\item[\ding{194}] $= \left \lbrace \v r \in \mathbb R^3 \vert z < -d \right \rbrace$.
\end{itemize}

Region \ding{193} might correspond to the 3D TI, its surfaces should be characterized by a QH-state with $\sigma_{xx}  =0$ and definite $\sigma_{xy}$. Equivalently, one can can describe the three regions $a \in \left\lbrace \right .$ \ding{192},\ding{193},\ding{194}$\left .\right \rbrace$ by definite bulk $\vartheta_a$. In addition, localized charges might contribute to non-trivial $\epsilon_a$ and $\mu_a$.

\subsection{Positions of the mirror charges}

Let $z_0 >0$ denote the position of the actual charge. We will need the quantity $\tilde z_0 = \left \lbrace \frac{z_0}{2d} \right \rbrace \times 2d $ (curly brackets denote the fractional part of a real number). We have to consider two separate cases
\begin{itemize}
\item[I.] Let $z_0 \in \left [2k d, \left (2k+1\right ) d\right ]$ with $k \in \mathbb N_0$\\
We define, according to Fig. \ref{fig:app:Sketchimagecharges}, $\Delta z_0 \equiv \tilde z_0 < d$.
\item[II.] Let $z_0 \in \left [\left (2k - 1\right ) d, 2k d\right ]$ with $k \in \mathbb N$\\
In this case, by definition and according to Fig. \ref{fig:app:Sketchimagecharges}, $\Delta z_0 \equiv 2d - \tilde z_0 < d$.
\end{itemize}
In both cases, the position of (mirror) charges is thus given by (see again Fig. \ref{fig:app:Sketchimagecharges})
\begin{equation}
z_m^\pm \equiv 2m d \pm  \Delta z_0 \; (m \in \mathbb Z).
\end{equation}
By convention, the defining tuples $\left (m,s\right )$ (with $m \in \mathbb Z$ and $s = \pm$) are ordered by the order implied of $z_m^s$ [e.g., since $z_2^- < z_2^+$ the following inequality holds: $(2,-) < (2,+)$].

Clearly, depending on their $(m,s)$-values, the charges reside in the following regions:
\begin{itemize}
\item[\ding{192}:] $(m,s) \geq (0,+)$,
\item[\ding{193}:] $(m,s) = (0,-)$,
\item[\ding{194}:] $(m,s) \leq (-1,+)$.
\end{itemize}
In case I, the actual charge sits at $z_0 = z_{\left \lfloor \frac{z_0}{2d} \right \rfloor}^+$ while for case II $z_0 = z_{\left \lceil \frac{z_0}{2d} \right \rceil}^-$ follows. The symbols $\lfloor \dots \rfloor$ and $\lceil \dots \rceil$ denote floor and ceiling functions.

\subsection{Solution of the image charge problem for the thin film}

Following Karch,\cite{Karch09} we use the unified description in terms of the vector $\left (\v D, 2\alpha \v B\right )$ which is connected to $\v E$ and $\v H$ via
\begin{equation}
\left (\begin{array}{c}
\v D \\ 2\alpha \v B
\end{array} \right ) = \mathcal M \left (\begin{array}{c}
2\alpha \v E \\ \v H
\end{array} \right )
\end{equation}
with the matrix 
$$\mathcal M = \frac{2\alpha}{c^2\epsilon} \left (\begin{array}{cc}
\frac{\vartheta^2}{4\pi^2} + (\frac{c\epsilon}{2\alpha})^2 & -\frac{\vartheta}{2\pi} \\
-\frac{\vartheta}{2\pi} & 1
\end{array} \right )$$ 
in each of the three regions $a \in \left\lbrace \right .$ \ding{192},\ding{193},\ding{194}$\left .\right \rbrace$.
We make the following Ansatz for the potential $\underline \Phi = \left (\Phi_E, 2\alpha \Phi_M\right )$ with $\left (\v D, 2\alpha \v B\right ) = - \nabla \underline \Phi$:

\begin{align}
\underline \Phi_\text{\ding{192}} &= \sum_{n = -\infty}^\infty \sum_{s = \pm} \frac{\underline A_n^{(s)}}{\vert \v x - z_n^{(s)} \hat e_z\vert}, \\
\underline \Phi_\text{\ding{193}} &= \sum_{n = -\infty}^\infty \sum_{s = \pm} \frac{\underline B_n^{(s)}}{\vert \v x - z_n^{(s)} \hat e_z\vert} ,\\
\underline \Phi_\text{\ding{194}} &= \sum_{n = -\infty}^\infty \sum_{s = \pm} \frac{\underline C_n^{(s)}}{\vert \v x - z_n^{(s)} \hat e_z\vert} .
\end{align}
In order to fulfill the Poisson/Laplace equation the series of (mirror) charges (defined each as $\underline A = \left (A_E, 2\alpha A_M\right )$ etc.) has the form
\begin{align}
\left (\underline B_n^{(s)}\right ) &= \left (\dots \underline B_{-1}^{-},\underline B_{-1}^{+};0;\underline B_{0}^{+},\underline B_{1}^{-},\dots  \right ) ,\\
\left (\underline C_n^{(s)}\right ) &= \left (\dots, 0,0;\underline C_{0}^{-};\underline C_{0}^{+},\underline C_{1}^{-}, \dots\right ) .
\end{align}
Further, in case I
\begin{equation}
\left (\underline A_n^{(s)}\right ) = \left (\dots \underline A_{-1}^{-},\underline A_{-1}^{+}; \underline A_{0}^{-};0, \dots, 0, \underline A_{\left \lfloor \frac{z_0}{2d} \right \rfloor}^+, 0, \dots \right )
\end{equation}
while in case II
\begin{equation}
\left (\underline A_n^{(s)}\right ) = \left (\dots \underline A_{-1}^{-},\underline A_{-1}^{+};\underline A_{0}^{-};0, \dots, 0, \underline A_{\left \lceil \frac{z_0}{2d} \right \rceil}^-, 0, \dots \right ).
\end{equation}
In these sequences, elements left to the first semicolon are associated to region \ding{194}, the element in between of the two semicola resides in region \ding{193}, while elements on its right are associated to charges in the positive half-plane, region \ding{192}.
Clearly $\underline A_{\left \lfloor \frac{z_0}{2d} \right \rfloor}^+$ and $\underline A_{\left \lceil \frac{z_0}{2d} \right \rceil}^-$ are given by the ``bare''(actual) value of the charge $\underline Q$ placed close to the interface. 

\subsubsection{Derivation of recursion relations}
The position of mirror charges is constructed such that a reflection at the interface \ding{192}-\ding{193} implies $z_m^{(s)}\rightarrow z_{-m}^{(-s)}$ and at a reflection at the interface \ding{193}-\ding{194} $z_m^{(s)}\rightarrow z_{-(m+1)}^{(-s)}$ ($-s = \mp$ for $s = \pm$). Then the continuity of perpendicular components of $\left (\v D , 2\alpha \v B \right )$ and parallel components of $\left (2\alpha \v E , \v H
\right )$ yields the following infinite series of conditions
\begin{subequations}
\begin{align}
\underline A_{n}^{(s)} - \underline A_{-n}^{(-s)} &= \underline B_{n}^{(s)} - \underline B_{-n}^{(-s)} ,\label{eq:condAB1}\\
\mathcal M_1^{-1} \left (\underline A_{n}^{(s)} + \underline A_{-n}^{(-s)}\right ) &= \mathcal M_2^{-1}\left ( \underline B_{n}^{(s)} + \underline B_{-n}^{(-s)} \right ), \label{eq:condAB2}
\end{align}
\end{subequations}
with $\left (n,s\right ) \in $ \ding{192}, and
\begin{subequations}
\begin{align}
\underline B_{n}^{(s)} - \underline B_{-(n+1)}^{(-s)} &= \underline C_{n}^{(s)} - \underline C_{-(n+1)}^{(-s)} ,\label{eq:condBC1}\\
\mathcal M_2^{-1} \left (\underline B_{n}^{(s)} + \underline B_{-(n+1)}^{(-s)}\right ) &= \mathcal M_3^{-1}\left ( \underline C_{n}^{(s)} + \underline C_{-(n+1)}^{(-s)} \right ), \label{eq:condBC2}
\end{align}
\end{subequations}
with $\left (n,s\right ) \in $ \ding{192} $\cup$ \ding{193}. In this region $\underline C_{-(n+1)}^{(-s)} = 0$ and therefore \eqref{eq:condBC1} and \eqref{eq:condBC2} lead to\
\begin{subequations}
\begin{align}
 0 &= \left (1+\mathcal M_3 \mathcal M_2^{-1}\right ) \underline B_{-1}^+ ,\\
 \left (1-\mathcal M_3 \mathcal M_2^{-1}\right ) \underline B_{n}^{(s)} &= \left (1+\mathcal M_3 \mathcal M_2^{-1}\right ) \underline B_{-(n+1)}^{(-s)} ,
\end{align}
\end{subequations}
where $\left (n,s\right ) \in \text{\ding{192}}$.
We can plug this knowledge on $\underline B$'s back into \eqref{eq:condAB1} and \eqref{eq:condAB2} leading to the following final relations:\\
\underline{{``Initial conditions'':}}
\begin{subequations}
\begin{align}
			R_{21}^+ \underline A_0^- &= R_{21}^- \underline A_0^+ ,\\
			R_{21}^+ \underline A_{-1}^+ &= R_{21}^- \underline A_1^-.
\end{align}
\end{subequations}
\underline{{	``Recursive relations'':}} [$(n,s) \in$ \ding{192}]
\begin{align}
		R_{32}^-R_{21}^- \underline A_{-n}^{(-s)}  &+ R_{32}^+R_{21}^+ \underline A_{-(n+1)}^{(-s)} \notag \\
		&= R_{32}^-R_{21}^+ \underline A_{n}^{(s)}  + R_{32}^+R_{21}^- \underline A_{n+1}^{(s)} 	.
\end{align}
Here we have defined $R_{ab}^\pm = 1 \pm \mathcal M_a \mathcal M_b^{-1}$.

\subsubsection{Solution of recursion relations}

The general solution of these relations for a charge sitting at $z_{n_0}^{(s_0)}$ is
\begin{subequations}
\begin{align}
\underline A_{n_0}^{(s_0)} &= \underline Q ,\\
\underline A_{n}^{(s_0)} &= 0 	\; \forall n\neq n_0 ,\\
\underline A_{n}^{(-s_0)} &=  0	\; \forall n > - n_0  ,\\
\underline A_{-n_0}^{(-s_0)} &= \left (R_{21}^+\right )^{-1}\left (R_{21}^-\right )\underline Q ,\\
\underline A_{-(n_0+l)}^{(-s_0)} &= \left (-\right )^{l-1} \left [\left (R_{21}^+\right )^{-1}\left (R_{32}^+\right )^{-1}\left (R_{32}^-\right )\left (R_{21}^-\right )\right ]^{l}  \notag \\
			&\times \left [\left (R_{21}^-\right )^{-1}\left (R_{21}^+\right ) - \left (R_{21}^+\right )^{-1}\left (R_{21}^-\right )\right ]\underline Q \; \forall l\geq 1.
\end{align}
\end{subequations}

\subsubsection{Limits and checks}
Two simple checks of the correctness of the result are in order.
\begin{itemize}
\item[1.] Let $\left (\epsilon_2,\mu_2, \vartheta_2\right ) = \left (\epsilon_3, \mu_3, \vartheta_3\right )$. Then $R_{23}^- = 0$. It follows that the only non-trivial mirror-charge is
\begin{eqnarray}
A_{-n_0}^{(-s_0)} &=& \left (R_{21}^+\right )^{-1}\left (R_{21}^-\right )\underline Q \notag \\ &=& \left (\mathcal M_1\mathcal M_2^{-1} + 1\right )^{-1}\left (\mathcal M_1\mathcal M_2^{-1} - 1\right )\underline Q, \notag\\
\end{eqnarray}
in accordance with Ref.~\onlinecite{Karch09}.
\item[2.] Let $\left (\epsilon_1, \mu_1, \vartheta_1\right )=\left (\epsilon_2,\mu_2, \vartheta_2\right ) $. Then $R_{21}^- = 0$ and $R_{21}^+ = 2$. It follows that the only non-trivial mirror-charge is
\begin{equation}
A_{-(n_0-1)}^{(-s_0)} = \left (R_{32}^+\right )^{-1}\left (R_{32}^-\right )\underline Q ,
\end{equation}
in accordance with the previous limit and Ref.~\onlinecite{Karch09}.
\end{itemize}

\subsubsection{The Potential and its Fourier transform}

We can thus write
\begin{eqnarray}
\underline \Phi_{\text{\ding{192}}} \left (\v x,z_0\right ) 	&=& \frac{1}{\vert \v x - z_0 \hat e_z\vert} \underline Q \notag \\
							&+& \frac{1}{\vert \v x + z_0 \hat e_z\vert} \left (R_{21}^+\right )^{-1}\left (R_{21}^-\right )\underline Q \notag \\
							&-&	\sum_{l=1}^{\infty} \frac{\left [-\left (R_{21}^+\right )^{-1}\left (R_{32}^+\right )^{-1}\left (R_{32}^-\right )\left (R_{21}^-\right )\right ]^{l}}{\vert \v x + \left (z_0 + 2ld\right ) \hat e_z\vert} \notag \\ &\times & \left [\left (R_{21}^-\right )^{-1}\left (R_{21}^+\right ) - \left (R_{21}^+\right )^{-1}\left (R_{21}^-\right )\right ]\underline Q .\notag
\end{eqnarray}
In Fourier space (Fourier transform only with respect to x and y coordinates) $\underline \Phi_{\text{\ding{192}}}$ simplifies
\begin{eqnarray}
\underline \Phi_{\text{\ding{192}}} \left (q,z,z_0\right ) &=& \frac{2\pi}{q} \bigg \lbrace e^{-\vert z-z_0 \vert q} \notag \\
												&+&  e^{-(z+z_0)q} T_{\rm eff} \bigg \rbrace \underline Q. \notag
\end{eqnarray}
We introduced the matrix
\begin{eqnarray}
T_{\rm eff} &=&  \left (R_{32}^+R_{21}^+e^{dq}+R_{32}^-R_{21}^-e^{-dq}\right )^{-1} \notag \\ &&\times \left (R_{32}^+R_{21}^-e^{dq}+R_{32}^-R_{21}^+e^{-dq}\right ) .
\end{eqnarray}
In view of 2D rotational invariance, the potential only depends on $q = \vert \vec q \vert$ (in this appendix 2D vectors are denoted by arrows). One can exploit this formula and Fourier transform back to real space
\begin{equation}
\underline \Phi_{\text{\ding{192}}} \left (\v x,z_0\right ) = \int_0^\infty dq \;\frac{q }{2\pi} \underline \Phi_{\text{\ding{192}}}  \left (q,z,z_0\right )J_0\left (q\rho\right )
\end{equation}
where $J_0\left (q\rho\right )$ is the zeroth Bessel function and $\rho = \vert \vec x \vert$ is the norm of the 2D component of $\v x$ perpendicular to $\hat e_z$.

This concludes the derivation of Eq.~\eqref{eq:DoubleQHpotentialFT} of the main text.

\subsubsection{Further limits and checks}

With the help of $\underline \Phi_{\text{\ding{192}}}$ in Fourier space and the matrix $T_{\rm eff}$, one can easily check the $d\rightarrow \infty$ and  $d\rightarrow 0$ limits.

First consider $d\rightarrow \infty$. As expected, we obtain a single mirror charge at $z = -z_0$ with charge $\left (R_{21}^+\right )^{-1}R_{21}^- \underline Q$.

Now consider $d\rightarrow 0$. After a bit of algebra exploiting the definition of $R_{ab}^{\pm}$, we obtain the expected result: a single mirror charge at $z = -z_0$ with charge $\left (R_{31}^+\right )^{-1}R_{31}^- \underline Q$. (The same result, as if region \ding{193} never existed.)


\end{document}